\documentclass[traditabstract,longauth]{aa}
\usepackage{epsfig}
\usepackage{graphicx}
\usepackage{url}
\usepackage[dvips]{color}
\usepackage{natbib}
\usepackage[breaklinks,colorlinks, citecolor=blue]{hyperref}
\usepackage{breakurl}
\usepackage{amssymb,,amsmath}
\usepackage{longtable}
\usepackage{multirow}
\bibpunct{(}{)}{;}{a}{}{,}

\newcommand{\etal}{{et~al.~}}
\newcommand{\Var}[1]{\mathrm{Var}(#1)}
\def\lsim{\mathrel{\raise .4ex\hbox{\rlap{$<$}\lower 1.2ex\hbox{$\sim$}}}}
\def\gsim{\mathrel{\raise .4ex\hbox{\rlap{$>$}\lower 1.2ex\hbox{$\sim$}}}}
\def\arcm{\ifmmode {^{\scriptscriptstyle\prime}}
          \else $^{\scriptscriptstyle\prime}$\fi}
\def\Planck{{\it Planck}}

\def\setsymbol#1#2{\expandafter\def\csname #1\endcsname{#2}}
\def\getsymbol#1{\csname #1\endcsname}

\def\Planck{{\it Planck\/}}

\def\allearlypapers{\nocite{planck2011-1.1, planck2011-1.3, planck2011-1.4, planck2011-1.5, planck2011-1.6, planck2011-1.7, planck2011-1.10, planck2011-1.10sup, planck2011-5.1a, planck2011-5.1b, planck2011-5.2a, planck2011-5.2b, planck2011-5.2c, planck2011-6.1, planck2011-6.2, planck2011-6.3a, planck2011-6.4a, planck2011-6.4b, planck2011-6.6, planck2011-7.0, planck2011-7.2, planck2011-7.3, planck2011-7.7a, planck2011-7.7b, planck2011-7.12, planck2011-7.13}}

\newbox\tablebox    \newdimen\tablewidth
\def\leaderfil{\leaders\hbox to 5pt{\hss.\hss}\hfil}
\def\endPlancktable{\tablewidth=\columnwidth 
    $$\hss\copy\tablebox\hss$$
    \vskip-\lastskip\vskip -2pt}

\def\tablenote#1 #2\par{\begingroup \parindent=0.8em
    \abovedisplayshortskip=0pt\belowdisplayshortskip=0pt
    \noindent
    $$\hss\vbox{\hsize\tablewidth \hangindent=\parindent \hangafter=1 \noindent
    \hbox to \parindent{\sup{\rm #1}\hss}\strut#2\strut\par}\hss$$
    \endgroup}
\def\doubleline{\vskip 3pt\hrule \vskip 1.5pt \hrule \vskip 5pt}

\def\L2{\ifmmode L_2\else $L_2$\fi}

\def\DeltaT{\ifmmode \Delta T\else $\Delta T$\fi}
\def\deltat{\ifmmode \Delta t\else $\Delta t$\fi}
\def\fknee{\ifmmode f_{\rm knee}\else $f_{\rm knee}$\fi}
\def\Fmax{\ifmmode F_{\rm max}\else $F_{\rm max}$\fi}
\def\solar{\ifmmode{\rm M}_{\mathord\odot}\else${\rm M}_{\mathord\odot}$\fi}

\def\inv{\ifmmode^{-1}\else$^{-1}$\fi}
\def\mo{\ifmmode^{-1}\else$^{-1}$\fi}
\def\sup#1{\ifmmode ^{\rm #1}\else $^{\rm #1}$\fi}
\def\expo#1{\ifmmode \times 10^{#1}\else $\times 10^{#1}$\fi}
\def\,{\thinspace}
\def\lsim{\mathrel{\raise .4ex\hbox{\rlap{$<$}\lower 1.2ex\hbox{$\sim$}}}}
\def\gsim{\mathrel{\raise .4ex\hbox{\rlap{$>$}\lower 1.2ex\hbox{$\sim$}}}}

\def\simprop{\mathrel{\raise .4ex\hbox{\rlap{$\propto$}\lower 1.2ex\hbox{$\sim$}}}}
\def\deg{\ifmmode^\circ\else$^\circ$\fi}
\def\pdeg{\ifmmode $\setbox0=\hbox{$^{\circ}$}\rlap{\hskip.11\wd0 .}$^{\circ}
          \else \setbox0=\hbox{$^{\circ}$}\rlap{\hskip.11\wd0 .}$^{\circ}$\fi}
\def\arcs{\ifmmode {^{\scriptstyle\prime\prime}}
          \else $^{\scriptstyle\prime\prime}$\fi}
\def\arcm{\ifmmode {^{\scriptstyle\prime}}
          \else $^{\scriptstyle\prime}$\fi}
\newdimen\sa  \newdimen\sb
\def\parcs{\sa=.07em \sb=.03em
     \ifmmode \hbox{\rlap{.}}^{\scriptstyle\prime\kern -\sb\prime}\hbox{\kern -\sa}
     \else \rlap{.}$^{\scriptstyle\prime\kern -\sb\prime}$\kern -\sa\fi}
\def\parcm{\sa=.08em \sb=.03em
     \ifmmode \hbox{\rlap{.}\kern\sa}^{\scriptstyle\prime}\hbox{\kern-\sb}
     \else \rlap{.}\kern\sa$^{\scriptstyle\prime}$\kern-\sb\fi}
\def\ra[#1 #2 #3.#4]{#1\sup{h}#2\sup{m}#3\sup{s}\llap.#4}
\def\dec[#1 #2 #3.#4]{#1\deg#2\arcm#3\arcs\llap.#4}
\def\deco[#1 #2 #3]{#1\deg#2\arcm#3\arcs}
\def\rra[#1 #2]{#1\sup{h}#2\sup{m}}

\def\dots{\relax\ifmmode \ldots\else $\ldots$\fi}
\def\WHzsr{\ifmmode $W\,Hz\mo\,sr\mo$\else W\,Hz\mo\,sr\mo\fi}
\def\mHz{\ifmmode $\,mHz$\else \,mHz\fi}
\def\GHz{\ifmmode $\,GHz$\else \,GHz\fi}
\def\mKs{\ifmmode $\,mK\,s$^{1/2}\else \,mK\,s$^{1/2}$\fi}
\def\muKs{\ifmmode \,\mu$K\,s$^{1/2}\else \,$\mu$K\,s$^{1/2}$\fi}
\def\muKRJs{\ifmmode \,\mu$K$_{\rm RJ}$\,s$^{1/2}\else \,$\mu$K$_{\rm RJ}$\,s$^{1/2}$\fi}
\def\muKHz{\ifmmode \,\mu$K\,Hz$^{-1/2}\else \,$\mu$K\,Hz$^{-1/2}$\fi}
\def\MJysr{\ifmmode \,$MJy\,sr\mo$\else \,MJy\,sr\mo\fi}
\def\MJysrmK{\ifmmode \,$MJy\,sr\mo$\,mK$_{\rm CMB}\mo\else \,MJy\,sr\mo\,mK$_{\rm CMB}\mo$\fi}
\def\microns{\ifmmode \,\mu$m$\else \,$\mu$m\fi}

\def\muK{\ifmmode \,\mu$K$\else \,$\mu$\hbox{K}\fi}
\def\microK{\ifmmode \,\mu$K$\else \,$\mu$\hbox{K}\fi}
\def\muW{\ifmmode \,\mu$W$\else \,$\mu$\hbox{W}\fi}
\def\kms{\ifmmode $\,km\,s$^{-1}\else \,km\,s$^{-1}$\fi}
\def\kmsMpc{\ifmmode $\,\kms\,Mpc\mo$\else \,\kms\,Mpc\mo\fi}
\setsymbol{LFI:center:frequency:70GHz:units}{70.3\,GHz}
\setsymbol{LFI:center:frequency:44GHz:units}{44.1\,GHz}
\setsymbol{LFI:center:frequency:30GHz:units}{28.5\,GHz}

\setsymbol{LFI:center:frequency:70GHz}{70.3}
\setsymbol{LFI:center:frequency:44GHz}{44.1}
\setsymbol{LFI:center:frequency:30GHz}{28.5}

\setsymbol{LFI:center:frequency:LFI18:Rad:M:units}{71.7\GHz}
\setsymbol{LFI:center:frequency:LFI19:Rad:M:units}{67.5\GHz}
\setsymbol{LFI:center:frequency:LFI20:Rad:M:units}{69.2\GHz}
\setsymbol{LFI:center:frequency:LFI21:Rad:M:units}{70.4\GHz}
\setsymbol{LFI:center:frequency:LFI22:Rad:M:units}{71.5\GHz}
\setsymbol{LFI:center:frequency:LFI23:Rad:M:units}{70.8\GHz}
\setsymbol{LFI:center:frequency:LFI24:Rad:M:units}{44.4\GHz}
\setsymbol{LFI:center:frequency:LFI25:Rad:M:units}{44.0\GHz}
\setsymbol{LFI:center:frequency:LFI26:Rad:M:units}{43.9\GHz}
\setsymbol{LFI:center:frequency:LFI27:Rad:M:units}{28.3\GHz}
\setsymbol{LFI:center:frequency:LFI28:Rad:M:units}{28.8\GHz}
\setsymbol{LFI:center:frequency:LFI18:Rad:S:units}{70.1\GHz}
\setsymbol{LFI:center:frequency:LFI19:Rad:S:units}{69.6\GHz}
\setsymbol{LFI:center:frequency:LFI20:Rad:S:units}{69.5\GHz}
\setsymbol{LFI:center:frequency:LFI21:Rad:S:units}{69.5\GHz}
\setsymbol{LFI:center:frequency:LFI22:Rad:S:units}{72.8\GHz}
\setsymbol{LFI:center:frequency:LFI23:Rad:S:units}{71.3\GHz}
\setsymbol{LFI:center:frequency:LFI24:Rad:S:units}{44.1\GHz}
\setsymbol{LFI:center:frequency:LFI25:Rad:S:units}{44.1\GHz}
\setsymbol{LFI:center:frequency:LFI26:Rad:S:units}{44.1\GHz}
\setsymbol{LFI:center:frequency:LFI27:Rad:S:units}{28.5\GHz}
\setsymbol{LFI:center:frequency:LFI28:Rad:S:units}{28.2\GHz}

\setsymbol{LFI:center:frequency:LFI18:Rad:M}{71.7}
\setsymbol{LFI:center:frequency:LFI19:Rad:M}{67.5}
\setsymbol{LFI:center:frequency:LFI20:Rad:M}{69.2}
\setsymbol{LFI:center:frequency:LFI21:Rad:M}{70.4}
\setsymbol{LFI:center:frequency:LFI22:Rad:M}{71.5}
\setsymbol{LFI:center:frequency:LFI23:Rad:M}{70.8}
\setsymbol{LFI:center:frequency:LFI24:Rad:M}{44.4}
\setsymbol{LFI:center:frequency:LFI25:Rad:M}{44.0}
\setsymbol{LFI:center:frequency:LFI26:Rad:M}{43.9}
\setsymbol{LFI:center:frequency:LFI27:Rad:M}{28.3}
\setsymbol{LFI:center:frequency:LFI28:Rad:M}{28.8}
\setsymbol{LFI:center:frequency:LFI18:Rad:S}{70.1}
\setsymbol{LFI:center:frequency:LFI19:Rad:S}{69.6}
\setsymbol{LFI:center:frequency:LFI20:Rad:S}{69.5}
\setsymbol{LFI:center:frequency:LFI21:Rad:S}{69.5}
\setsymbol{LFI:center:frequency:LFI22:Rad:S}{72.8}
\setsymbol{LFI:center:frequency:LFI23:Rad:S}{71.3}
\setsymbol{LFI:center:frequency:LFI24:Rad:S}{44.1}
\setsymbol{LFI:center:frequency:LFI25:Rad:S}{44.1}
\setsymbol{LFI:center:frequency:LFI26:Rad:S}{44.1}
\setsymbol{LFI:center:frequency:LFI27:Rad:S}{28.5}
\setsymbol{LFI:center:frequency:LFI28:Rad:S}{28.2}

\setsymbol{LFI:white:noise:sensitivity:70GHz:units}{152.6\muKs}
\setsymbol{LFI:white:noise:sensitivity:44GHz:units}{173.1\muKs}
\setsymbol{LFI:white:noise:sensitivity:30GHz:units}{146.8\muKs}

\setsymbol{LFI:white:noise:sensitivity:70GHz}{152.6}
\setsymbol{LFI:white:noise:sensitivity:44GHz}{173.1}
\setsymbol{LFI:white:noise:sensitivity:30GHz}{146.8}

\setsymbol{LFI:white:noise:sensitivity:LFI18:Rad:M:units}{512.0\muKs}
\setsymbol{LFI:white:noise:sensitivity:LFI19:Rad:M:units}{581.4\muKs}
\setsymbol{LFI:white:noise:sensitivity:LFI20:Rad:M:units}{590.8\muKs}
\setsymbol{LFI:white:noise:sensitivity:LFI21:Rad:M:units}{455.2\muKs}
\setsymbol{LFI:white:noise:sensitivity:LFI22:Rad:M:units}{492.0\muKs}
\setsymbol{LFI:white:noise:sensitivity:LFI23:Rad:M:units}{507.7\muKs}
\setsymbol{LFI:white:noise:sensitivity:LFI24:Rad:M:units}{462.2\muKs}
\setsymbol{LFI:white:noise:sensitivity:LFI25:Rad:M:units}{413.6\muKs}
\setsymbol{LFI:white:noise:sensitivity:LFI26:Rad:M:units}{478.6\muKs}
\setsymbol{LFI:white:noise:sensitivity:LFI27:Rad:M:units}{277.7\muKs}
\setsymbol{LFI:white:noise:sensitivity:LFI28:Rad:M:units}{312.3\muKs}
\setsymbol{LFI:white:noise:sensitivity:LFI18:Rad:S:units}{465.7\muKs}
\setsymbol{LFI:white:noise:sensitivity:LFI19:Rad:S:units}{555.6\muKs}
\setsymbol{LFI:white:noise:sensitivity:LFI20:Rad:S:units}{623.2\muKs}
\setsymbol{LFI:white:noise:sensitivity:LFI21:Rad:S:units}{564.1\muKs}
\setsymbol{LFI:white:noise:sensitivity:LFI22:Rad:S:units}{534.4\muKs}
\setsymbol{LFI:white:noise:sensitivity:LFI23:Rad:S:units}{542.4\muKs}
\setsymbol{LFI:white:noise:sensitivity:LFI24:Rad:S:units}{399.2\muKs}
\setsymbol{LFI:white:noise:sensitivity:LFI25:Rad:S:units}{392.6\muKs}
\setsymbol{LFI:white:noise:sensitivity:LFI26:Rad:S:units}{418.6\muKs}
\setsymbol{LFI:white:noise:sensitivity:LFI27:Rad:S:units}{302.9\muKs}
\setsymbol{LFI:white:noise:sensitivity:LFI28:Rad:S:units}{285.3\muKs}

\setsymbol{LFI:white:noise:sensitivity:LFI18:Rad:M}{512.0}
\setsymbol{LFI:white:noise:sensitivity:LFI19:Rad:M}{581.4}
\setsymbol{LFI:white:noise:sensitivity:LFI20:Rad:M}{590.8}
\setsymbol{LFI:white:noise:sensitivity:LFI21:Rad:M}{455.2}
\setsymbol{LFI:white:noise:sensitivity:LFI22:Rad:M}{492.0}
\setsymbol{LFI:white:noise:sensitivity:LFI23:Rad:M}{507.7}
\setsymbol{LFI:white:noise:sensitivity:LFI24:Rad:M}{462.2}
\setsymbol{LFI:white:noise:sensitivity:LFI25:Rad:M}{413.6}
\setsymbol{LFI:white:noise:sensitivity:LFI26:Rad:M}{478.6}
\setsymbol{LFI:white:noise:sensitivity:LFI27:Rad:M}{277.7}
\setsymbol{LFI:white:noise:sensitivity:LFI28:Rad:M}{312.3}
\setsymbol{LFI:white:noise:sensitivity:LFI18:Rad:S}{465.7}
\setsymbol{LFI:white:noise:sensitivity:LFI19:Rad:S}{555.6}
\setsymbol{LFI:white:noise:sensitivity:LFI20:Rad:S}{623.2}
\setsymbol{LFI:white:noise:sensitivity:LFI21:Rad:S}{564.1}
\setsymbol{LFI:white:noise:sensitivity:LFI22:Rad:S}{534.4}
\setsymbol{LFI:white:noise:sensitivity:LFI23:Rad:S}{542.4}
\setsymbol{LFI:white:noise:sensitivity:LFI24:Rad:S}{399.2}
\setsymbol{LFI:white:noise:sensitivity:LFI25:Rad:S}{392.6}
\setsymbol{LFI:white:noise:sensitivity:LFI26:Rad:S}{418.6}
\setsymbol{LFI:white:noise:sensitivity:LFI27:Rad:S}{302.9}
\setsymbol{LFI:white:noise:sensitivity:LFI28:Rad:S}{285.3}

\setsymbol{LFI:knee:frequency:70GHz:units}{29.5\mHz}
\setsymbol{LFI:knee:frequency:44GHz:units}{56.2\mHz}
\setsymbol{LFI:knee:frequency:30GHz:units}{113.7\mHz}

\setsymbol{LFI:knee:frequency:70GHz}{29.5}
\setsymbol{LFI:knee:frequency:44GHz}{56.2}
\setsymbol{LFI:knee:frequency:30GHz}{113.7}

\setsymbol{LFI:knee:frequency:LFI18:Rad:M:units}{16.3\mHz}
\setsymbol{LFI:knee:frequency:LFI19:Rad:M:units}{15.1\mHz}
\setsymbol{LFI:knee:frequency:LFI20:Rad:M:units}{18.7\mHz}
\setsymbol{LFI:knee:frequency:LFI21:Rad:M:units}{37.2\mHz}
\setsymbol{LFI:knee:frequency:LFI22:Rad:M:units}{12.7\mHz}
\setsymbol{LFI:knee:frequency:LFI23:Rad:M:units}{34.6\mHz}
\setsymbol{LFI:knee:frequency:LFI24:Rad:M:units}{46.2\mHz}
\setsymbol{LFI:knee:frequency:LFI25:Rad:M:units}{24.9\mHz}
\setsymbol{LFI:knee:frequency:LFI26:Rad:M:units}{67.6\mHz}
\setsymbol{LFI:knee:frequency:LFI27:Rad:M:units}{187.4\mHz}
\setsymbol{LFI:knee:frequency:LFI28:Rad:M:units}{122.2\mHz}
\setsymbol{LFI:knee:frequency:LFI18:Rad:S:units}{17.7\mHz}
\setsymbol{LFI:knee:frequency:LFI19:Rad:S:units}{22.0\mHz}
\setsymbol{LFI:knee:frequency:LFI20:Rad:S:units}{8.7\mHz}
\setsymbol{LFI:knee:frequency:LFI21:Rad:S:units}{25.9\mHz}
\setsymbol{LFI:knee:frequency:LFI22:Rad:S:units}{15.8\mHz}
\setsymbol{LFI:knee:frequency:LFI23:Rad:S:units}{129.8\mHz}
\setsymbol{LFI:knee:frequency:LFI24:Rad:S:units}{100.9\mHz}
\setsymbol{LFI:knee:frequency:LFI25:Rad:S:units}{38.9\mHz}
\setsymbol{LFI:knee:frequency:LFI26:Rad:S:units}{58.9\mHz}
\setsymbol{LFI:knee:frequency:LFI27:Rad:S:units}{104.4\mHz}
\setsymbol{LFI:knee:frequency:LFI28:Rad:S:units}{40.7\mHz}

\setsymbol{LFI:knee:frequency:LFI18:Rad:M}{16.3}
\setsymbol{LFI:knee:frequency:LFI19:Rad:M}{15.1}
\setsymbol{LFI:knee:frequency:LFI20:Rad:M}{18.7}
\setsymbol{LFI:knee:frequency:LFI21:Rad:M}{37.2}
\setsymbol{LFI:knee:frequency:LFI22:Rad:M}{12.7}
\setsymbol{LFI:knee:frequency:LFI23:Rad:M}{34.6}
\setsymbol{LFI:knee:frequency:LFI24:Rad:M}{46.2}
\setsymbol{LFI:knee:frequency:LFI25:Rad:M}{24.9}
\setsymbol{LFI:knee:frequency:LFI26:Rad:M}{67.6}
\setsymbol{LFI:knee:frequency:LFI27:Rad:M}{187.4}
\setsymbol{LFI:knee:frequency:LFI28:Rad:M}{122.2}
\setsymbol{LFI:knee:frequency:LFI18:Rad:S}{17.7}
\setsymbol{LFI:knee:frequency:LFI19:Rad:S}{22.0}
\setsymbol{LFI:knee:frequency:LFI20:Rad:S}{8.7}
\setsymbol{LFI:knee:frequency:LFI21:Rad:S}{25.9}
\setsymbol{LFI:knee:frequency:LFI22:Rad:S}{15.8}
\setsymbol{LFI:knee:frequency:LFI23:Rad:S}{129.8}
\setsymbol{LFI:knee:frequency:LFI24:Rad:S}{100.9}
\setsymbol{LFI:knee:frequency:LFI25:Rad:S}{38.9}
\setsymbol{LFI:knee:frequency:LFI26:Rad:S}{58.9}
\setsymbol{LFI:knee:frequency:LFI27:Rad:S}{104.4}
\setsymbol{LFI:knee:frequency:LFI28:Rad:S}{40.7}

\setsymbol{LFI:slope:70GHz:units}{$-1.03$\mHz}
\setsymbol{LFI:slope:44GHz:units}{$-0.89$\mHz}
\setsymbol{LFI:slope:30GHz:units}{$-0.87$\mHz}

\setsymbol{LFI:slope:70GHz}{$-1.03$}
\setsymbol{LFI:slope:44GHz}{$-0.89$}
\setsymbol{LFI:slope:30GHz}{$-0.87$}

\setsymbol{LFI:slope:LFI18:Rad:M:units}{$-1.04$\mHz}
\setsymbol{LFI:slope:LFI19:Rad:M:units}{$-1.09$\mHz}
\setsymbol{LFI:slope:LFI20:Rad:M:units}{$-0.69$\mHz}
\setsymbol{LFI:slope:LFI21:Rad:M:units}{$-1.56$\mHz}
\setsymbol{LFI:slope:LFI22:Rad:M:units}{$-1.01$\mHz}
\setsymbol{LFI:slope:LFI23:Rad:M:units}{$-0.96$\mHz}
\setsymbol{LFI:slope:LFI24:Rad:M:units}{$-0.83$\mHz}
\setsymbol{LFI:slope:LFI25:Rad:M:units}{$-0.91$\mHz}
\setsymbol{LFI:slope:LFI26:Rad:M:units}{$-0.95$\mHz}
\setsymbol{LFI:slope:LFI27:Rad:M:units}{$-0.87$\mHz}
\setsymbol{LFI:slope:LFI28:Rad:M:units}{$-0.88$\mHz}
\setsymbol{LFI:slope:LFI18:Rad:S:units}{$-1.15$\mHz}
\setsymbol{LFI:slope:LFI19:Rad:S:units}{$-1.00$\mHz}
\setsymbol{LFI:slope:LFI20:Rad:S:units}{$-0.95$\mHz}
\setsymbol{LFI:slope:LFI21:Rad:S:units}{$-0.92$\mHz}
\setsymbol{LFI:slope:LFI22:Rad:S:units}{$-1.01$\mHz}
\setsymbol{LFI:slope:LFI23:Rad:S:units}{$-0.95$\mHz}
\setsymbol{LFI:slope:LFI24:Rad:S:units}{$-0.73$\mHz}
\setsymbol{LFI:slope:LFI25:Rad:S:units}{$-1.16$\mHz}
\setsymbol{LFI:slope:LFI26:Rad:S:units}{$-0.79$\mHz}
\setsymbol{LFI:slope:LFI27:Rad:S:units}{$-0.82$\mHz}
\setsymbol{LFI:slope:LFI28:Rad:S:units}{$-0.91$\mHz}

\setsymbol{LFI:slope:LFI18:Rad:M}{$-1.04$}
\setsymbol{LFI:slope:LFI19:Rad:M}{$-1.09$}
\setsymbol{LFI:slope:LFI20:Rad:M}{$-0.69$}
\setsymbol{LFI:slope:LFI21:Rad:M}{$-1.56$}
\setsymbol{LFI:slope:LFI22:Rad:M}{$-1.01$}
\setsymbol{LFI:slope:LFI23:Rad:M}{$-0.96$}
\setsymbol{LFI:slope:LFI24:Rad:M}{$-0.83$}
\setsymbol{LFI:slope:LFI25:Rad:M}{$-0.91$}
\setsymbol{LFI:slope:LFI26:Rad:M}{$-0.95$}
\setsymbol{LFI:slope:LFI27:Rad:M}{$-0.87$}
\setsymbol{LFI:slope:LFI28:Rad:M}{$-0.88$}
\setsymbol{LFI:slope:LFI18:Rad:S}{$-1.15$}
\setsymbol{LFI:slope:LFI19:Rad:S}{$-1.00$}
\setsymbol{LFI:slope:LFI20:Rad:S}{$-0.95$}
\setsymbol{LFI:slope:LFI21:Rad:S}{$-0.92$}
\setsymbol{LFI:slope:LFI22:Rad:S}{$-1.01$}
\setsymbol{LFI:slope:LFI23:Rad:S}{$-0.95$}
\setsymbol{LFI:slope:LFI24:Rad:S}{$-0.73$}
\setsymbol{LFI:slope:LFI25:Rad:S}{$-1.16$}
\setsymbol{LFI:slope:LFI26:Rad:S}{$-0.79$}
\setsymbol{LFI:slope:LFI27:Rad:S}{$-0.82$}
\setsymbol{LFI:slope:LFI28:Rad:S}{$-0.91$}

\setsymbol{LFI:FWHM:70GHz:units}{13\parcm01}
\setsymbol{LFI:FWHM:44GHz:units}{27\parcm92}
\setsymbol{LFI:FWHM:30GHz:units}{32\parcm65}

\setsymbol{LFI:FWHM:70GHz}{13.01}
\setsymbol{LFI:FWHM:44GHz}{27.92}
\setsymbol{LFI:FWHM:30GHz}{32.65}

\setsymbol{LFI:FWHM:LFI18:units}{13\parcm39}
\setsymbol{LFI:FWHM:LFI19:units}{13\parcm01}
\setsymbol{LFI:FWHM:LFI20:units}{12\parcm75}
\setsymbol{LFI:FWHM:LFI21:units}{12\parcm74}
\setsymbol{LFI:FWHM:LFI22:units}{12\parcm87}
\setsymbol{LFI:FWHM:LFI23:units}{13\parcm27}
\setsymbol{LFI:FWHM:LFI24:units}{22\parcm98}
\setsymbol{LFI:FWHM:LFI25:units}{30\parcm46}
\setsymbol{LFI:FWHM:LFI26:units}{30\parcm31}
\setsymbol{LFI:FWHM:LFI27:units}{32\parcm65}
\setsymbol{LFI:FWHM:LFI28:units}{32\parcm66}

\setsymbol{LFI:FWHM:LFI18}{13.39}
\setsymbol{LFI:FWHM:LFI19}{13.01}
\setsymbol{LFI:FWHM:LFI20}{12.75}
\setsymbol{LFI:FWHM:LFI21}{12.74}
\setsymbol{LFI:FWHM:LFI22}{12.87}
\setsymbol{LFI:FWHM:LFI23}{13.27}
\setsymbol{LFI:FWHM:LFI24}{22.98}
\setsymbol{LFI:FWHM:LFI25}{30.46}
\setsymbol{LFI:FWHM:LFI26}{30.31}
\setsymbol{LFI:FWHM:LFI27}{32.65}
\setsymbol{LFI:FWHM:LFI28}{32.66}

\setsymbol{LFI:FWHM:uncertainty:LFI18:units}{0.170\arcm}
\setsymbol{LFI:FWHM:uncertainty:LFI19:units}{0.174\arcm}
\setsymbol{LFI:FWHM:uncertainty:LFI20:units}{0.170\arcm}
\setsymbol{LFI:FWHM:uncertainty:LFI21:units}{0.156\arcm}
\setsymbol{LFI:FWHM:uncertainty:LFI22:units}{0.164\arcm}
\setsymbol{LFI:FWHM:uncertainty:LFI23:units}{0.171\arcm}
\setsymbol{LFI:FWHM:uncertainty:LFI24:units}{0.652\arcm}
\setsymbol{LFI:FWHM:uncertainty:LFI25:units}{1.075\arcm}
\setsymbol{LFI:FWHM:uncertainty:LFI26:units}{1.131\arcm}
\setsymbol{LFI:FWHM:uncertainty:LFI27:units}{1.266\arcm}
\setsymbol{LFI:FWHM:uncertainty:LFI28:units}{1.287\arcm}

\setsymbol{LFI:FWHM:uncertainty:LFI18}{0.170}
\setsymbol{LFI:FWHM:uncertainty:LFI19}{0.174}
\setsymbol{LFI:FWHM:uncertainty:LFI20}{0.170}
\setsymbol{LFI:FWHM:uncertainty:LFI21}{0.156}
\setsymbol{LFI:FWHM:uncertainty:LFI22}{0.164}
\setsymbol{LFI:FWHM:uncertainty:LFI23}{0.171}
\setsymbol{LFI:FWHM:uncertainty:LFI24}{0.652}
\setsymbol{LFI:FWHM:uncertainty:LFI25}{1.075}
\setsymbol{LFI:FWHM:uncertainty:LFI26}{1.131}
\setsymbol{LFI:FWHM:uncertainty:LFI27}{1.266}
\setsymbol{LFI:FWHM:uncertainty:LFI28}{1.287}

\setsymbol{HFI:center:frequency:100GHz:units}{100\,GHz}
\setsymbol{HFI:center:frequency:143GHz:units}{143\,GHz}
\setsymbol{HFI:center:frequency:217GHz:units}{217\,GHz}
\setsymbol{HFI:center:frequency:353GHz:units}{353\,GHz}
\setsymbol{HFI:center:frequency:545GHz:units}{545\,GHz}
\setsymbol{HFI:center:frequency:857GHz:units}{857\,GHz}

\setsymbol{HFI:center:frequency:100GHz}{100}
\setsymbol{HFI:center:frequency:143GHz}{143}
\setsymbol{HFI:center:frequency:217GHz}{217}
\setsymbol{HFI:center:frequency:353GHz}{353}
\setsymbol{HFI:center:frequency:545GHz}{545}
\setsymbol{HFI:center:frequency:857GHz}{857}

\setsymbol{HFI:Ndetectors:100GHz}{8}
\setsymbol{HFI:Ndetectors:143GHz}{11}
\setsymbol{HFI:Ndetectors:217GHz}{12}
\setsymbol{HFI:Ndetectors:353GHz}{12}
\setsymbol{HFI:Ndetectors:545GHz}{3}
\setsymbol{HFI:Ndetectors:857GHz}{4}

\setsymbol{HFI:FWHM:Maps:100GHz:units}{9\parcm88}
\setsymbol{HFI:FWHM:Maps:143GHz:units}{7\parcm18}
\setsymbol{HFI:FWHM:Maps:217GHz:units}{4\parcm87}
\setsymbol{HFI:FWHM:Maps:353GHz:units}{4\parcm65}
\setsymbol{HFI:FWHM:Maps:545GHz:units}{4\parcm72}
\setsymbol{HFI:FWHM:Maps:857GHz:units}{4\parcm39}
\setsymbol{HFI:FWHM:Maps:100GHz}{9.88}
\setsymbol{HFI:FWHM:Maps:143GHz}{7.18}
\setsymbol{HFI:FWHM:Maps:217GHz}{4.87}
\setsymbol{HFI:FWHM:Maps:353GHz}{4.65}
\setsymbol{HFI:FWHM:Maps:545GHz}{4.72}
\setsymbol{HFI:FWHM:Maps:857GHz}{4.39}

\setsymbol{HFI:beam:ellipticity:Maps:100GHz}{1.15}
\setsymbol{HFI:beam:ellipticity:Maps:143GHz}{1.01}
\setsymbol{HFI:beam:ellipticity:Maps:217GHz}{1.06}
\setsymbol{HFI:beam:ellipticity:Maps:353GHz}{1.05}
\setsymbol{HFI:beam:ellipticity:Maps:545GHz}{1.14}
\setsymbol{HFI:beam:ellipticity:Maps:857GHz}{1.19}

\setsymbol{HFI:FWHM:Mars:100GHz:units}{9\parcm37}
\setsymbol{HFI:FWHM:Mars:143GHz:units}{7\parcm04}
\setsymbol{HFI:FWHM:Mars:217GHz:units}{4\parcm68}
\setsymbol{HFI:FWHM:Mars:353GHz:units}{4\parcm43}
\setsymbol{HFI:FWHM:Mars:545GHz:units}{3\parcm80}
\setsymbol{HFI:FWHM:Mars:857GHz:units}{3\parcm67}

\setsymbol{HFI:FWHM:Mars:100GHz}{9.37}
\setsymbol{HFI:FWHM:Mars:143GHz}{7.04}
\setsymbol{HFI:FWHM:Mars:217GHz}{4.68}
\setsymbol{HFI:FWHM:Mars:353GHz}{4.43}
\setsymbol{HFI:FWHM:Mars:545GHz}{3.80}
\setsymbol{HFI:FWHM:Mars:857GHz}{3.67}

\setsymbol{HFI:beam:ellipticity:Mars:100GHz}{1.18}
\setsymbol{HFI:beam:ellipticity:Mars:143GHz}{1.03}
\setsymbol{HFI:beam:ellipticity:Mars:217GHz}{1.14}
\setsymbol{HFI:beam:ellipticity:Mars:353GHz}{1.09}
\setsymbol{HFI:beam:ellipticity:Mars:545GHz}{1.25}
\setsymbol{HFI:beam:ellipticity:Mars:857GHz}{1.03}

\setsymbol{HFI:CMB:relative:calibration:100GHz}{$\lsim 1\%$}
\setsymbol{HFI:CMB:relative:calibration:143GHz}{$\lsim 1\%$}
\setsymbol{HFI:CMB:relative:calibration:217GHz}{$\lsim 1\%$}
\setsymbol{HFI:CMB:relative:calibration:353GHz}{$\lsim 1\%$}
\setsymbol{HFI:CMB:relative:calibration:545GHz}{}
\setsymbol{HFI:CMB:relative:calibration:857GHz}{}

\setsymbol{HFI:CMB:absolute:calibration:100GHz}{$\lsim 2\%$}
\setsymbol{HFI:CMB:absolute:calibration:143GHz}{$\lsim 2\%$}
\setsymbol{HFI:CMB:absolute:calibration:217GHz}{$\lsim 2\%$}
\setsymbol{HFI:CMB:absolute:calibration:353GHz}{$\lsim 2\%$}
\setsymbol{HFI:CMB:absolute:calibration:545GHz}{}
\setsymbol{HFI:CMB:absolute:calibration:857GHz}{}

\setsymbol{HFI:FIRAS:gain:calibration:accuracy:statistical:100GHz}{}
\setsymbol{HFI:FIRAS:gain:calibration:accuracy:statistical:143GHz}{}
\setsymbol{HFI:FIRAS:gain:calibration:accuracy:statistical:217GHz}{}
\setsymbol{HFI:FIRAS:gain:calibration:accuracy:statistical:353GHz}{2.5\%}
\setsymbol{HFI:FIRAS:gain:calibration:accuracy:statistical:545GHz}{1\%}
\setsymbol{HFI:FIRAS:gain:calibration:accuracy:statistical:857GHz}{0.5\%}

\setsymbol{HFI:FIRAS:gain:calibration:accuracy:systematic:100GHz}{}
\setsymbol{HFI:FIRAS:gain:calibration:accuracy:systematic:143GHz}{}
\setsymbol{HFI:FIRAS:gain:calibration:accuracy:systematic:217GHz}{}
\setsymbol{HFI:FIRAS:gain:calibration:accuracy:systematic:353GHz}{}
\setsymbol{HFI:FIRAS:gain:calibration:accuracy:systematic:545GHz}{7\%}
\setsymbol{HFI:FIRAS:gain:calibration:accuracy:systematic:857GHz}{7\%}

\setsymbol{HFI:FIRAS:zero:point:accuracy:100GHz:units}{0.8\MJysr}
\setsymbol{HFI:FIRAS:zero:point:accuracy:143GHz:units}{}
\setsymbol{HFI:FIRAS:zero:point:accuracy:217GHz:units}{}
\setsymbol{HFI:FIRAS:zero:point:accuracy:353GHz:units}{1.4\MJysr}
\setsymbol{HFI:FIRAS:zero:point:accuracy:545GHz:units}{2.2\MJysr}
\setsymbol{HFI:FIRAS:zero:point:accuracy:857GHz:units}{1.7\MJysr}

\setsymbol{HFI:FIRAS:zero:point:accuracy:100GHz}{0.8}
\setsymbol{HFI:FIRAS:zero:point:accuracy:143GHz}{}
\setsymbol{HFI:FIRAS:zero:point:accuracy:217GHz}{}
\setsymbol{HFI:FIRAS:zero:point:accuracy:353GHz}{1.4}
\setsymbol{HFI:FIRAS:zero:point:accuracy:545GHz}{2.2}
\setsymbol{HFI:FIRAS:zero:point:accuracy:857GHz}{1.7}

\setsymbol{HFI:unit:conversion:100GHz:units}{0.2415\MJysrmK}
\setsymbol{HFI:unit:conversion:143GHz:units}{0.3694\MJysrmK}
\setsymbol{HFI:unit:conversion:217GHz:units}{0.4811\MJysrmK}
\setsymbol{HFI:unit:conversion:353GHz:units}{0.2883\MJysrmK}
\setsymbol{HFI:unit:conversion:545GHz:units}{0.05826\MJysrmK}
\setsymbol{HFI:unit:conversion:857GHz:units}{0.002238\MJysrmK}

\setsymbol{HFI:unit:conversion:100GHz}{0.2415}
\setsymbol{HFI:unit:conversion:143GHz}{0.3694}
\setsymbol{HFI:unit:conversion:217GHz}{0.4811}
\setsymbol{HFI:unit:conversion:353GHz}{0.2883}
\setsymbol{HFI:unit:conversion:545GHz}{0.05826}
\setsymbol{HFI:unit:conversion:857GHz}{0.002238}

\setsymbol{HFI:colour:correction:alpha=-2:V1.01:100GHz}{0.9893}
\setsymbol{HFI:colour:correction:alpha=-2:V1.01:143GHz}{0.9759}
\setsymbol{HFI:colour:correction:alpha=-2:V1.01:217GHz}{1.0007}
\setsymbol{HFI:colour:correction:alpha=-2:V1.01:353GHz}{1.0028}
\setsymbol{HFI:colour:correction:alpha=-2:V1.01:545GHz}{1.0019}
\setsymbol{HFI:colour:correction:alpha=-2:V1.01:857GHz}{0.9889}

\setsymbol{HFI:colour:correction:alpha=0:V1.01:100GHz}{1.0008}
\setsymbol{HFI:colour:correction:alpha=0:V1.01:143GHz}{1.0148}
\setsymbol{HFI:colour:correction:alpha=0:V1.01:217GHz}{0.9909}
\setsymbol{HFI:colour:correction:alpha=0:V1.01:353GHz}{0.9888}
\setsymbol{HFI:colour:correction:alpha=0:V1.01:545GHz}{0.9878}
\setsymbol{HFI:colour:correction:alpha=0:V1.01:857GHz}{1.0014}

\begin{document}

\title{\Planck\ early results. V. The Low Frequency Instrument data processing}

\titlerunning{LFI data processing}
\authorrunning{Zacchei A., \etal}

\author{\small
A.~Zacchei\inst{36}
\and
D.~Maino\inst{22, 38}
\and
C.~Baccigalupi\inst{54}
\and
M.~Bersanelli\inst{22, 38}
\and
A.~Bonaldi\inst{34}
\and
L.~Bonavera\inst{54, 5}
\and
C.~Burigana\inst{37}
\and
R.~C.~Butler\inst{37}
\and
F.~Cuttaia\inst{37}
\and
G.~de Zotti\inst{34, 54}
% ...
}
\institute{\small
Aalto University Mets\"{a}hovi Radio Observatory, Mets\"{a}hovintie 114, FIN-02540 Kylm\"{a}l\"{a}, Finland\\
\and
Agenzia Spaziale Italiana Science Data Center, c/o ESRIN, via Galileo Galilei, Frascati, Italy\\
\and
Agenzia Spaziale Italiana, Viale Liegi 26, Roma, Italy\\
\and
Astroparticule et Cosmologie, CNRS (UMR7164), Universit\'{e} Denis Diderot Paris 7, B\^{a}timent Condorcet, 10 rue A. Domon et L\'{e}onie Duquet, Paris, France\\
\and
Australia Telescope National Facility, CSIRO, P.O. Box 76, Epping, NSW 1710, Australia\\
\and
CNR - ISTI, Area della Ricerca, via G. Moruzzi 1, Pisa, Italy\\
\and
CNRS, IRAP, 9 Av. colonel Roche, BP 44346, F-31028 Toulouse cedex 4, France\\
\and
California Institute of Technology, Pasadena, California, U.S.A.\\
\and
Centre of Mathematics for Applications, University of Oslo, Blindern, Oslo, Norway\\
\and
DTU Space, National Space Institute, Juliane Mariesvej 30, Copenhagen, Denmark\\
\and
Departamento de F\'{\i}sica, Universidad de Oviedo, Avda. Calvo Sotelo s/n, Oviedo, Spain\\
\and
Departamento de Matem\'{a}ticas, Estad\'{\i}stica y Computaci\'{o}n, Universidad de Cantabria, Avda. de los Castros s/n, Santander, Spain\\
\and
Department of Physics \& Astronomy, University of British Columbia, 6224 Agricultural Road, Vancouver, British Columbia, Canada\\
\and
Department of Physics and Astronomy, University of Southern California, Los Angeles, California, U.S.A.\\
\and
Department of Physics, Gustaf H\"{a}llstr\"{o}min katu 2a, University of Helsinki, Helsinki, Finland\\
\and
Department of Physics, University of California, Berkeley, California, U.S.A.\\
\and
Department of Physics, University of California, One Shields Avenue, Davis, California, U.S.A.\\
\and
Department of Physics, University of California, Santa Barbara, California, U.S.A.\\
\and
Department of Physics, University of Oxford, 1 Keble Road, Oxford, U.K.\\
\and
Dipartimento di Fisica G. Galilei, Universit\`{a} degli Studi di Padova, via Marzolo 8, 35131 Padova, Italy\\
\and
Dipartimento di Fisica, Universit\`{a} La Sapienza, P. le A. Moro 2, Roma, Italy\\
\and
Dipartimento di Fisica, Universit\`{a} degli Studi di Milano, Via Celoria, 16, Milano, Italy\\
\and
Dipartimento di Fisica, Universit\`{a} degli Studi di Trieste, via A. Valerio 2, Trieste, Italy\\
\and
Dipartimento di Fisica, Universit\`{a} di Ferrara, Via Saragat 1, 44122 Ferrara, Italy\\
\and
Dipartimento di Fisica, Universit\`{a} di Roma Tor Vergata, Via della Ricerca Scientifica, 1, Roma, Italy\\
\and
Dipartimento di Matematica, Universit\`{a} di Roma Tor Vergata, Via della Ricerca Scientifica, 1, Roma, Italy\\
\and
Dpto. Astrof\'{i}sica, Universidad de La Laguna (ULL), E-38206 La Laguna, Tenerife, Spain\\
\and
European Space Agency, ESAC, Planck Science Office, Camino bajo del Castillo, s/n, Urbanizaci\'{o}n Villafranca del Castillo, Villanueva de la Ca\~{n}ada, Madrid, Spain\\
\and
European Space Agency, ESTEC, Keplerlaan 1, 2201 AZ Noordwijk, The Netherlands\\
\and
Haverford College Astronomy Department, 370 Lancaster Avenue, Haverford, Pennsylvania, U.S.A.\\
\and
Helsinki Institute of Physics, Gustaf H\"{a}llstr\"{o}min katu 2, University of Helsinki, Helsinki, Finland\\
\and
INAF - Osservatorio Astrofisico di Arcetri, Largo Enrico Fermi 5, Firenze, Italy\\
\and
INAF - Osservatorio Astrofisico di Catania, Via S. Sofia 78, Catania, Italy\\
\and
INAF - Osservatorio Astronomico di Padova, Vicolo dell'Osservatorio 5, Padova, Italy\\
\and
INAF - Osservatorio Astronomico di Roma, via di Frascati 33, Monte Porzio Catone, Italy\\
\and
INAF - Osservatorio Astronomico di Trieste, Via G.B. Tiepolo 11, Trieste, Italy\\
\and
INAF/IASF Bologna, Via Gobetti 101, Bologna, Italy\\
\and
INAF/IASF Milano, Via E. Bassini 15, Milano, Italy\\
\and
INRIA, Laboratoire de Recherche en Informatique, Universit\'{e} Paris-Sud 11, B\^{a}timent 490, 91405 Orsay Cedex, France\\
\and
ISDC Data Centre for Astrophysics, University of Geneva, ch. d'Ecogia 16, Versoix, Switzerland\\
\and
Infrared Processing and Analysis Center, California Institute of Technology, Pasadena, CA 91125, U.S.A.\\
\and
Institut d'Astrophysique Spatiale, CNRS (UMR8617) Universit\'{e} Paris-Sud 11, B\^{a}timent 121, Orsay, France\\
\and
Institut d'Astrophysique de Paris, CNRS UMR7095, Universit\'{e} Pierre \& Marie Curie, 98 bis boulevard Arago, Paris, France\\
\and
Institute of Astronomy, University of Cambridge, Madingley Road, Cambridge CB3 0HA, U.K.\\
\and
Institute of Theoretical Astrophysics, University of Oslo, Blindern, Oslo, Norway\\
\and
Instituto de Astrof\'{\i}sica de Canarias, C/V\'{\i}a L\'{a}ctea s/n, La Laguna, Tenerife, Spain\\
\and
Instituto de F\'{\i}sica de Cantabria (CSIC-Universidad de Cantabria), Avda. de los Castros s/n, Santander, Spain\\
\and
Jet Propulsion Laboratory, California Institute of Technology, 4800 Oak Grove Drive, Pasadena, California, U.S.A.\\
\and
Jodrell Bank Centre for Astrophysics, Alan Turing Building, School of Physics and Astronomy, The University of Manchester, Oxford Road, Manchester, M13 9PL, U.K.\\
\and
LERMA, CNRS, Observatoire de Paris, 61 Avenue de l'Observatoire, Paris, France\\
\and
Lawrence Berkeley National Laboratory, Berkeley, California, U.S.A.\\
\and
Max-Planck-Institut f\"{u}r Astrophysik, Karl-Schwarzschild-Str. 1, 85741 Garching, Germany\\
\and
MilliLab, VTT Technical Research Centre of Finland, Tietotie 3, Espoo, Finland\\
\and
SISSA, Astrophysics Sector, via Bonomea 265, 34136, Trieste, Italy\\
% ...
}


\abstract{We describe the processing of data from the Low
  Frequency Instrument (LFI) used in production of the \Planck\ Early Release Compact Source
  Catalogue (ERCSC).  In particular, we discuss the steps involved in reducing the data from telemetry
  packets to cleaned, calibrated, time-ordered data (TOD) and frequency maps. Data are continuously
  calibrated
  using the modulation of the temperature of the cosmic microwave background radiation induced by the motion
  of the spacecraft.  Noise properties are estimated from TOD from which the sky signal has been removed
  using a generalized least square map-making algorithm.  Measured $1/f$ noise knee-frequencies range
  from $\sim 100$\,mHz at 30\,GHz to a few tens of mHz at 70\,GHz. A destriping code ({\tt Madam})
  is employed to combine radiometric data and pointing information into sky maps,
  minimizing the variance of correlated noise.  Noise covariance matrices required to
  compute statistical uncertainties on LFI and \Planck\ products are also produced.  Main
  beams are estimated down to the $\approx -10$\,dB level using Jupiter
  transits, which are also used for  geometrical calibration of the focal plane.}

 \maketitle

\section{Introduction}

\Planck\footnote{\Planck\ (http://www.esa.int/\Planck) is a
project of the European Space Agency (ESA) with instruments
provided by two scientific consortia funded by ESA member states
(in particular the lead countries France and Italy), with
contributions from NASA (USA) and telescope reflectors provided by
a collaboration between ESA and a scientific consortium led and
funded by Denmark.} \citep{tauber2010a,planck2011-1.1} is a third
generation space mission to measure the anisotropy of the Cosmic
Microwave Background (CMB).  It observes the sky in nine frequency
bands covering 30--857\,GHz with high sensitivity and angular
resolution from 31\arcm to 5\arcm.  The Low Frequency Instrument
(LFI) \citep{mandolesi2010, bersanelli2010, planck2011-1.4} covers
the 30, 44, and 70\,GHz bands with amplifiers cooled to
20\,\hbox{K}.  The High Frequency Instrument (HFI)
\citep{lamarre2010,planck2011-1.5} covers the 100, 143, 217, 353,
545, and 857\,GHz bands with bolometers cooled to 0.1\,\hbox{K}.
Polarization is measured in all but the highest two bands
\citep{leahy2010,rosset2010}.  A combination of radiative cooling
and three mechanical coolers produces the temperatures needed for
the detectors and optics \citep{planck2011-1.3}.  Two Data
Processing Centres (DPCs), conceived as interacting and
complementary since the earliest design of the \Planck\ scientific
ground segment \citep{pasian2000}; check and calibrate the data
and make maps of the sky, this paper and  \citep{planck2011-1.7}.
\Planck's sensitivity, angular resolution, and frequency coverage
make it a powerful instrument for galactic and extragalactic
astrophysics, as well as cosmology.  Early astrophysics results
are given in Planck Collaboration, 2011c--z.

The Low Frequency Instrument LFI on \Planck\ comprises a
set of 11 radiometer chain assemblies (RCAs), each composed
of two independent, pseudo-correlation radiometers.  There are two
RCAs at 30\,GHz, three at 44\,GHz, and six at 70\,GHz.
Each radiometer has two independent diodes
for detection, integration, and conversion from radio frequency signal to
DC voltage.  The LFI is cryogenically cooled to 20\,K to reduce
noise, while the pseudo-correlation design with reference
loads at $\approx$ 4\,K ensures good suppression of $1/f$ noise
\citep{planck2011-1.4}.

LFI produces full-sky maps centered near 30, 44, and 70\,GHz, with
significant improvements with respect to current CMB data in the
same frequency range.  Careful data
processing is required in order to realize the full potential of LFI and
the ambitious science goals of \Planck, which require that systematic effects be limited to a few
$\mu$K per resolution element.

In this paper we describe the processing
steps implemented to create LFI data products, with particular attention to the needs of
the first set of astrophysics results.

The structure of the paper follows the flow of the data
through the analysis pipeline.  Section~\ref{toicreation}
describes the creation of time ordered information (TOI) from
telemetry packets, time stamping, pointing reconstruction, and data flagging.
Section~\ref{toiprocessing} describes the main operations
performed on the TOI, including removal of frequency spikes,
creation of differenced data, determination of the gain
modulation factor, and diode combination.
Beam reconstruction is discussed in Sect.~\ref{beamrecovery}, calibration
in Sect.~\ref{calibration}, and noise in Sect.~\ref{noise}.
Map-making, covariance matrices, and tests based on jackknife analysis and Monte Carlo
simulations are described in Sect.~\ref{mmaking}.
Section~\ref{colcorr} reports on colour corrections.
Section~\ref{cmbremoval} describes how the CMB was removed from LFI and HFI maps.
Finally, Sect.~\ref{Infrast} gives an overview of the software
infrastructure at the LFI DPC.

\section{Creation of time ordered information}
\label{toicreation}

The task of the Level~1 DPC pipeline is to retrieve all necessary information from packets received each day from the Mission Operations Centre (MOC) and to transform the scientific TOI and housekeeping (H/K) data into a form that is manageable by the scientific pipeline.

During the $\sim$3\,h daily telecommunication period (DTCP), the
MOC receives telemetry from the previous day, archived in on-board
mass memory, together with real-time telemetry.  Additional
auxiliary files, such as the attitude history file (AHF) of the
satellite, are produced.

The MOC consolidates the data for each day, checking for gaps or corrupted telemetry packets, then provides
the data, together with additional auxiliary data, to the DPCs through a client/server application called the data disposition system (DDS).

The data are received at the DPC as a stream of packets, which are
handled automatically by four Level~1 pipelines: Data Receipt,
Telemetry Handling, Auxiliary Data, and Command History.

The Data Receipt pipeline implements the client side of the interface with the \hbox{DDS}. It requests a subset of data provided through this interface.  A finite-state machine model has been used in the design of
this pipeline for better formalization of the actions required during interaction with the DDS server.

The Telemetry Handling pipeline is triggered when a new segment of telemetry data is received.  The first task (Telemetry Unscrambler) discriminates between scientific and housekeeping telemetry packets.  Scientific packets are grouped according to radiometer, detector source, and processing type, then uncompressed and decoded (see next paragraph).  The on-board time of each sample is computed based on the packet on-board time and the detector sampling frequency.  Housekeeping telemetry packets are also grouped
according to packet type, and each housekeeping parameter within the packet is extracted and saved into \hbox{TOI}.  Subsequent tasks of the pipeline perform calibration of housekeeping and scientific TOIs together with additional quality checks (e.g., out of limits, time correlation). The last task, FITS2DMC, ingests the TOIs into the Data Management Component (DMC), making them available to the Level~2 and Level~3 pipelines.

The Auxiliary Data pipeline ingests the AHF provided by Flight Dynamics into the \hbox{DMC}.  Finally the Command History pipeline requests and stores the list of telecommands sent to the satellite during the
DTCP.

The four pipelines are implemented as {\tt perl} scripts, scheduled every 5\,min.  Trigger files are created to activate the processing in the Auxiliary Data and Command History pipelines, and a pipeline monitoring facility displays information about the status of each pipeline.  The entire Level~1 pipeline was heavily
tested and validated before the start of \Planck\ operations \citep[see][for more details]{frailis2009}.

\subsection{Scientific data processing}

When creating TOI, the Level~1 pipeline must recover accurately
the values of the original (averaged) sky and load samples acquired
on-board. The instrument can acquire scientific
data in several modes or ``PTypes''; we describe here only the
nominal one (PType~5) \citep[see][]{zacchei2009}.  The key feature is
that two independent differenced time streams are created from the sky
and load signals with two different gain modulation factors (GMFs).

Data of PType~5 are first uncompressed. The lossless
compression applied on-board is inverted, and the number of
samples obtained is checked against auxiliary packet information.
Decompressed data $Q_{i=1,2}$ are then subject to a
dequantization step to recover the original signals $P_{i}$ according to
 \begin{equation}
   P_i =\frac{Q_i}{\mathrm{SECOND\_QUANT}}-{\mathrm{OFFSET\_ADJUST}}\, ,
 \end{equation}
where $\mathrm{SECOND\_QUANT}$ and $\mathrm{OFFSET\_ADJUST}$ are
parameters of the readout electronics box assembly (REBA),
calibration of which is described by \citet{maris2009}.

After dequantization, data are demixed to obtain $\overline{S}_{\rm sky}$ and
$\overline{S}_{\rm load}$ using as inputs the gain modulation factors $R_1$ and $R_2$
determined during REBA calibration \citep{maris2009}:
 \begin{equation}
    \overline{S}_{\rm sky}=\frac{R_2 \cdot {P_{1}} - R_1 \cdot {P_{2}} }{R_2 -
    R_1},
 \end{equation}
 \begin{equation}
   \overline{S}_{\rm load}=\frac{{P_{1}} - {P_{2}}}{R_2-R_1}.
 \end{equation}

Conversion from ADU (analog-to-digital units) to volts is achieved by
\begin{equation}
  \overline{V}_{i} = \frac{\overline{S}_{i} -
  {Z_{\mathrm{DAE}}}}{G_{\mathrm{DAE}}}- {O_{\mathrm{DAE}}}\, ,
\end{equation}
where $G_{\rm DAE}$, $O_{\rm DAE}$, and $Z_{\rm DAE}$ are data
acquisition electronics (DAE) gain, offset, and small tunable offset, respectively,
whose optimal values were determined during ground tests \citep{maris2009}.

\subsection{On-Board Time reconstruction}

A time stamp is assigned to each data sample.  If the phase switch \citep{planck2011-1.4} is off (not switching), the packet contains consecutive values of either sky or load samples.  Then
\begin{equation}
  t^{\mathrm{obt}}_{i_{\mathrm{smp}}} = t^{\mathrm{obt}}_{0} + i_{\mathrm{smp}}\frac{N_{\mathrm{aver}} }{f_{\mathrm{samp}}}\, ,
\end{equation}
where ${i_{\mathrm{smp}}} \ge 0$ is the sample index within the
packet and $t^{\mathrm{obt}}_{0}$ is the mean time stamp of the
first averaged sample.  $N_{\rm aver}$ is the number of fast
samples averaged together to obtain a single detector sample, and
$f_{\rm samp} \simeq 4$\,kHz is the detector sampling frequency.

If the phase switch is on (nominal case),
consecutive pairs of either sky$-$load or load$-$sky samples are stored in
the packet.  Then consecutive pairs of samples have the same time
stamp and
 \begin{equation}
  t^{\mathrm{obt}}_{i_{\mathrm{smp}}} = t^{\mathrm{obt}}_{0} + 2\;{{\mathrm{trunc}}}\left[\frac{i_{\mathrm{smp}}}{2}\right]\frac{N_{\mathrm{aver}}}{f_{\mathrm{samp}}}.
 \end{equation}
On-board time information is stored in the form of TOI and
 directly linked to its scientific sample.

\subsection{Data flagging}

For each sample we define a 32-bit flag
mask to identify potential inconsistencies in the data and to enable the
pipeline to skip that sample or handle it differently during further processing.
Currently flags that are checked include: those
identifying the stable pointing period (determined from the AHF);
science data that cannot be recovered (e.g., because of
saturation); samples artificially created to fill data gaps; and
samples affected by planet transits and moving objects within the
Solar system.

\section{TOI processing}
\label{toiprocessing}

\subsection{Electronic spikes}

\begin{figure}[here]
\centerline{
\includegraphics[width=\columnwidth]{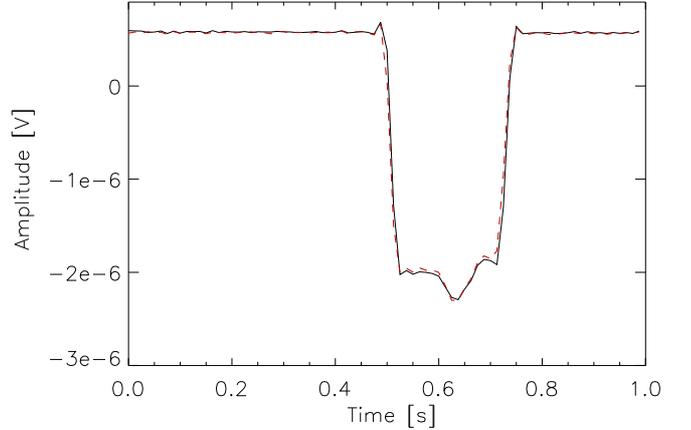}
}
\caption{A square wave template for both sky (black) and load
(red/dashed line) for one of the 44\,GHz detectors, computed by adding
data between Operational Day (OD) 91 and 389 in phase over a 1-hr interval. Individual
templates are directly subtracted from the un-differenced data.}
\label{squarewave}
\end{figure}

The clock in the housekeeping electronics is inadequately shielded from the data lines, resulting in noise ``spikes'' in the frequency domain at multiples of 1\,Hz \citep{meinhold2009}.  The spikes are synchronous with the on-board time, with no change in phase over the entire survey, allowing construction of a  piecewise-continuous template by summing the data for a given detector onto a one second interval (Fig.~\ref{squarewave})  The amplitude and shape differ from detector to detector; differences between detectors of different frequency tend to be larger than between those of the same frequency.  The amplitude also varies with time.  This variation is estimated by constructing templates like Fig.~\ref{squarewave} summed over the entire survey to obtain the shape of the signal, and then fitting the amplitude of a signal of this shape for each hour of data.   This amplitude is smoothed with a 20-day boxcar window function to reduce the noise.  Because of noise, this is likely to be an overestimate of the true variations.

\begin{figure*}[!ht]
\centerline{\hglue -3mm
\includegraphics[width=9.7cm]{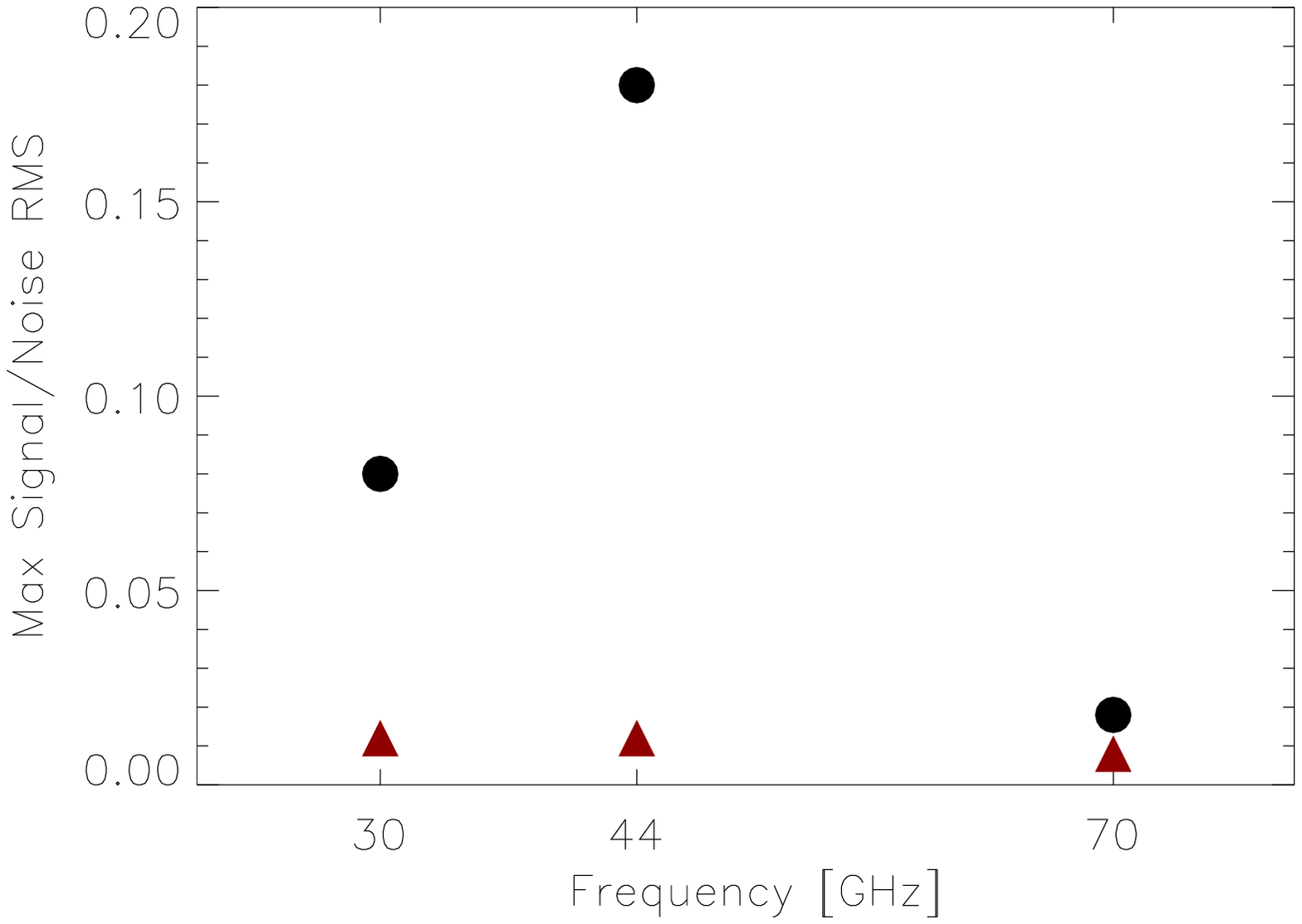}\hspace{-0.2cm}\includegraphics[width=9.7cm]{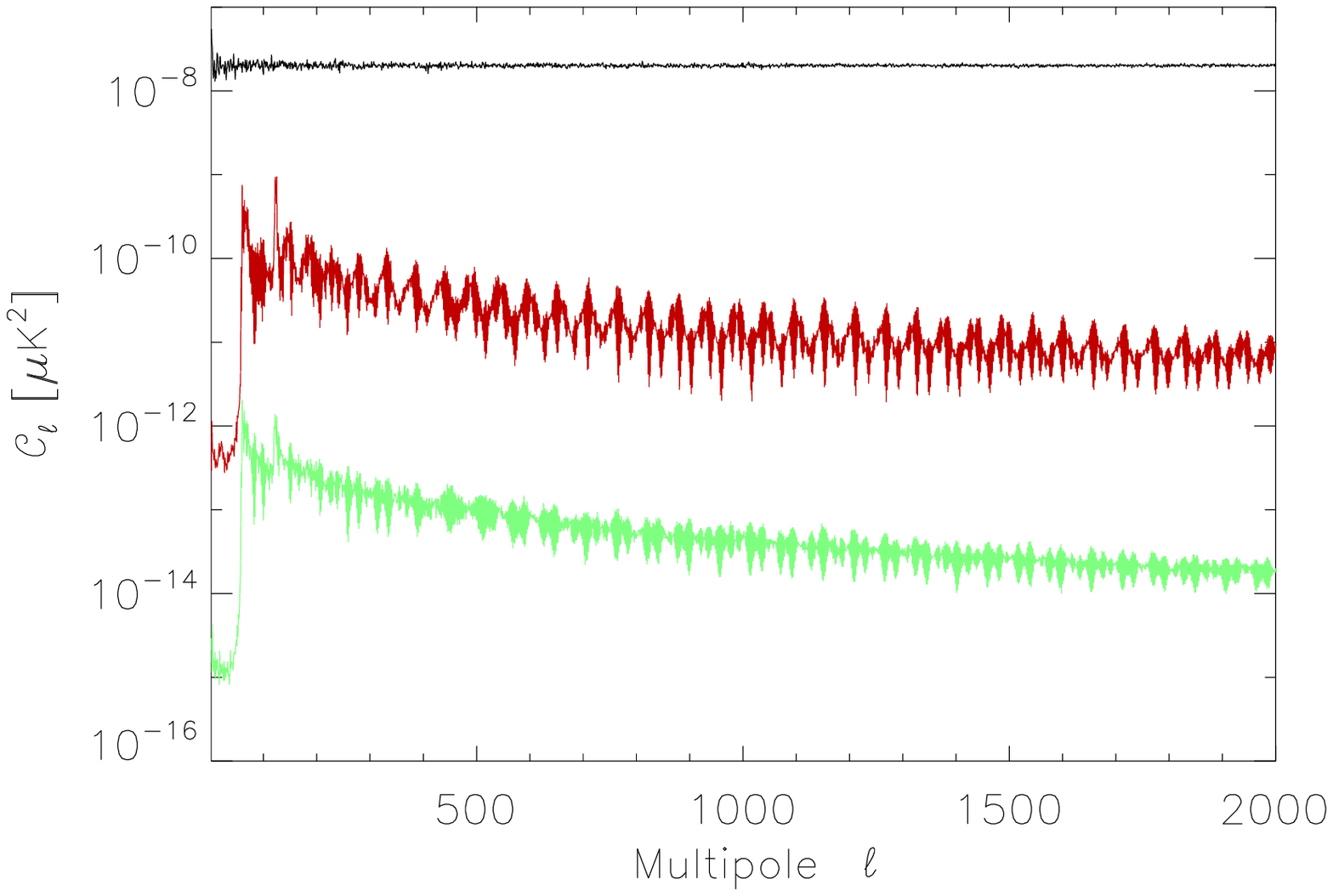}
}
\caption{The effect of electronic spikes on the data.  \textit{Left --- }Maximum pixel value in the simulated ``spike" (black dots) and ``spike-subtracted'' maps (red triangles; see text), scaled to the local pixel noise.  In our data processing, the square wave signal is subtracted only at 44\,GHz detectors.  The black circles therefore represent the estimated highest spike signal level in the 30 GHz and 70\,GHz maps, while the red triangle represents the estimated highest residual spike signal level in the 44\,GHz map.  \textit{Right --- }Angular power spectra of the 44 GHz simulations: the red (middle) line shows the power spectrum of the simulated spike map, and the green (bottom) line shows the power spectrum of the simulated spike-subtracted map.  Subtraction reduces the power by a factor of about $100$, from a small to an insignificant fraction of the white noise power, shown by the black (top) line.}
\label{fig:spikes_noise_comp}
\end{figure*}

To estimate the effect of spikes on the science data, we generate three simulated maps at each frequency.  The first is a noise map, generated from the instrument white noise levels as measured in the data and the scan strategy of \Planck, but no spikes or correlated noise.   This is a best case scenario, with the lowest noise level possible in a real map.  The second is a ``spike'' map, calculated assuming the square wave template for each detector modulated by a time-varying amplitude measured from the data, as described above.  Because the variation of amplitudes is an overestimate, as described above, this is a worst-case scenario of the effect of spikes.  The third map is a ``spike-subtracted'' map, the same as the second, but with a constant spike template subtracted.   This gives an estimate of the residual effect of electronic spikes that would be left in the maps if the spike template were subtracted.  The 30\,GHz maps are at HEALPix resolution $N_{\rm side}=512$; the 44\,GHz and 70\,GHz maps were produced at $N_{\rm side}=1024$.

We scale the spike and spike subtracted maps to the noise, i.e.,  Map2/rms(Map3) and Map2/rms(Map3), where the rms is calculated from the global rms of the noise map scaled as appropriate for the relative number of observations (``hits'') in that pixel.  Figure~\ref{fig:spikes_noise_comp} (left) shows the maximum value of these ratios over the whole sky.  At 44\,GHz, the most affected frequency, the effect in the worst pixel is less than 20\% of the noise.   At 70\,GHz the effect in the worst pixel is an insignificant 2\% of the noise.

Figure~\ref{fig:spikes_noise_comp} (right) gives angular power spectra of the three 44\,GHz maps.  The effect is everywhere well below the noise, and subtraction of a constant amplitude square-wave template reduces the effect by almost two orders of magnitude.

We decided to remove a square-wave template only at 44\,GHz.  This reduces the spike residual from 20\% of the noise to 1\% of the noise.  At 70\,GHz the effect of spikes is extremely small without correction, and at 30\,GHz uncertainty in the template combined with the small size of the effect argued against removal.

\subsection{Gain modulation factor and differenced data}
\label{diff_data}

The output of each detector (diode) switches at 4096\,Hz \citep{mennella2010} between the sky,  $V_{\rm sky}$, and the 4\,K reference load, $V_{\rm load}$.  $V_{\rm sky}$ and $V_{\rm load}$ are dominated by $1/f$ noise, with knee frequencies of tens of hertz.  This noise is highly correlated between the two streams, a result of the pseudo-correlation design \citep{bersanelli2010}, and differencing the streams results in a dramatic reduction of the $1/f$ noise.  The two arms of the radiometer are slightly unbalanced, as one looks at the 2.7\,K sky and the other looks at the $\sim 4.5$\,K reference load.  To force the mean of the difference to zero, the load signal is multiplied by the GMF, $R$, which can be  computed in several ways \citep{mennella2003}. The simplest method, and the one implemented in the processing pipeline, is to take the ratio of DC levels from sky and load outputs obtained by averaging the two time streams, i.e., $R = \langle V_{\rm sky}\rangle/\langle V_{\rm load}\rangle$.  Then
\begin{equation}
\Delta V(t) = V_{\rm sky}(t) - \frac{\langle V_{\rm sky}\rangle}{\langle V_{\rm load}\rangle} V_{\rm load}(t)\, .
\label{requation}
\end{equation}
We compute $R$ from unflagged data for each pointing period identified from the AHF information.

To verify the accuracy of this approach, we started with a time
stream of real differenced data, then generated two time streams
of undifferenced data using a constant (typical) value of $R$.  We
then ran these two time streams through the pipeline, and compared
the results with the original time stream.  Deviations between the
pipeline values of $R$ and the constant input value used to
generate the undifferenced data were at the 0.002\% level.

The $R$ factor has been stable over the mission so far,  with overall variations of 0.03--0.04\%.  To keep the pipeline simple, we apply a single value of $R$ to each pointing period.

\begin{figure*}
\hspace{1.5cm}
\includegraphics[width=12cm,angle=90]{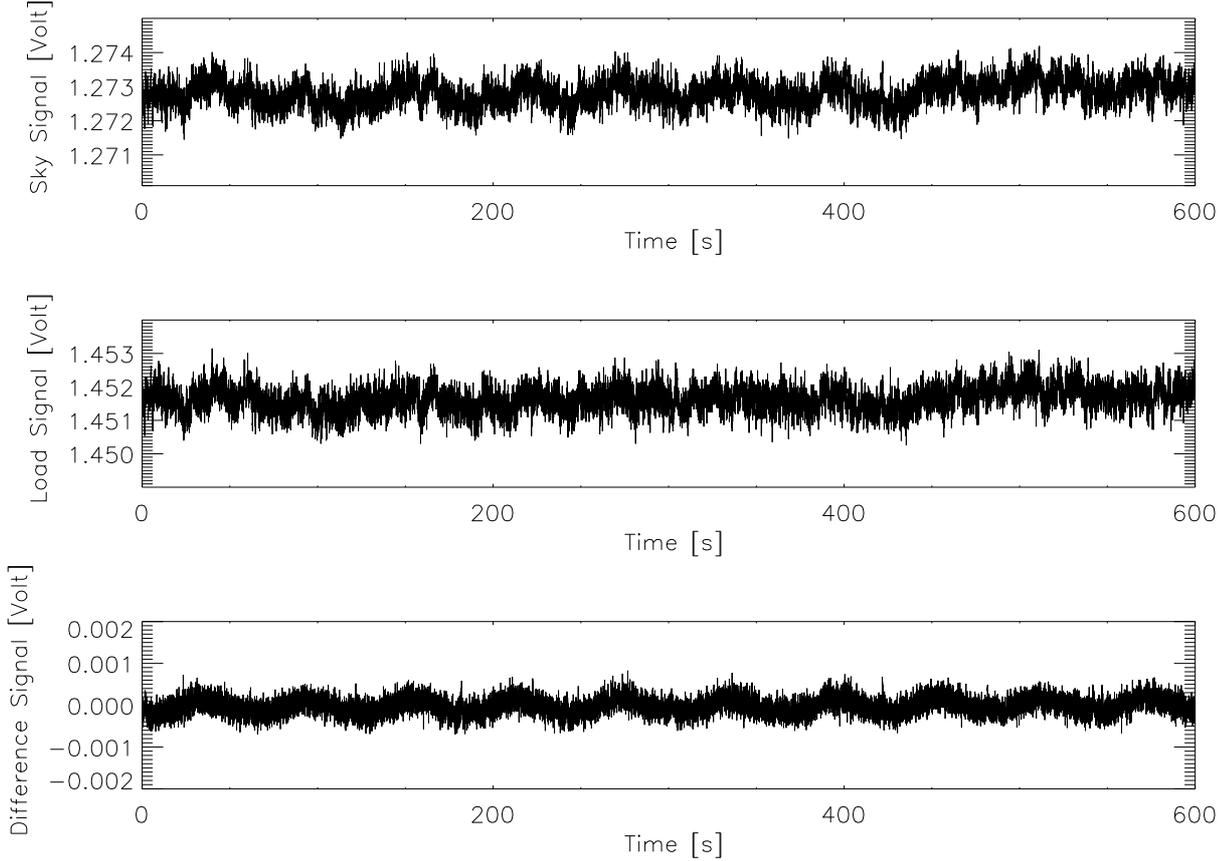}
\caption{Effect of the gain modulation factor (GMF) on sky and load signals for flight data. The upper and middle panels show 10\,min of sky and load signals of the {\tt LFI27S-11} detector: they are highly correlated with clear signatures of low-frequency noise.  After application of the GMF in taking the difference (Eq.~\ref{requation}), such fluctuations are dramatically reduced, revealing the presence of a sky signal dominated by the CMB dipole (lower panel).  Note the change in the y-axis scale.}
\label{skyloaddatadiff}
\end{figure*}

Figure~\ref{skyloaddatadiff} shows the effect of applying Eq.\ref{requation} with the $R$ factor to flight data.  The correlated $1/f$ noise in sky and load streams (evident in the two upper plots of the figure) is reduced dramatically.  The residual $1/f$ noise has a knee frequencies of 25\,mHz, and little effect on maps of the sky, as described in Sect.~\ref{mmaking}.

\subsection{The diode combination}

Having two diodes for each radiometer enables observation of both sky and load with a combined duty cycle of almost 100\%.  In combining the outputs, however, we must take into account the effects of imperfect isolation and differences in noise between the two diodes.

Isolation between diodes was measured for each radiometer in ground tests and verified in flight using the CMB dipole, planets, and Galactic plane crossings.  Typical values range from $-13$ to $-20$\,\hbox{dB}.  This is within specifications, and does not compromise LFI sensitivity.  It does, however, produce a small  anti-correlation of the white noise of the two diodes of a given radiometer.  When data from the two diodes are averaged, the white noise of the resulting TOI is lower than would be the case if they were statistically independent.  A complete mathematical description of this behaviour is given in \cite{planck2011-1.4}.  This causes no difficulty in subsequent calibration and further processing; however, the effect must be taken into account in inferring the noise properties of individual detector chains from the combined outputs.

To take account of differences in noise in combining the diode outputs, we assign relative weights to the
uncalibrated diode time-streams based on their calibrated noise.  Specifically, we make a first order calibration of the timelines, $G_{0}$ and $G_{1}$, subtract a signal estimate, and calculate the calibrated white noise levels, $\sigma_0$ and $\sigma_1$, for the two diodes.  The weights for the two diodes ($i = 0$ or 1) are
\begin{equation}
W_i = \frac{\sigma_i^2}{G_{01}}\frac{1}{\sigma_0^2 + \sigma_1^2}\,
\end{equation}
where the weighted calibration constant $G_{01}$ is given by
\begin{equation}
G_{01} = \frac{1}{\sigma_0^2 + \sigma_1^2}\left[G_0\sigma_1^2 + G_1\sigma_0^2\right]\, ,
\end{equation}
and is the same for each diode.

The weights are fixed to a single value per radiometer for the entire dataset.  Since all calibrations, noise estimation, and other tests are done on these combined data streams, small errors in the weights  cause inconsequential losses in sensitivity, and no systematic errors.

\subsection{Detector pointing}
\label{detectorpointing}

Detector pointing is a fundamental ingredient in data processing
that requires knowledge of the spacecraft attitude and the
location of the horns in the focal plane.  The AHF gives the
orientation of the spacecraft spin axis in quaternions sampled at
8\,Hz, as well as beginning and ending times for a single pointing
period.  It specifies with appropriate flags the periods of
spin-axis maneuvers during which star tracker positions are
unreliable.  Horn locations within the focal plane are determined
from both ground measurements and planet crossings.

The orientation of the spacecraft spin axis at the time of each data sample is determined by linear spherical interpolation of the 8\,Hz quaternions.  Individual detector pointings are determined by simple rotations from the spin-axis reference frame to the telescope optical axis, then to the relevant horn position, with an additional rotation to account for the orientation of the horn in the focal plane.

In some cases small extrapolations of the quaternions are necessary at the end of a pointing period.  Simulations verify that these introduce no significant degradation of the pointing accuracy.

\section{Main beams and the geometrical calibration of the focal plane}
\label{beamrecovery}

Knowledge of the beams is of paramount importance in CMB experiments.  Errors and uncertainties, and the details of complex non-Gaussian shapes, directly affect cosmological parameters.

We determine the main beam parameters and the position of each horn in the focal plane from planet observations.  Jupiter gives the best results, but other planets and bright celestial sources have been used as well.  Inputs to the calculations include TOI from each radiometer throughout the planet crossing, the AHF for the same period, and the time-dependent position of the planet as seen by \Planck, provided by the JPL Horizons system, which accounts for both spacecraft and planet motion.

\subsection{Algorithm and testing}
\label{algo}

We create a 2D map of the footprint of the focal plane on the sky by selecting data within 10\deg\ of the telescope line of sight.  This comprises the whole extension of the LFI focal plane.   To minimize the effects of $1/f$ noise on weak sources, we use TOI from which offsets per ring derived by the {\tt Madam} destriper (Sect.~7) have already been removed.

We fit a bivariate Gaussian beam model to these data \citep{burigana2001}:
\begin{eqnarray}
\label{eq:any-reference}
B(x_i,y_i) &=& \frac{A}{d^2} {\rm exp}\left\{-\frac{1}{2}\left[\frac{(\Delta x_i\,{\rm cos}\alpha + \Delta y_i\,{\rm sin}\alpha)^2}{\sigma_{x}^2}\, +\right . \right . \nonumber \\
        &\,& \left . \left .    \quad\quad\quad\quad\,\,\,\,\,  \frac{(-\Delta x_i\,{\rm sin}\alpha + \Delta y_i\,{\rm cos}\alpha)^2}{\sigma_{y}^2}\right]\right\}.
\end{eqnarray}
Here $A$ is an overall amplitude.  $x_i$ and $y_i$ are Cartesian coordinates, with $x_{\mathrm c}$, $y_{\mathrm c}$ the position of the centre of the beam, and $\Delta x_i \equiv x_i - x_{\mathrm c}$ and $\Delta y_i \equiv y_i - y_{\mathrm c}$.  \quad$\sigma_x$ and $\sigma_y$ are the beamwidth parameters of the elliptical approximation of the beam shape, and the angle $\alpha$ is
the reconstructed orientation of the beam in the focal plane and
$d$ is the actual distance (in astronomical units) of the planet.

We tested our technique with simulations using the measured beam
patterns together with a detailed model of the \Planck\ telescope.
The simulations included the nominal main and far beam patterns,
the effects of smearing caused by the motion of the satellite, and
pointing uncertainties.  Using these simulations of Jupiter
crossings (including instrumental noise and complete sky signal),
we are able to reconstruct the main beam shape down to $-20$\,dB
and to recover the main beam properties at the 1\% level or better
for all LFI beams.  Table 1 reports results for the main beam
properties for a sample of the LFI beams. These figures are
representative of our expected accuracy for in-flight beam and
focal plane reconstruction.

\begin{table}
\begingroup
\newdimen\tblskip \tblskip=5pt
\caption{Simulation of the reconstruction of the beams and focal plane geometry from observations of Jupiter, including realistic models of the beams, instrument noise, beam smearing, and star tracker uncertainties.}
\label{beamres}
\nointerlineskip
\vskip -6mm
\footnotesize
\setbox\tablebox=\vbox{
   \newdimen\digitwidth
   \setbox0=\hbox{\rm 0}
   \digitwidth=\wd0
   \catcode`*=\active
   \def*{\kern\digitwidth}
   \newdimen\signwidth
   \setbox0=\hbox{+}
   \signwidth=\wd0
   \catcode`!=\active
   \def!{\kern\signwidth}
\halign{\hbox to 1.3in{#\leaderfil}\tabskip=1.6em&
        \hfil#\hfil&
        \hfil#\hfil&
        \hfil#\hfil\tabskip=0pt\cr
\noalign{\doubleline}
\omit&&&$\delta$\cr
\noalign{\vskip 4pt}
\omit\hfil Parameter\hfil&Input&Reconstruction&[\%]\cr
\noalign{\vskip 3pt\hrule\vskip 5pt}
\multispan2{\bf LFI19S --- 70\,GHz}\hfil\cr
\noalign{\vskip 2pt}
\hglue 2em FWHM & 12\parcm83& 12\parcm97& 1.12\cr
\hglue 2em ellipticity & 1.280& 1.276& 0.98\cr
\hglue 2em $x_0$&$-2\pdeg8715$& $-2\pdeg8704$& 0.36\cr
\hglue 2em $y_0$&$-1\pdeg5678$& $-1\pdeg5829$& 0.96\cr
\noalign{\vskip 7pt}
\multispan2{\bf LFI25S --- 44\,GHz}\hfil\cr
\noalign{\vskip 2pt}
\hglue 2em FWHM& 29\parcm33& 30\parcm23& 3.07\cr
\hglue 2em ellipticity & 1.170& 1.230& 5.12\cr
\hglue 2em $x_0$& $-2\pdeg8227$& $-2\pdeg8293$& 0.23\cr
\hglue 2em $y_0$& $-5\pdeg1369$& $-5\pdeg0844$& 1.02\cr
\noalign{\vskip 7pt}
\multispan2{\bf LFI27M --- 30\,GHz}\hfil\cr
\noalign{\vskip 2pt}
\hglue 2em FWHM & 32\parcm42& 32\parcm89& 1.45\cr
\hglue 2em ellipticity & 1.380& 1.384& 0.32\cr
\hglue 2em $x_0$&$-4\pdeg7788$&$-4\pdeg7798$& 0.02\cr
\hglue 2em $y_0$&  !2\pdeg4903&  !2\pdeg3958& 3.79\cr
\noalign{\vskip 5pt\hrule\vskip 3pt}}}
\endPlancktable
\endgroup
\end{table}

\begin{figure}[!ht]
\centerline{
\includegraphics[width=\columnwidth]{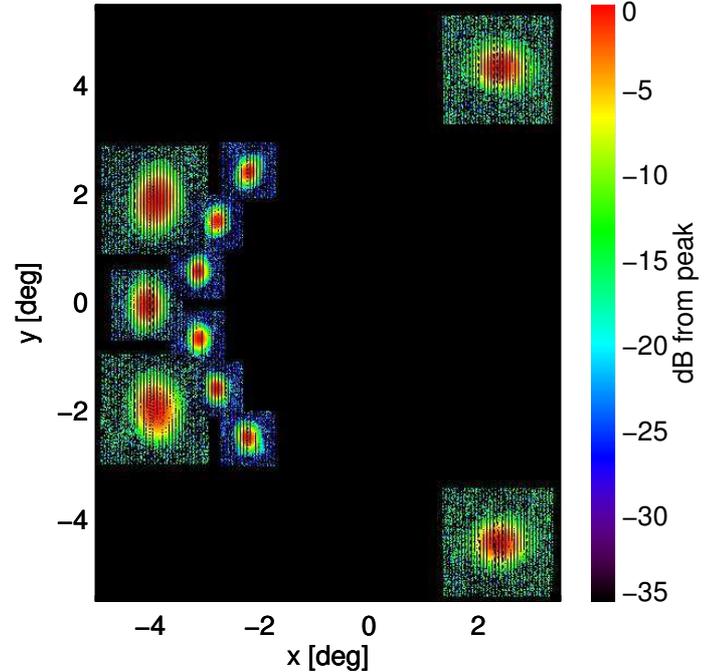}}
\caption{LFI focal plane as determined from the first season of
Jupiter observations, 24~October to 1~November 2009.  Contour
levels are in dB from the peak.  All beams are well approximated
by an elliptical Gaussian down to the $-10$\,dB level.}
\label{fp_jup1}
\end{figure}

\begin{figure*}[!ht]
\centerline{
\includegraphics[width=5cm,angle=270]{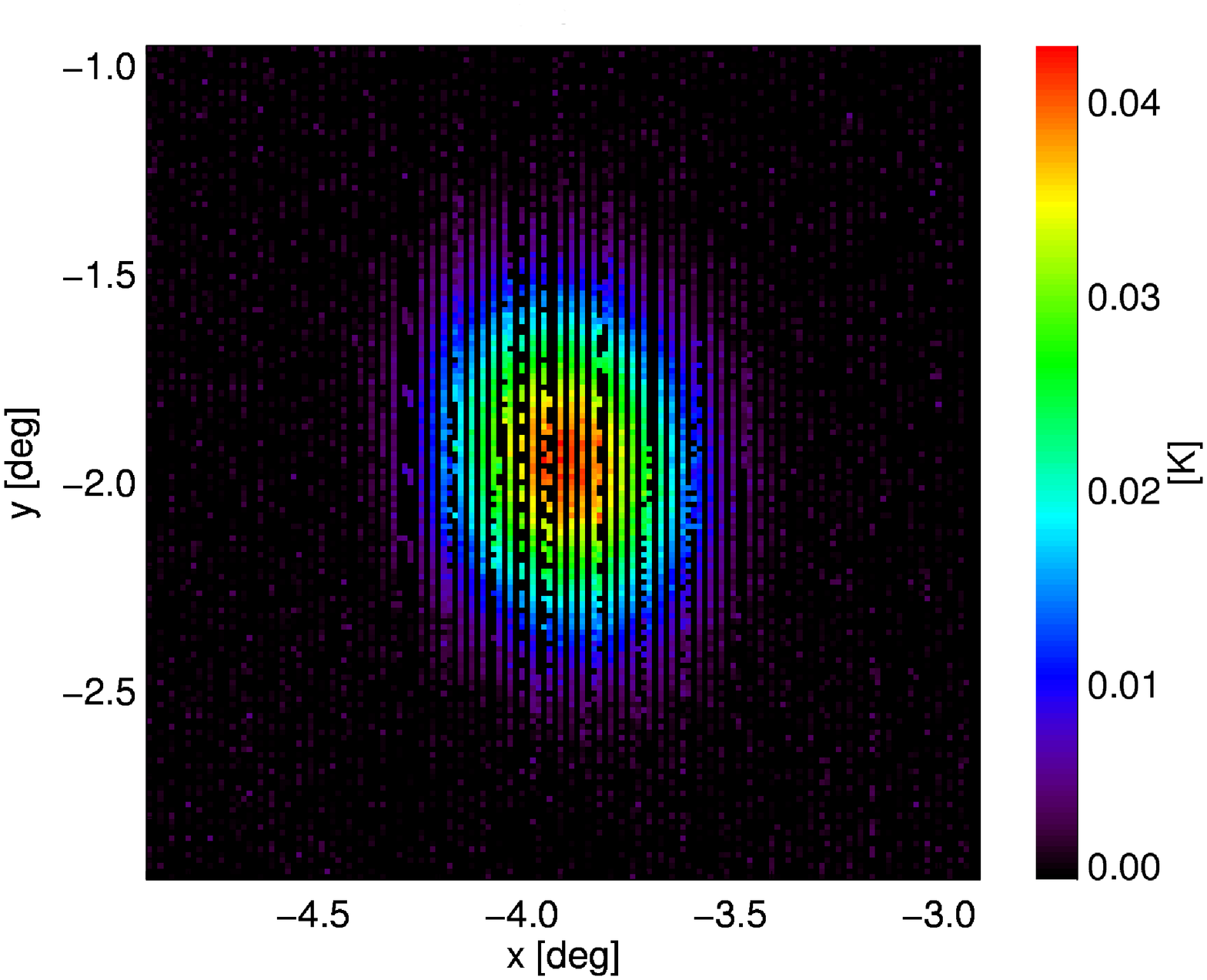}\includegraphics[width=5cm,angle=270]{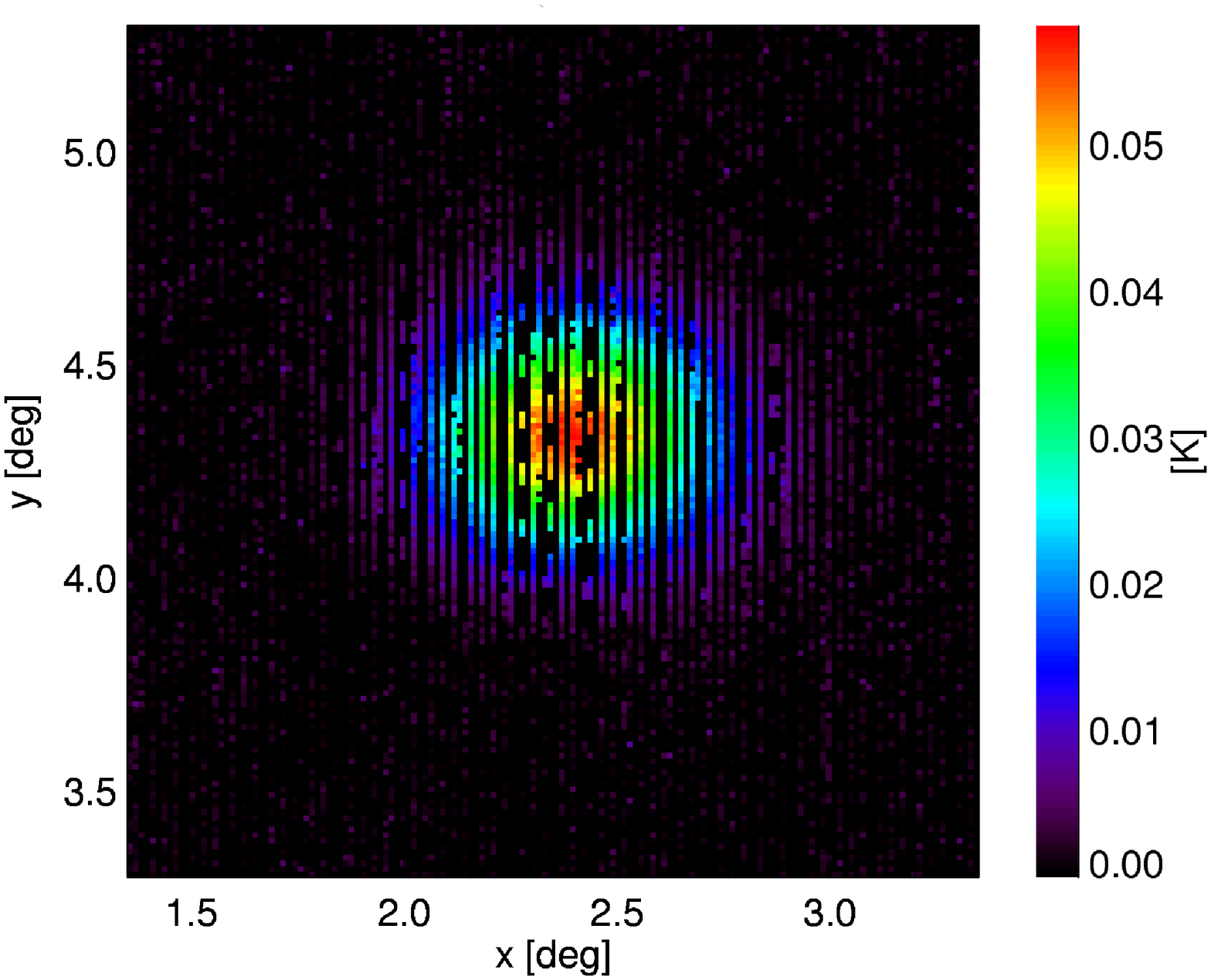}
\includegraphics[width=5cm,angle=270]{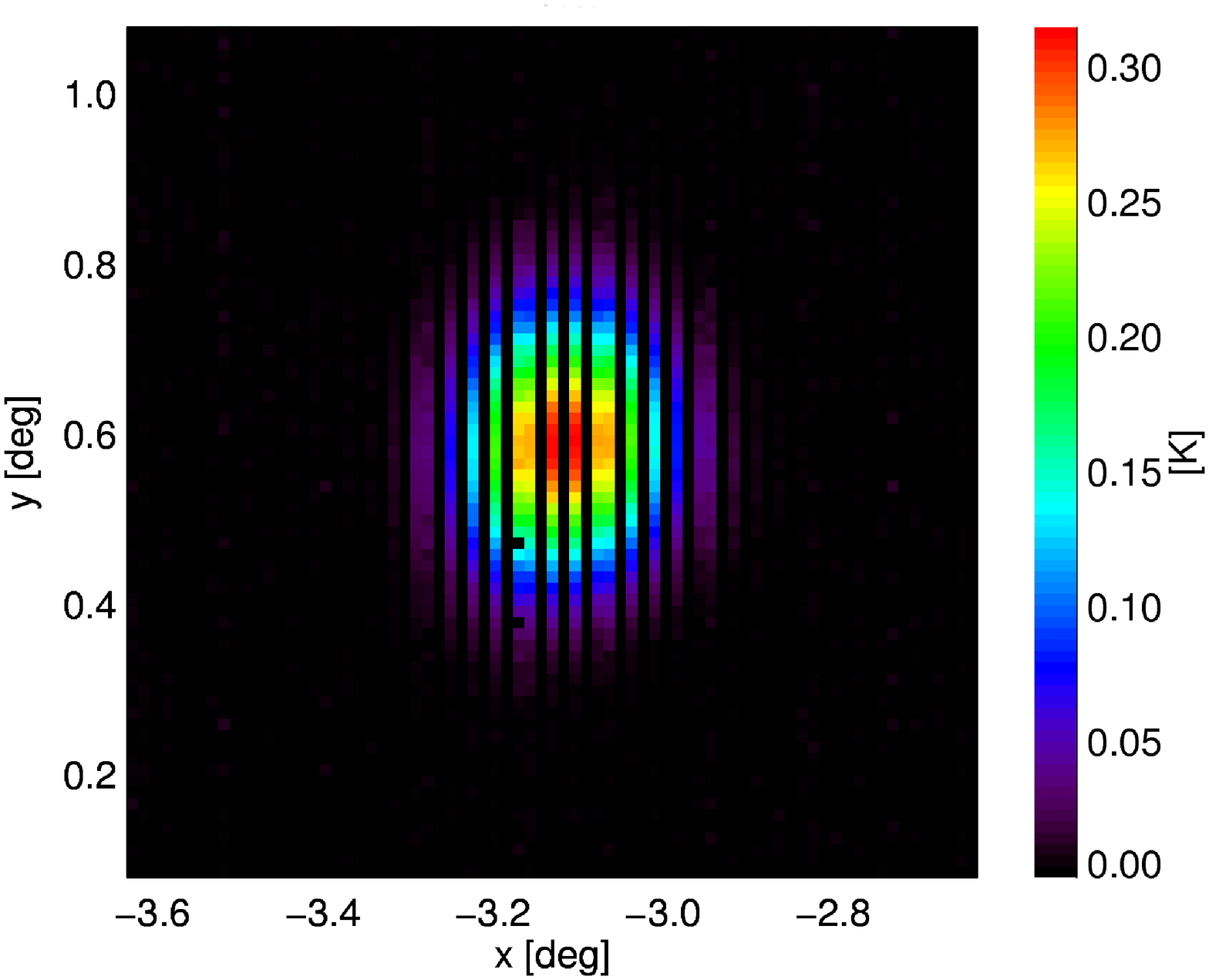}
} \caption{Beam map of LFI data around the Jupiter observations
(24 October--1 November 2009) for LFI horns {\tt LFI28M} (left),
{\tt LFI25M} (middle) and {\tt LFI21M} (right).} \label{jup2}
\end{figure*}

Figure~\ref{fp_jup1} shows the footprint of the LFI focal plane obtained during the first season of Jupiter observations, from 24 October to 1 November 2009.  Figure~\ref{jup2} shows  beam images for {\tt LFI28M}, {\tt LFI25M} and {\tt LFI21M} from those observations.  As expected, all beams are asymmetric but  with no significant departures from an elliptical shape visible down to the $\sim -10$ dB level.  For lower levels, aberration starts to distort the beam response, creating non-elliptical shapes.

We also constructed a planet mask, including Jupiter, Mars, and Saturn, the most luminous planets at LFI frequencies. The planet mask is radiometer-dependent, since each horn observes a planet at different times. The planet masking algorithm assigns an appropriate flag to data that lie within an ellipse, centred at the position of the planet and with an orientation that matches the beam orientation, with axes $\sim 3$ times larger than the beam widths derived from beam fitting.  These flags are used in the map-making and ensuing data analysis to discard samples affected by planet transits

\section{Photometric calibration}
\label{calibration}

\subsection{First steps}

The ideal source for photometric calibration, i.e., conversion of
the data from volts to kelvin, should be constant, perfectly
known, present during all observations, and have the same
frequency spectrum as the \hbox{CMB}.  In the frequency range of
the LFI, the CMB dipole, caused by the motion of the Solar system
with respect to the CMB reference frame, satisfies nearly all of
these requirements, lacking only in that it is well, but not
perfectly, known.  The modulation induced on the CMB dipole by the
orbital motion of \Planck\ around the Sun satisfies even this last
requirement, and will be the ultimate calibration source for the
LFI; however, it cannot be used effectively until data for a full
orbit of the Sun are available.  For this paper, therefore, we
must use the CMB dipole. We follow essentially the calibration
procedure used for the $WMAP$ first year data
\citep{hinshaw2003b}.  For the $k^{\rm th}$ pointing period, the
signal from each detector can be written as
\begin{equation}
{\Delta V}_{k} = g_k({\Delta T}_{\rm sky} + {n}) + b_k\, ,
\end{equation}
where ${\Delta T}_{\rm sky}$ is the sky signal,  ${n}$ is the noise, and $g_k$ and $b_k$ are the gain and baseline solution.  The dominant sky signal on short time scales is the CMB dipole (Galactic plane crossings produce a localized spike that is easy to exclude).   This is modeled as
\begin{equation}
{\Delta  V}_{\mathrm m}(g_k,b_k) = g_k({\Delta T}_d + {\Delta T}_{\mathrm v}) + b_k\, ,
\end{equation}
where we have considered both the cosmological dipole ${\Delta
T}_d$ and the modulation from the spacecraft motion ${\Delta
T}_{\mathrm v}$. We fit for $g_k$ and $b_k$ for each pointing
period $k$ by minimising
\begin{equation}
\chi^2 = \sum_{i\in k}\frac{\left[\mathbf{\Delta V}(t_i) - \mathbf{\Delta V}_m(t_i|g_k,b_k)\right]^2}
{{\rm rms}_i^2}\,.
\label{caliblaw}
\end{equation}
The sum includes unflagged samples within a given pointing period $k$ that lie outside a Galactic mask.

The mask is created from simulations of microwave emission provided by the Planck Sky Model (PSM)%
\footnote{The \Planck\ Sky Model is available at:
    {\tt http://www.apc.univ-paris7.fr/APC\_CS/Recherche/Adamis/\\PSM/psky-en.html}}.
Of the LFI frequencies, 30\,GHz has the strongest diffuse foreground emission.  The mask excludes all pixels that in the 30\,GHz PSM are more than $5\times 10^{-4}$ times the expected rms of the \hbox{CMB}.  It also excludes point sources brighter than 1 Jy found in a compilation of all radio catalogues available at high frequencies (the \Planck\ Input Catalogue, see \citealt{massardi2006}).  The Galactic and point source masks preserve $\sim 82\%$ of the sky.

The {\it Planck} scan strategy is such that the instrument field of view describes nearly great circles on the sky.  The signal mean is therefore almost zero and nearly constant from one circle to the next.
This reduces the correlation between the gain and baseline solutions, a feature also taken advantage of by WMAP \citep{hinshaw2003b}.

\begin{figure*}[!ht]
\centerline{
\includegraphics[width=14cm]{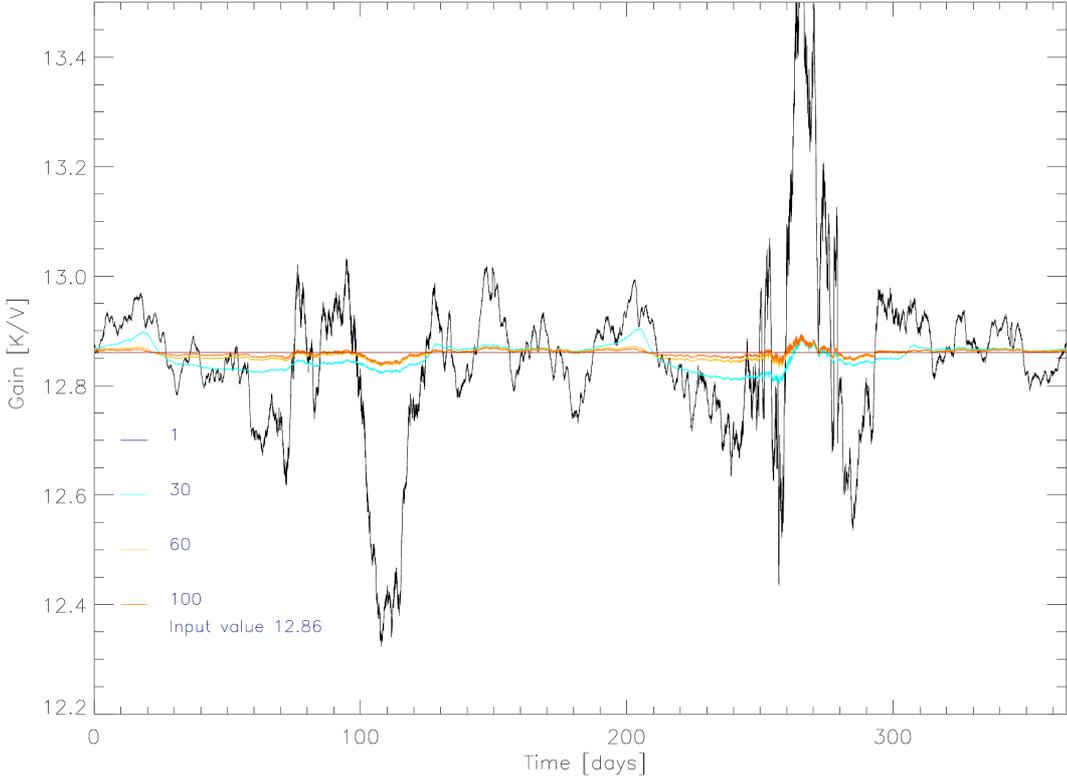}
} \caption{Simulation showing convergence of the gain solution for
one year of observations of one 30\,GHz detector.  The simulations
include CMB dipole(s), CMB anisotropies, and Galactic emission.
The input gain was 12.86\,K/V and the sky included all diffuse
components as well as nominal instrument noise. The first
iteration shows large errors caused by Galactic and CMB anisotropy
emissions; however, after one hundred iterations, convergence is
achieved with an overall deviation from the input value of less
than 0.01\%. The various curves show the solution after $1$, $30$,
$60$ and $100$ iterations.} \label{calibplot}
\end{figure*}

As pointed out by \citet{hinshaw2003b} and \citet{cappellini2003}, the largest source of error in Eq.~\ref{caliblaw} arises from unmodelled sky signal $\mathbf{\Delta T}_a$ from CMB anisotropy and emission from the Galaxy.  To correct this, we solve iteratively for both $g'_k$ and ${\Delta T}'_{\mathrm a}$.  If $g'_k$ is the solution at a certain iteration, the next solution is derived using Eq.~\ref{caliblaw} with
\begin{equation}
{\Delta V}' = {\Delta V} - g'_k{\Delta T}'_{\mathrm a}\, ,
\end{equation}
where ${\Delta T}'_{\mathrm a}$ is the sky signal (minus dipole
components) estimated from a sky map built from the previous
iteration step.  This is repeated to convergence, typically after
a few tens of iterations.  Figure~\ref{calibplot} shows the gain
error induced by unmodeled sky signal in a one-year simulation of
one 30\,GHz detector.  The simulation includes CMB anisotropies,
the CMB dipole, and Galactic emission.  Gain errors in this
example are $\sim 5\%$ after one iteration.  After a few tens of
iterations, the residual errors are $<0.01\%$ over the entire
year.

\begin{figure*}[!ht]
\centerline{
    \includegraphics[width=14cm]{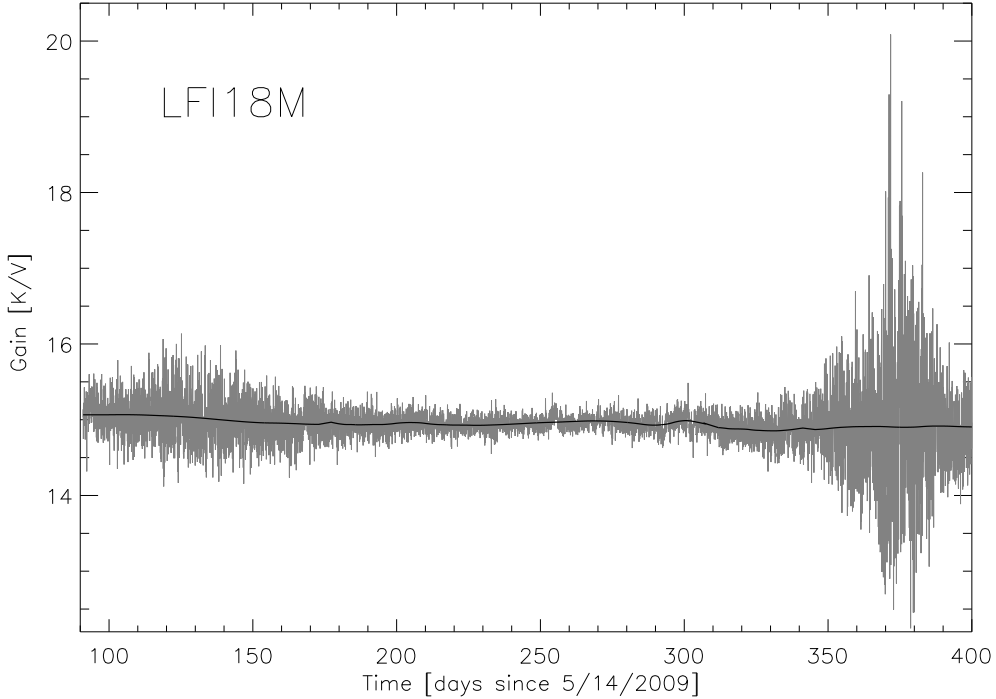}
  }
\caption{Hourly gain solution (gray line) from flight data for
{\tt LFI18M}, as derived from our iterative calibration algorithm.
The gain is quite stable over the observing period, although there
is a lot of scatter due to noise, especially during dipole minima.
The thick black line is the refined gain solution (see text)
applied to create calibrated TOI and sky maps.}
\label{realgain_18M}
\end{figure*}

The algorithm alternates between dipole fitting and map-making.
Maps are made with ({\tt Madam} Sect.~\ref{mmaking}) ignoring
polarisation, with no noise prior and baseline length equal to the
pointing period length. To improve calibration and reduce noise,
calibration is performed simultaneously for both radiometers of
each single horn.
In the presence of real noise, the actual
scatter from one gain solution to the other is quite large.
Figure~\ref{realgain_18M} shows an example of the hourly gain
solution (grey line) derived from the iterative scheme described
above for {\tt LFI18M}, one of the 70\,GHz radiometers.  Apart
from the scatter induced by instrument noise, the gain solution is
quite stable throughout the observation period.  Around the dipole
maxima, typical noise-induced variations are $\sim$ 0.8\% (rms).
Nonetheless, the stability of the gain solution is poor compared
to the stability of the instrument itself, as indicated by the
stability of the uncalibrated white noise level of both
differenced and undifferenced data, or the stability of the total
power from both sky and load signals.  This is particularly
evident during the minima of the dipole signal (see
\citealt{planck2011-1.4} for further details).

There are also specific things that affect the gain solution. To
the extent that they can be measured and understood, their effects
on the gain can be corrected directly.  For example, a
non-linearity in the analogue-to-digital converters (ADCs),
discovered during data analysis, produces a multiplicative effect
on the data that is recovered (erroneously) by the calibration
pipeline as a gain variation.  We have developed two independent,
complementary methods to correct for this.  In the first, we
calibrate the data using the gain solution that follows the
induced ADC gain variation.  In the second,  we model the
nonlinearity and remove the effect at the raw TOI level.

Alternatively, temperature variations of the amplifiers can induce
real gain variations on short time scales.  For example, during
the first 259 days after launch the downlink transponder was
powered up only for downlinks.  This induced rather sharp daily
variations in the temperature and gain of the amplifiers in the
back-end unit (BEM).  Starting on day 259, the transponder has
been powered up continuously, eliminating this source of gain
variations.

In the next section we discuss additional steps taken in the calibration procedure to deal with the effects of noise and gain changes induced by events such as the transponder cycle change.

\subsection{Improving calibration accuracy}

As shown in Fig.~\ref{realgain_18M}, the hourly gain solutions are strongly affected by noise.  To reduce the effects of noise and recover more accurately the true and quite stable gains of the instrument, we process the hourly gain solution as follows:

\begin{itemize}

\item calculate running averages of length 5 and 30 days.  The 5-day averages are still noisy during dipole minima, while the 30-day averages do not follow real but rapid gain changes accurately.

\item further smooth the 5- and 30-day curves with wavelets;

\item use the 30-day wavelet-smoothed curve during dipole minima;

\item use the 5-day unsmoothed curve around day 259 (the downlink transponder change) to trace real gain variations;

\item use the 5-day wavelet smoothed curve elsewhere.

\end{itemize}

A typical gain solution is plotted in Fig.~\ref{realgain_18M} as the solid black line.  From the 5- and 30-day gain curves we infer information on the actual gain stability of the instrument as the mission progresses, and also on the overall uncertainty in the gain reconstruction.  Specifically, the rms of the $g_k$ over a period of $N$ pointings is
\begin{equation}
\label{eq_gain_raw_uncertainty}
\delta g = \sqrt{\frac{\sum_{k=1}^N \bigl(g_k - \left<g\right>\bigr)^2}{N - 1}},
\end{equation}
where $\left<g\right>$ is the average of the $N$ gains.  The effect of the wavelet smoothing filter is to average over a number of consecutive pointings.  Ignoring the different weights in the average, the overall uncertainty can be approximated as
\begin{equation}
\left.\delta g\right|_\mathrm{stat} \simeq \frac{\delta g}{\sqrt{M}} = \frac1{\sqrt{M}}\,\sqrt{\frac{\sum_{k=1}^N  \bigl(g_k - \left<g\right>\bigr)^2}{N - 1}}.
\label{eq_gain_statistical_uncertainty}
\end{equation}

Table~\ref{gain_stats} lists the largest statistical uncertainties
and their associated mean gains out of four time windows
(days\,100--140, 280--320, 205--245, 349--389, the first two
corresponding to minimum and the second two to maximum dipole
response), for the main and side arms of the LFI radiometers.  In
order to provide conservative estimates, we have always chosen a
value for $M$ corresponding to the number of pointings in 5 days,
even in cases where a 30-day smoothing window was used.
Equations~\ref{eq_gain_raw_uncertainty} and
\ref{eq_gain_statistical_uncertainty} and Table~\ref{gain_stats}
are the same as equations 12 and 13 and Table~9 of
\cite{planck2011-1.4}.  Peak-to-peak variations in the daily gains
reach 10\% (with mean 7\%); however,  the rms of the smoothed gain
solution is generally in the $\sim 0.3-0.4\%$ range. This can be
taken as the current level of LFI calibration accuracy.

\begin{table}
\begingroup
\newdimen\tblskip \tblskip=5pt
\caption{Summary of dipole-based gain statistics}\label{beamres}
\label{gain_stats}
\nointerlineskip
\vskip -6mm
\footnotesize
\setbox\tablebox=\vbox{
   \newdimen\digitwidth
   \setbox0=\hbox{\rm 0}
   \digitwidth=\wd0
   \catcode`*=\active
   \def*{\kern\digitwidth}
   \newdimen\signwidth
   \setbox0=\hbox{+}
   \signwidth=\wd0
   \catcode`!=\active
   \def!{\kern\signwidth}
\halign{\hbox to 1.1in{#\leaderfil}\tabskip=2em&
        \hfil#\hfil\tabskip=1em&
        \hfil#\hfil\tabskip=2.2em&
        \hfil#\hfil\tabskip=1em&
        \hfil#\hfil\tabskip=0pt\cr
\noalign{\doubleline} \omit&\multispan2\hfil \sc Main
Arm\hfil&\multispan2\hfil \sc Side Arm\hfil\cr \noalign{\vskip
-4pt} \omit&\multispan2\hrulefill&\multispan2\hrulefill\cr
\noalign{\vskip 0pt} \omit&$\langle
g_k\rangle$&$\sigma_{g_k}$&$\langle g_k\rangle$&$\sigma_{g_k}$\cr
\noalign{\vskip 3pt} \omit\hfil \sc
Detector\hfil&[K/V]&[\%]&[K/V]&[\%]\cr \noalign{\vskip
3pt\hrule\vskip 5pt} \multispan2{\bf 70\,GHz}\hfil\cr
\noalign{\vskip 2pt} \hglue 1em LFI 18 &  14.935  & 0.279 & 22.932
&  0.243\cr \hglue 1em LFI 19 &  27.434  & 0.141 &   41.843 &
0.228\cr \hglue 1em LFI 20 &  25.572  & 0.253 &   29.581 &
0.261\cr \hglue 1em LFI 21 &  41.629  & 0.367 &   41.999 &
1.038\cr \hglue 1em LFI 22 &  64.275  & 0.367 &   62.504 &
0.185\cr \hglue 1em LFI 23 &  36.492  & 0.290 &   54.121 &
0.382\cr \noalign{\vskip 7pt} \multispan2{\bf 44\,GHz}\hfil\cr
\noalign{\vskip 2pt} \hglue 1em LFI 24 &282.295  & 0.349 &175.728
& 0.306\cr \hglue 1em LFI 25 &123.141  & 0.358 &123.958  &
0.279\cr \hglue 1em LFI 26 &167.364  & 0.398 &142.061  & 0.411\cr
\noalign{\vskip 7pt} \multispan2{\bf 30\,GHz}\hfil\cr
\noalign{\vskip 2pt} \hglue 1em LFI 27 &  12.875  & 0.314 & 15.320
& 0.349\cr \hglue 1em LFI 28 &  15.802  & 0.225 &  19.225 &
0.379\cr \noalign{\vskip 5pt\hrule\vskip 3pt}}}
\endPlancktable
\endgroup
\end{table}

Although the current pipeline provides results approaching those
expected from the stability of the instrument, we are working to
improve it as much as possible.  In particular, we would like to
trace gain variations on time scales shorter than the pointing
period.  To achieve this, we are developing a detailed gain model
(currently under test) based on calibration constants estimated
from the pipeline and instrument parameters (temperature sensors,
total power data), see \cite{planck2011-1.4} for further
information.

\section{Noise estimation}
\label{noise}

Once data are calibrated, we evaluate the noise properties of each
radiometer. We select data in chunks of 5~days each and then
compute noise properties.  This is done using the {\tt roma}
Iterative Generalized Least Square (IGLS) map-making algorithm
\citep{natoli2001,degasperis2005} which includes a noise
estimation tool based on the iterative approach described in
\citet{prunet2001}.  IGLS map-making is time and resource
intensive and cannot be run over the whole data set within the
current DPC system. However since the TOD length considered here
is only 5 ODs, it is possible to use the {\tt roma} implementation
of this algorithm which has a noise estimator built-in. The method
implemented here is summarized as follows.  Model the calibrated
TOD as
\begin{equation}
\mathbf{\Delta T} = \mathbf{P} \mathbf{m} +  \mathbf{n},
\end{equation}
where $\mathbf{n}$ is the noise vector, and $\mathbf{P}$ is a projection matrix
that relates a map pixel $\mathbf{m}$ to a TOD measurement $\mathbf{\Delta T}$.
We obtain a zeroth order estimate of the signal through a rebinned map and then
iterate noise and signal estimation:
\begin{equation}
\mathbf{\hat{n}_i} = \mathbf{\Delta T} - \mathbf{P\hat{m}_i},
\end{equation}
\begin{equation}
\mathbf{\hat{m}_{i+1}} = \mathbf{(P^T\hat{N}_i^{-1}P)^{-1}P^T\hat{N}_i^{-1}\Delta T},
\end{equation}
where $\mathbf{\hat{N}_i}$ is the noise covariance matrix in time
domain estimated at iteration $i$. We have verified that
convergence is reached in a few, usually three, iterations.

\begin{figure*}[!ht]
\centerline{\includegraphics[width=7cm,angle=90]{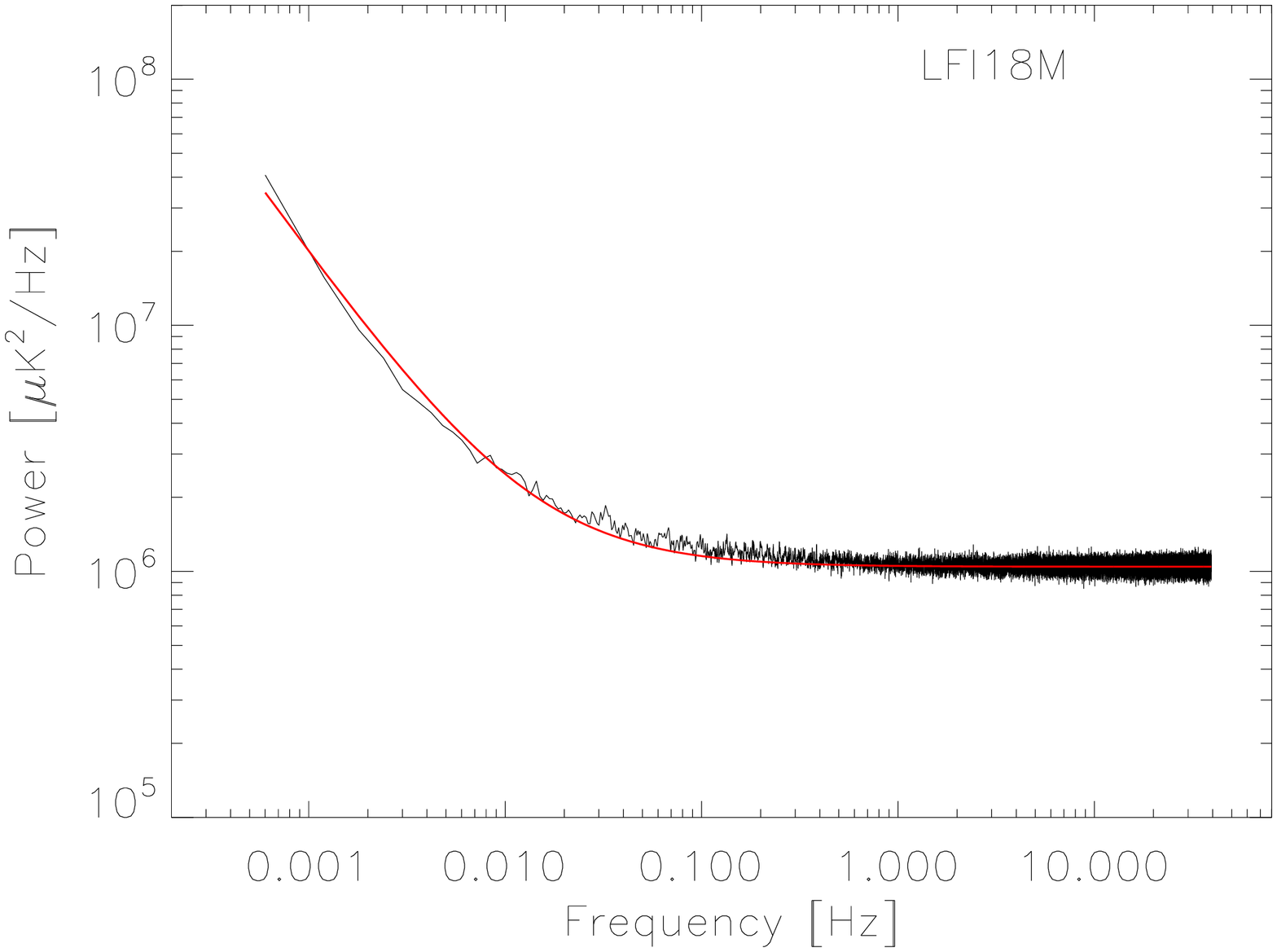}
            \includegraphics[width=7cm,angle=90]{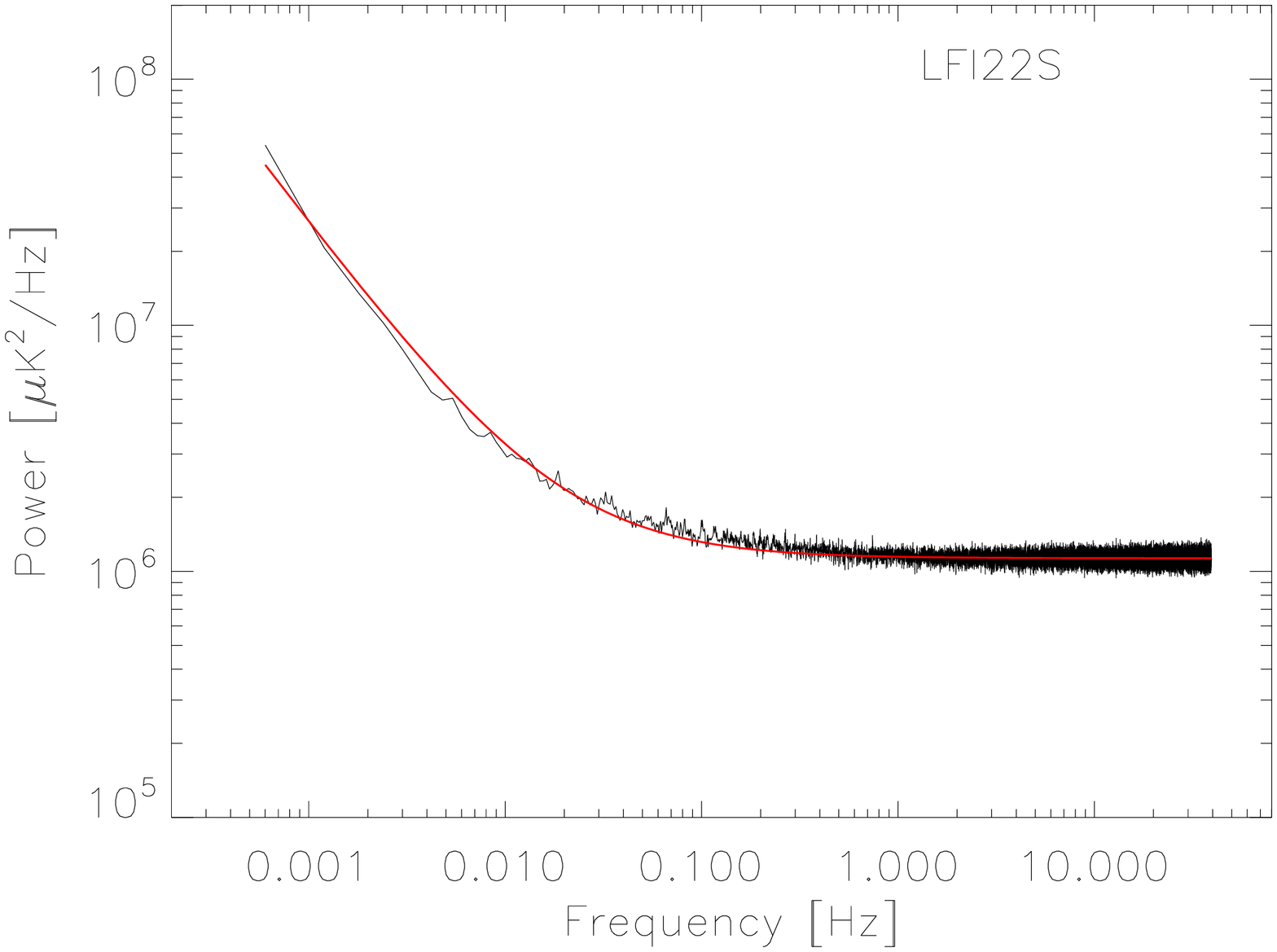}}
\centerline{\includegraphics[width=7cm,angle=90]{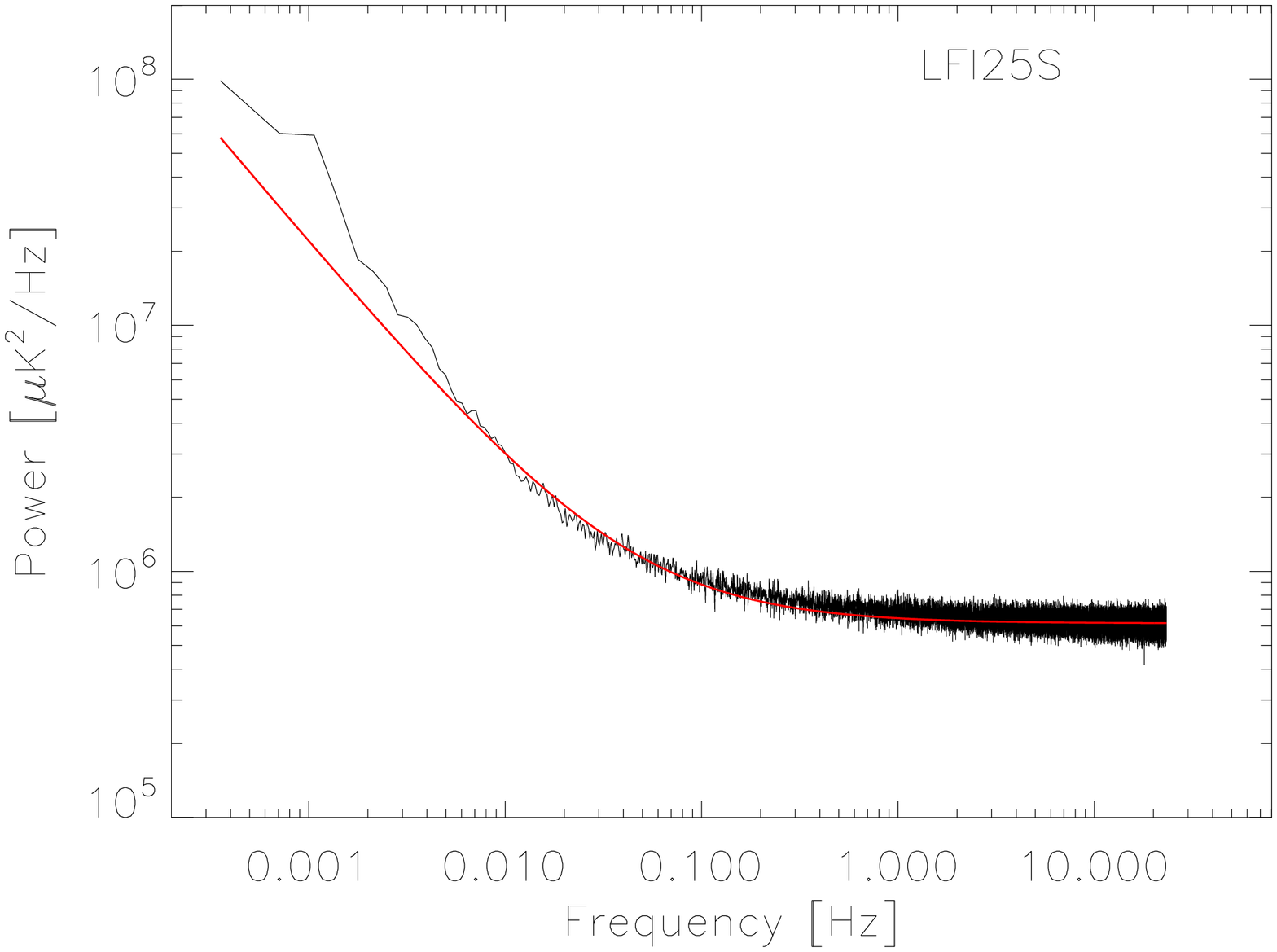}
            \includegraphics[width=7cm,angle=90]{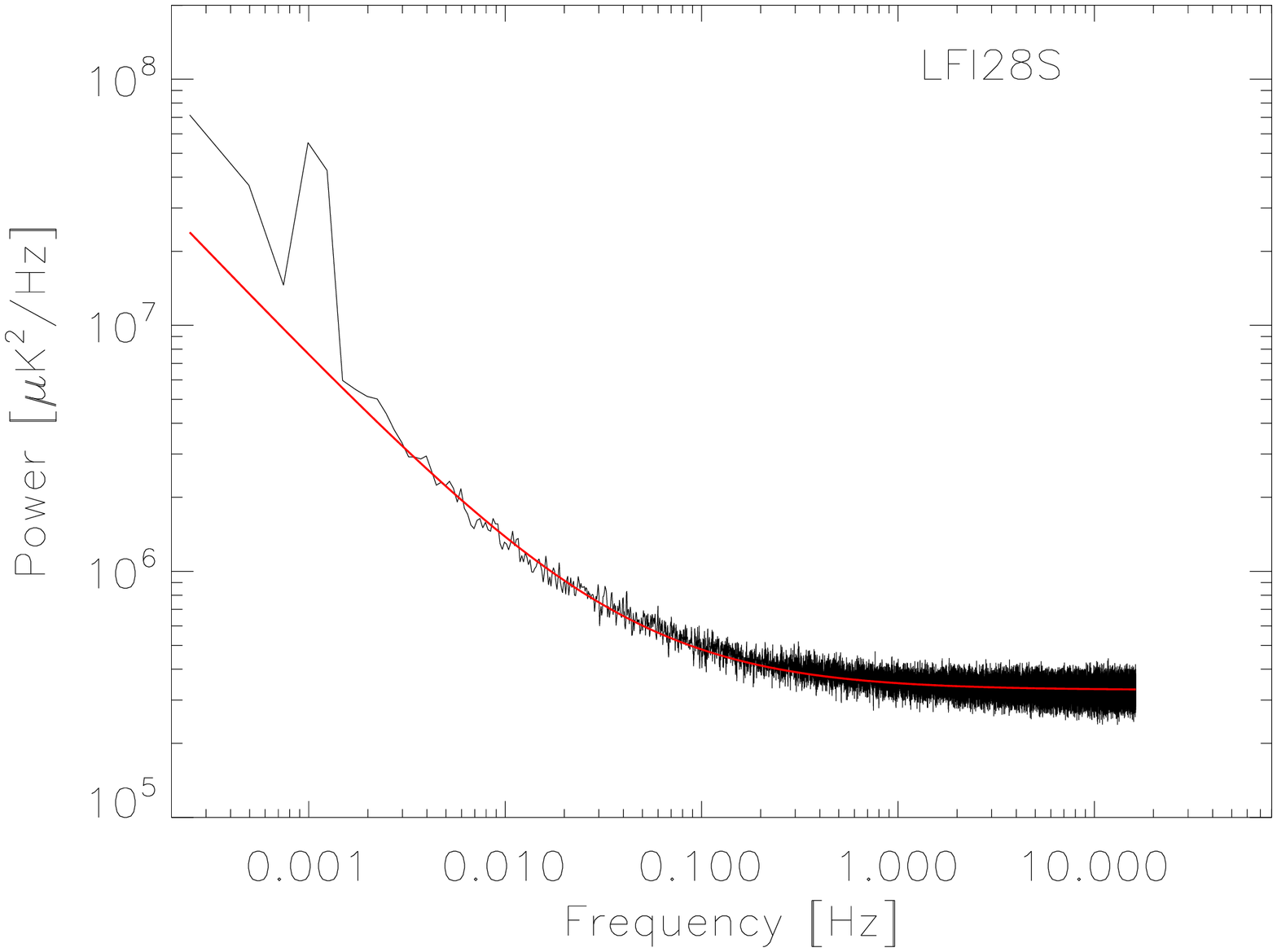}}
\caption{Noise spectra of radiometers {\tt LFI18M}, {\tt LFI22S},
{\tt LFI25S}, and {\tt LFI28S}) estimated by the noise pipeline
(black lines).  All spectra are well-fit by Eq.~\ref{noisefig}
with a single knee frequency and slope (red lines).  An excess
near 1\,mHz is visible in {\tt LFI25S} and {\tt LFI28S}.  This is
approximately the bed-switching frequency of the sorption cooler,
and the different slopes in {\tt LFI28S} and {\tt LFI25S} on the
low-frequency side of the spectrum are possibly indications of
thermal effects on the radiometer output.} \label{noisefig}
\end{figure*}

We calculate the Fourier transform of the noise time stream (with
an FFT algorithm) and fit the resulting spectrum for the three
parameters, the white noise level, the knee-frequency, and the
slope of the $1/f$ noise part:
\begin{equation}
P(f) = \sigma_{\rm WN}^2\left[1 + \left(\frac{f}{f_{\rm k}}\right)^\beta \right]\, .
\label{noise_spectrum}
\end{equation}
The white noise level is taken as the average of the last few
percent of frequency bins.  A linear fit to the log-log spectrum
low frequency tail gives the slope of the $1/f$ noise.  The knee
frequency, $f_{\rm k}$, is the frequency at which these two
straight lines intersect. We tested the accuracy with simulations
that included sky signal and instrumental noise with known
properties.  The noise properties were recovered with typical
deviations from input values of $\sim 10\%$ for knee-frequency and
slope, and less for white noise level. Examples of noise spectra
and corresponding fits are shown in Fig.~\ref{noisefig}.

\subsection{Noise constrained realizations and gap filling}
\label{gapfilling}

\begin{figure*}[!ht]
\centerline{\includegraphics[width=8cm]{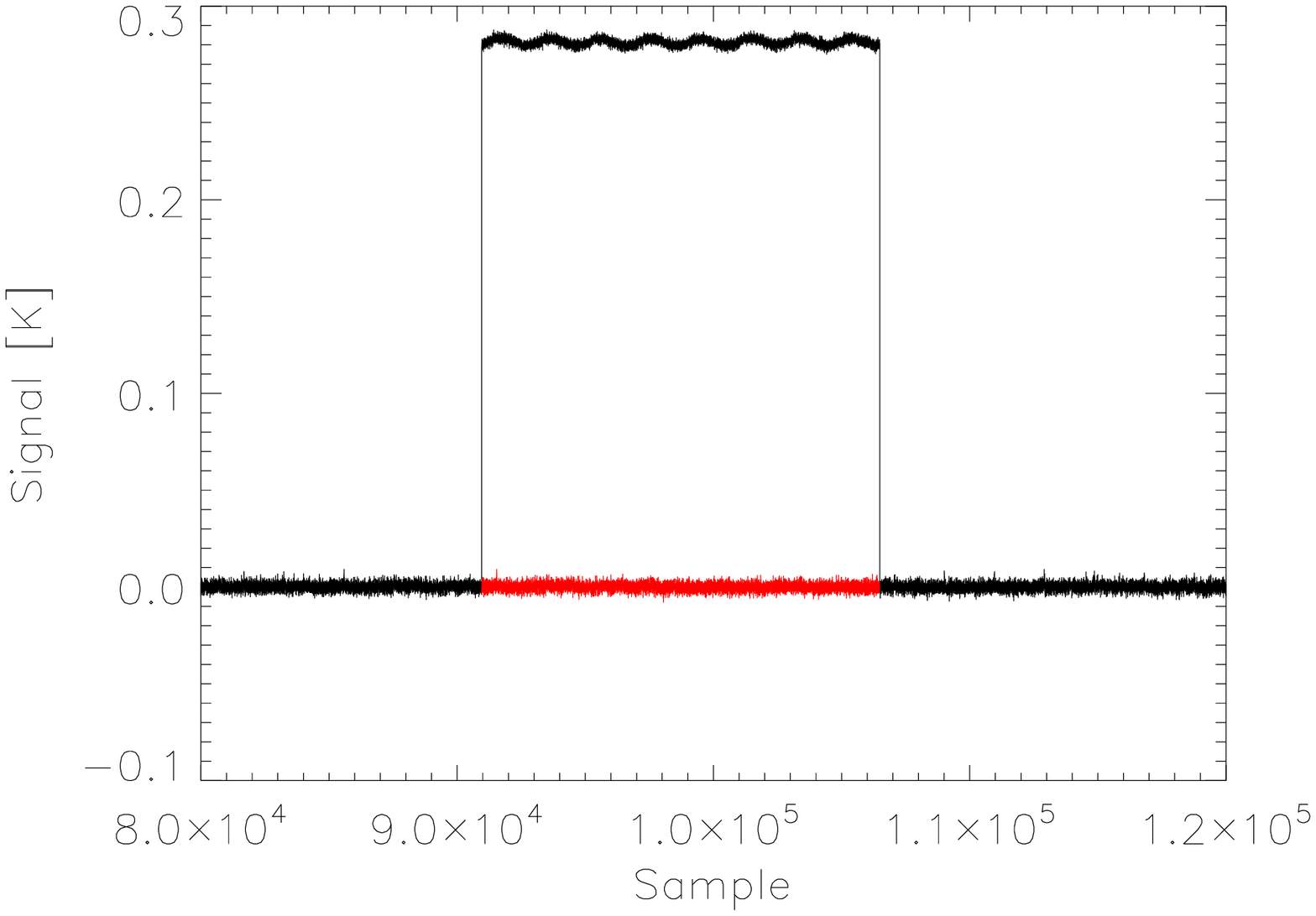}
            \includegraphics[width=8cm]{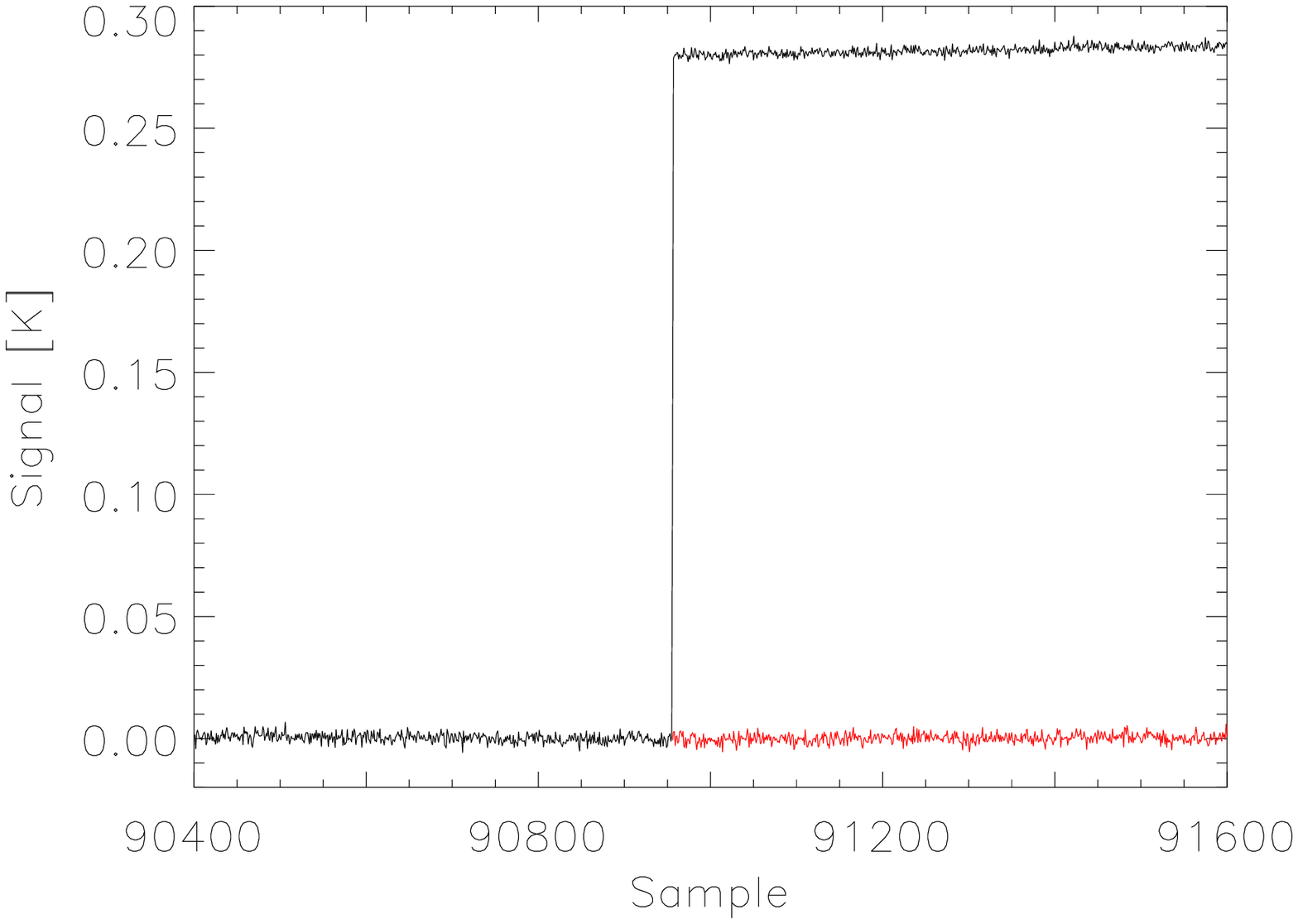}}
\caption{Gap filling procedure applied to {\tt LFI28M} for day 239. The upper panel shows the original TOI (black) where a step is caused by a DAE gain change that produces saturated data.  The (red) lines show the constrained noise realization used to replace those data. The lower panel shows a zoom around the position of the step to highlight the consistency of the gap filling data with the unflagged part of the TOI.}
\label{cnrfig}
\end{figure*}

The FFT-based noise power spectrum estimation method requires
continuity of the noise time stream.  As discussed in
Sect.~\ref{toiprocessing},  we identify bad data (e.g., unstable
spacecraft pointing, data saturation effects) and gaps in the data
with appropriate flags.  We fill in the flagged data with a
Gaussian noise realization constrained by data outside the gap
\citep{hoffman91}. Although in principle this method requires a
pure noise time stream outside the gap, we have verified that
given the low signal-to-noise ratio in the LFI TOD the procedure
is not affected by the signal present in the time streams.  We
fill the gap with Gaussian noise whose properties match those of
the noise power spectrum computed over the day immediately before
the one with flagged data.  An example is shown in
Fig.~\ref{cnrfig}.

\section{The Map-Making pipeline}
\label{mmaking}
\subsection{Frequency maps}
\label{sec:FrequencyMaps}

The map-making pipeline produces sky maps of temperature and polarisation for each frequency channel.  It takes as input the calibrated timelines and pointing information in the form of three angles ($\theta,\phi,\psi$) describing the orientation of the feed horns for each data sample.  An essential part of the map-making process is the reduction of correlated $1/f$ noise, a large part which can be removed by exploiting  redundancies in the scanning strategy.  While the underlying sky signal remains the same, the observed signal varies due to noise.  Statistical analysis of the signal variations allows one to distinguish between true sky signals and noise.

Among several map-making codes tested with simulated \Planck\ data
(see \citealt{ashdown2007a,ashdown2007b,ashdown2009}) the LFI
baseline \citep{mandolesi2010} is to use the {\tt Madam}
destriping code \citep{maino02}.  The algorithm and the underlying
theory are described in detail in \citet{keihanen2010,
kurki-suonio2009, keihanen2005}.  The basic idea is to model the
correlated noise component by a sequence of constant offsets,
called baselines.  A key parameter in the code is the length of
the baseline to be fitted to the data.  {\tt Madam} allows the use
of an optional noise prior, if the noise spectrum can be reliably
estimated, which further improves the accuracy of the output map.
Without the noise prior, the optimal baseline length is of the
order of the satellite spin period ($\approx$ 1 minute).  With an
accurate noise prior, a much shorter baseline can be used. The
shorter the baseline, the closer the {\tt Madam} solution will be
to the optimal Generalized Least Square solution (see Fig.~16 of
\citealt{ashdown2009}).

We are continually improving our knowledge of the instrument and its noise characteristics, and this information will eventually be used in the {\tt Madam} algorithm.  However, at this stage in the processing we decided to make two simplifications when running our map-making pipeline:  no noise prior was used, and  all radiometers were weighted equally.  These choices lead to a simpler and faster map-making algorithm,
which is sufficiently accurate for the \Planck\ Early Results and avoids using detailed parameters describing the instrument which are under continual revision.

With these simplifications, the map-making equations can be written in a  concise form.  Technically, we are neglecting the baseline covariance, $\mathbf{C_a}$, and setting the white noise variance $\mathbf{C_n}$ to unity.  The basic model behind the algorithm is %
\begin{equation}
\mathbf{\Delta T} = \mathbf{Pm} + \mathbf{n'}\, ,
\label{eq:madammodel1}
\end{equation}
where  $\mathbf{\Delta T}$ is the calibrated TOD, $\mathbf{P}$ is the pointing matrix, $\mathbf{m}$ is the pixelized sky map, and  $\mathbf{n'}$ is the instrumental noise. This last term can be written as
\begin{equation}
\mathbf{n'} = \mathbf{Fb} + \mathbf{n}\, ,
\label{eq:madammodel2}
\end{equation}
where $\mathbf{b}$ is the vector of unknown base function amplitudes  and the matrix $\mathbf{F}$ projects  these amplitudes  into the TOD. Since {\tt Madam} uses uniform baselines, the matrix  $\mathbf{F}$ consists of ones and zeros, indicating which TOD sample belongs to which baseline. Finally $\mathbf{n}$ is a pure
white noise stream assumed to be statistically independent of the baselines.

The maximum likelihood solution is obtained by minimizing
\begin{equation}
\chi^2 = (\mathbf{\Delta T} - \mathbf{Fb} - \mathbf{Pm})^T (\mathbf{\Delta T} - \mathbf{Fb} -\mathbf{Pm}),
 \label{eq:maxlikemadam}
 \end{equation}
with respect to the quantities $\mathbf{b}$ and $\mathbf{m}$.  The baseline amplitudes $\mathbf{b}$ are detemined by solving
\begin{equation}
(\mathbf{F}^T\mathbf{ZF})\mathbf{b} = \mathbf{F}^T \mathbf{Zy}\, ,
\label{eq:basesolution}
\end{equation}
where
\begin{equation}
 \mathbf{Z} \equiv \mathbf{I} - \mathbf{P}(\mathbf{P}^T\mathbf{P})^{-1}\mathbf{P}^T\, .
 \label{eq:z_definition}
\end{equation}
{\tt Madam} uses an iterative conjugate-gradient method to solve Eq. (\ref{eq:basesolution}).  An estimate for the map is finally obtained as
\begin{equation}
\mathbf{m} = (\mathbf{P}^T\mathbf{P})^{-1} \mathbf{P}^T
(\mathbf{\Delta T}- \mathbf{Fb})\, .
\label{eq:mapsolution}
\end{equation}

The map $\mathbf{m}$ has as many elements as pixels in the sky.  Each element is a Stokes parameter triplet $(I, Q, U)$ for a pixel $p$. The matrix $\mathbf{P}^T\mathbf{P}$ is a 3$\times$3 block diagonal matrix that operates on map space.  There is a block for each pixel $p$.  A block can only be inverted if the pixel $p$ is sampled with a sufficient number of different polarisation directions to allow determination of the three Stokes parameters for that pixel.  This is gauged by the condition number of the block.  For the present analysis, if the inverse condition number $rcond$ (ratio of the smallest to largest eigenvalue) was less than 0.01, the pixel $p$ was excluded from the $(I, Q, U)$ map.

The $\mathbf{P}^T\mathbf{P}$ blocks must be inverted when Eqs.~(\ref{eq:basesolution}) and (\ref{eq:z_definition}) are solved for the baselines.  These inversions are computed by eigenvalue decomposition. Eigenvalues whose magnitudes are less than 10$^{-6}$ times the largest eigenvalue are discarded; only the remaining part is inverted.

\begin{figure*}[!ht]
\centerline{
\includegraphics[width=6cm,angle=90]{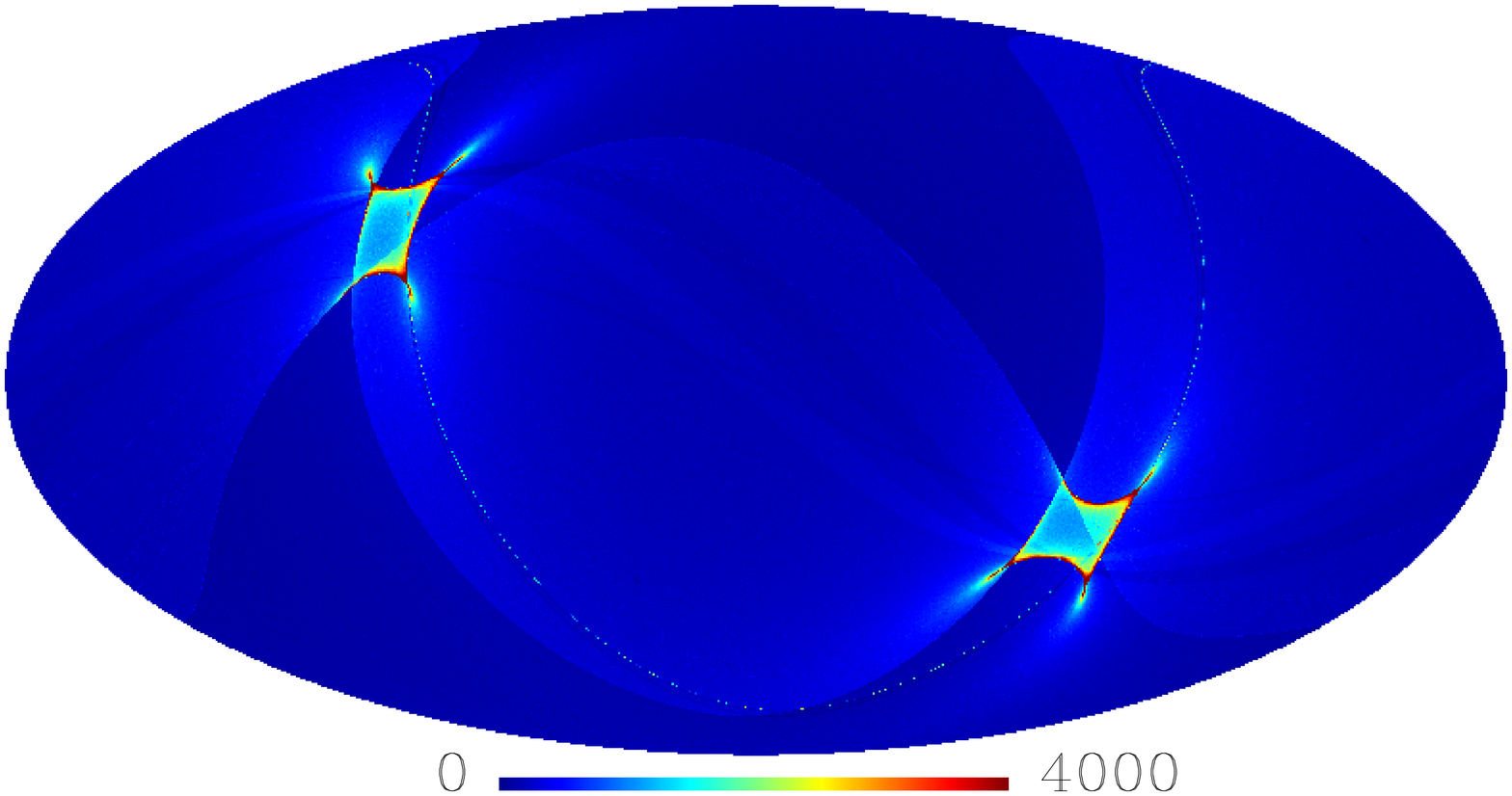}\includegraphics[width=6cm,angle=90]{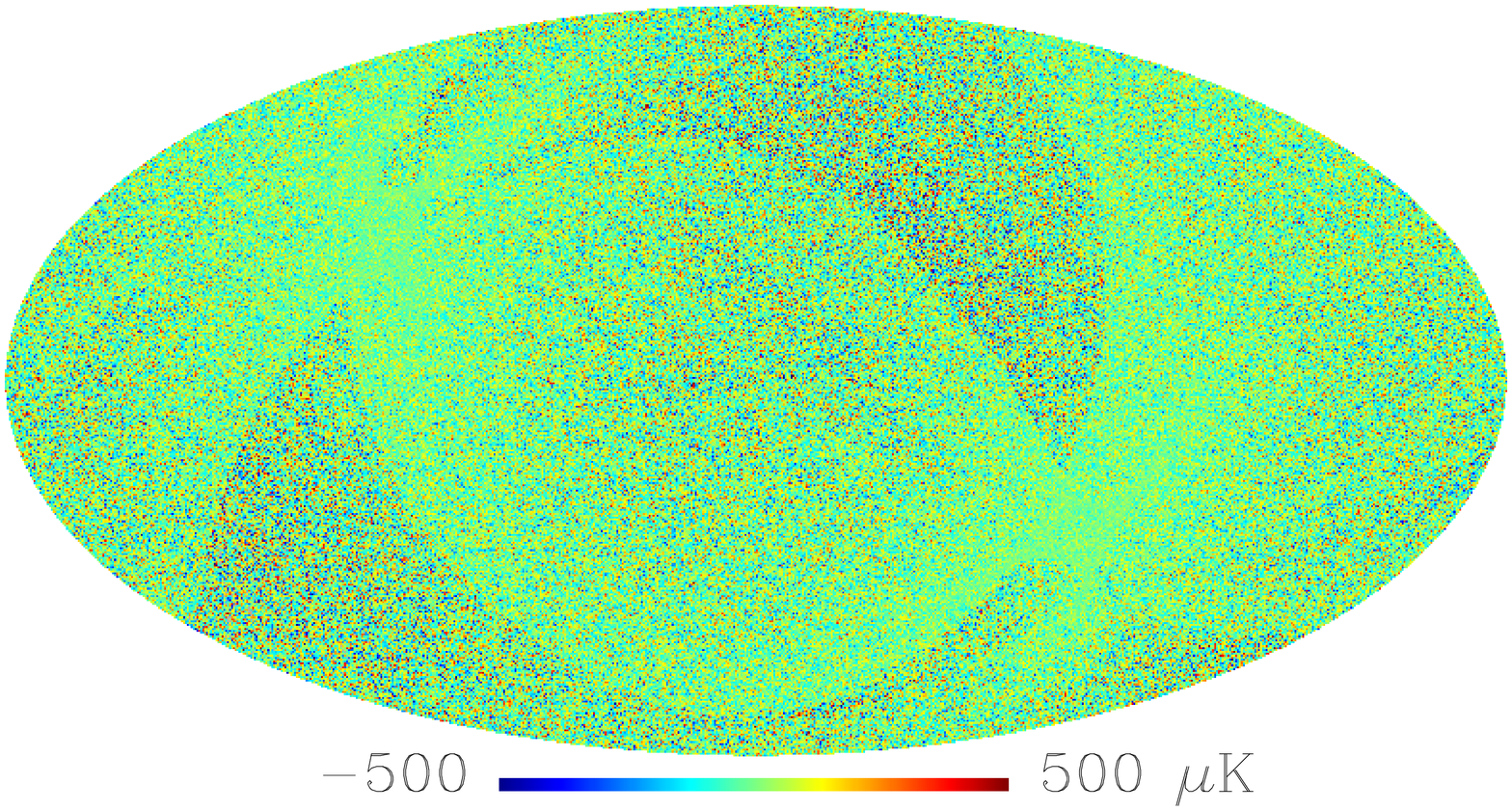}
} \centerline{\includegraphics[width=6cm,angle=90]{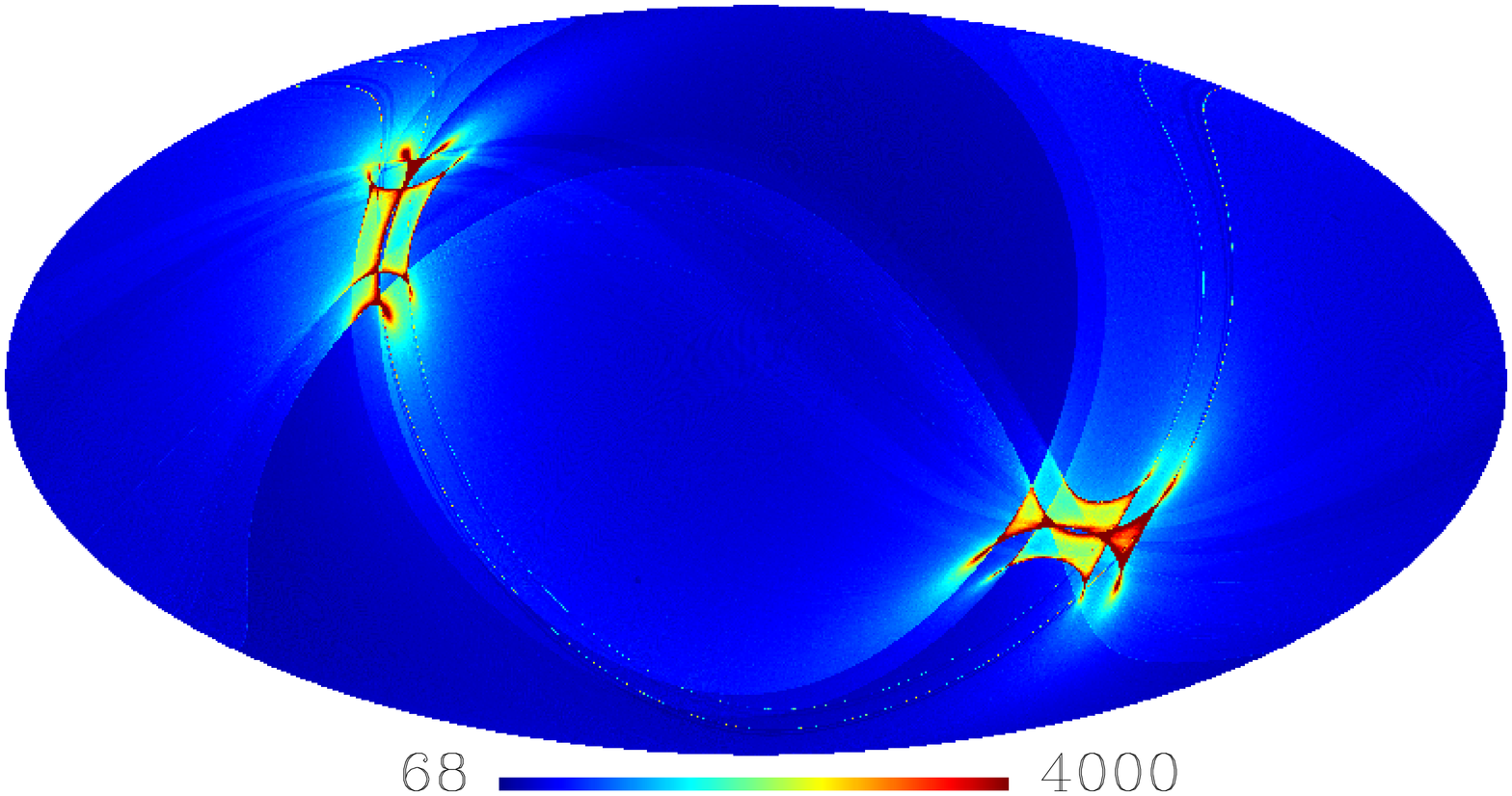}
            \includegraphics[width=6cm,angle=90]{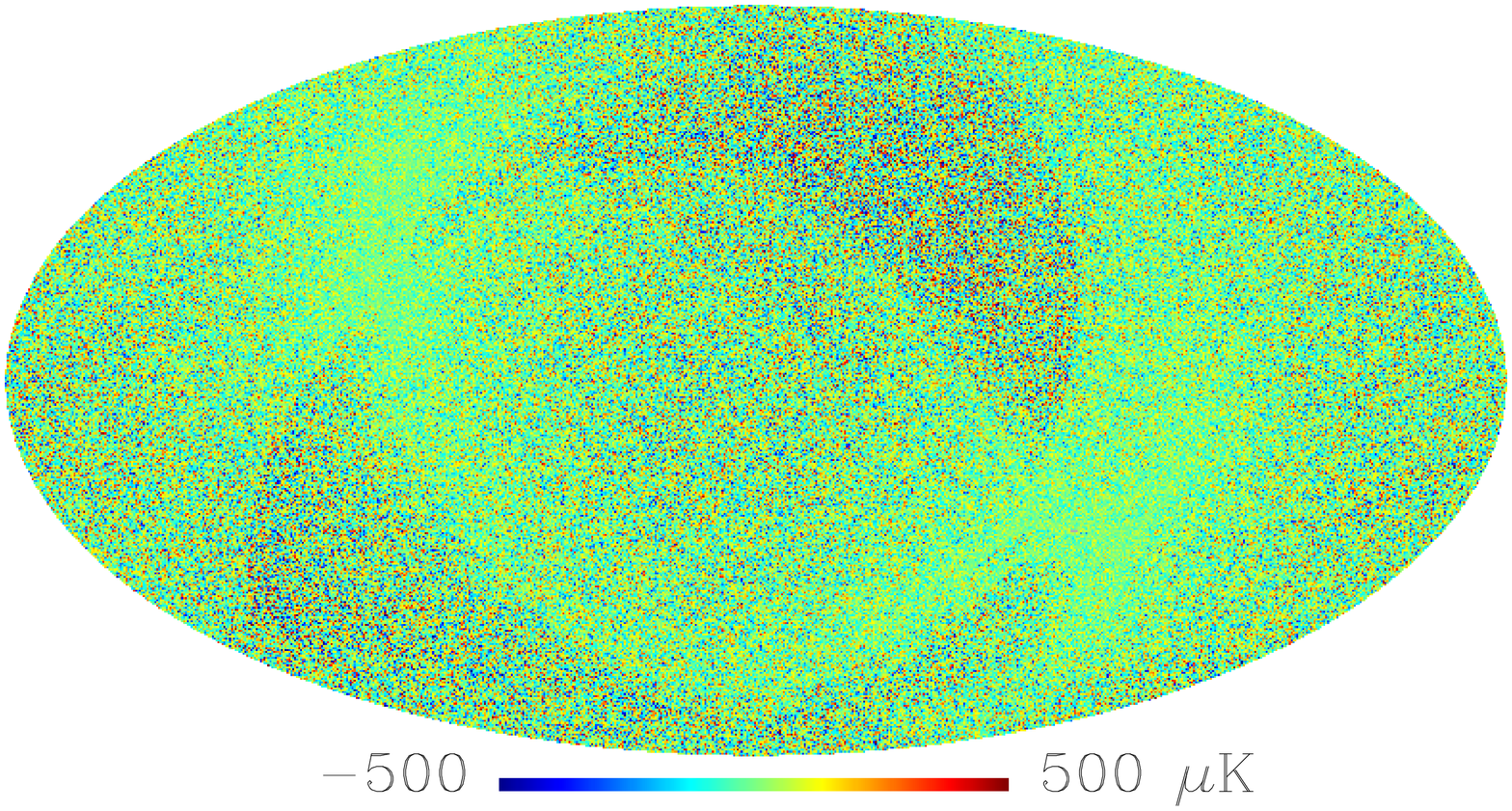}}
\centerline{\includegraphics[width=6cm,angle=90]{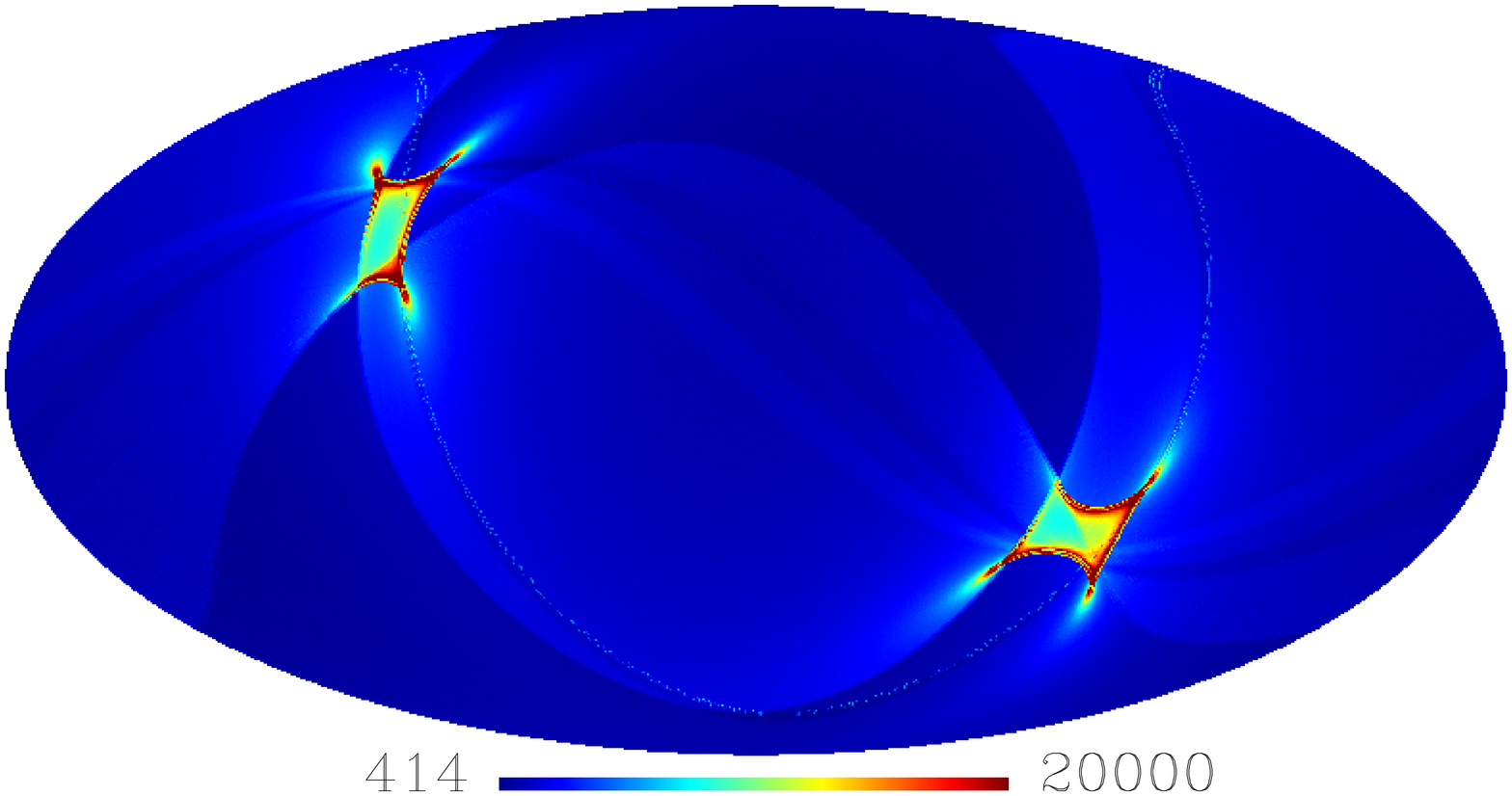}
            \includegraphics[width=6cm,angle=90]{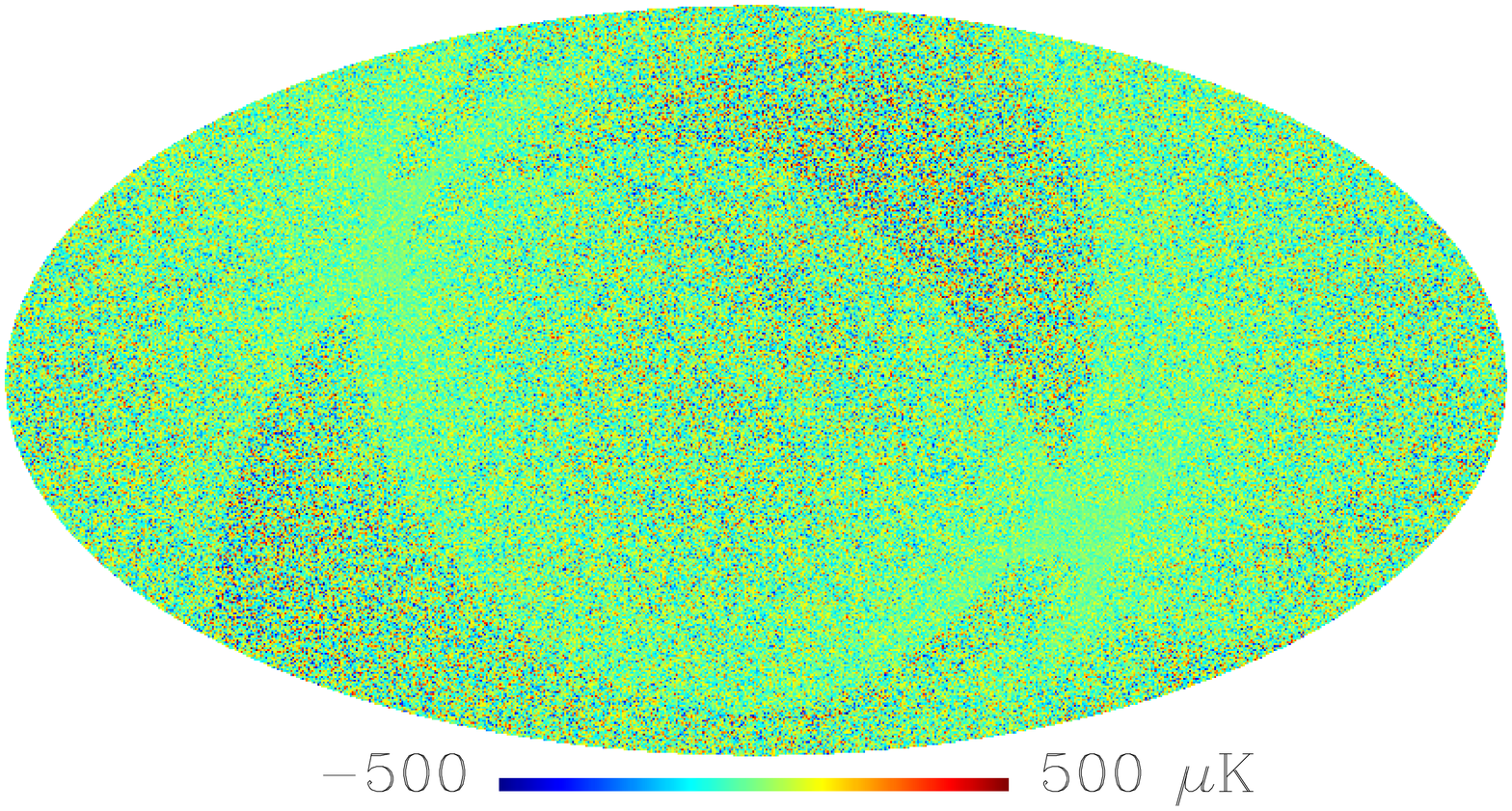}}
\caption{Hit count (left) and noise maps (right) at 30\,GHz (top),
44\,GHz (middle), and 70\,GHz (bottom). The complex distribution
around the ecliptic poles in the 44\,GHz hit map is caused by the
location of 44\,GHz horns on the focal plane.  The noise maps are
derived from half-ring jackknife tests described in the text.}
\label{hits}
\end{figure*}

For the present analysis, we need only the $I$-component maps at the three LFI nominal frequencies, combining observations of all radiometers at a given frequency.  Figure~\ref{hits} shows (left column) the hit count maps by frequency.  In addition, we produce maps from horn pairs scanning the same
path in the sky \citep[see][for details on the LFI focal plane arrangement]{planck2011-1.4}.  We have produced 30\,GHz maps at HEALPix resolution $N_{\rm side}$ = 512, and 44 and 70\,GHz maps at $N_{\rm side}$ = 1024.  All maps are in the NESTED scheme, in Galactic coordinates, with units of thermodynamic kelvin.  The baseline length in {\tt Madam} was one minute\footnote{One minute baselines for 30\,GHz, 44\,GHz, and 70\,GHz are 1950, 2792, and 4726 samples respectively.}.

\subsection{White noise covariance matrices}
\label{sec:wnc}

If we bin the pure white noise stream $\mathbf{n}$ to a map using the pointing $\mathbf{P}$, we obtain a binned white noise map,
\begin{equation}
 \mathbf{w} = (\mathbf{P}^T\mathbf{P})^{-1} \mathbf{P}^T \mathbf{n}\,.
 \label{eq:white_noise_map}
\end{equation}
This map is a theoretical concept because we do not have access to the radiometer white noise streams. Its covariance matrix, however, is important because it provides an estimate of both the white noise power in each pixel and white noise correlations between Stokes parameters at a given pixel.

This \textit{white noise covariance matrix} (WNC) is computed as (Eq.~ \ref{eq:white_noise_map})
\begin{equation}
\mathbf{C}_w = \langle\mathbf{w}\mathbf{w}^T\rangle = (\mathbf{P}^T\mathbf{P})^{-1}
                    (\mathbf{P}^T \mathbf{C}_n \mathbf{P}) (\mathbf{P}^T\mathbf{P})^{-1}\,.
\label{eq:wnc}
\end{equation}
Here $\mathbf{C}_n \equiv \langle \mathbf{n}\mathbf{n}^T\rangle$, $\mathbf{C}_n$ is a matrix that operates in the TOD domain, and angle brackets denote the ensemble mean.  Because the radiometers have independent white noise, $\mathbf{C}_n$ is diagonal.  We assume that each radiometer has a uniform white noise variance
$\sigma_{\rm WN}^2$ (see Sect.~\ref{noise}), but that each radiometer has its own variance.  The radiometer $\sigma_{\rm WN}$ values that we used in the WNC computation are reported in \citet{planck2011-1.4}.

\subsection{Half-ring jackknife noise maps}
\label{halfringjack}

For noise estimation purposes we divided the time ordered data into two halves and produced jackknife maps  as follows.  Each pointing period lasts typically $\approx$ 44\,min (median 43.5\,min, standard deviation 10\,min).  Typically, during the first 4\,min the pointing is unstable, so these data are not used for
science.  During the remaining stable 40\,min, each horn scans a ring on the sky. This ring consists of scan circles. One full scan circle takes 1\,min.  Therefore, each ring has about 40 scan circles. We made half-ring jackknife maps  $\mathbf{j_1}$ (and $\mathbf{j_2}$) with the same pipeline as described in Sect.~\ref{sec:FrequencyMaps}, but using stable data only from the first or the second half of each pointing period. Specificaly, this is implemented by marking the other half of each ring as a gap in the data.  {\tt Madam} knows that for any given pointing period the first-used scan circle/sample of any
half-ring is far apart in time (typically 25\,min) from the last used scan circle/sample of the previous half-ring.

At each pixel $p$, the jackknife maps $\mathbf{j_1}$ and $\mathbf{j_2}$ contain the same sky signal (as long no time-varying sources or moving objects cross $p$ at the time of observation), since they result from the same scanning pattern on the sky.  However, because of instrumental noise, the maps $\mathbf{j_1}$ and $\mathbf{j_2}$ are not identical.

We can estimate the sky signal+noise as
\begin{equation}
   \mathbf{m_{1+2}}(p) =  [\mathbf{j_1}(p) + \mathbf{j_2}(p)] / 2,
\label{eq:summap}
\end{equation}
and the noise in  map $\mathbf{m_{1+2}}$ as
\begin{equation}
   \mathbf{n_{1+2}}(p) =  [\mathbf{j_1}(p) - \mathbf{j_2}(p)] / 2.
\label{eq:summapnoise}
\end{equation}
This noise map includes noise that is not correlated on timescales longer than 20 minutes. In particular, $ \mathbf{n_{1+2}}$ gives a good estimate of the white noise in  $\mathbf{m_{1+2}}$.

However, we are interested in the noise level in the full map $\mathbf{m}$, (see Eq.~\ref{eq:mapsolution}). To estimate this, we construct another noise map
\begin{equation}
   \mathbf{n_{m}}(p) =  \frac{\mathbf{j_1}(p) - \mathbf{j_2}(p)}{\mathbf{w_{\rm hit}}(p)}\,,
\label{eq:noisemap}
\end{equation}
with weights
\begin{equation}
   \mathbf{w_{\rm hit}}(p) = \sqrt{ \mathbf{\mbox{\bf hit}_{\rm full}}(p)
     \left[ \frac{1}{\mathbf{\mbox{\bf hit}_1}(p)} +
       \frac{1}{\mathbf{\mbox{\bf hit}_2}(p)} \right]}\, .
\label{eq:hitweight}
\end{equation}
Here $\mathbf{\mbox{\bf hit}_{\rm full}}(p) = \mathbf{\mbox{\bf hit}_1}(p) + \mathbf{\mbox{\bf hit}_2}(p)$
is the hit count at pixel $p$ in the full map $\mathbf{m}$, while $\mathbf{\mbox{\bf hit}_1}$ and $\mathbf{\mbox{\bf hit}_2}$ are the hit counts of  $\mathbf{j_1}$ and $\mathbf{j_2}$, respectively.  The weight factor  $\mathbf{w_{\rm hit}}(p)$ is equal to $2$ only in those pixels where $\mathbf{\mbox{\bf hit}_1}(p) = \mathbf{\mbox{\bf hit}_2}(p)$.  In a typical pixel, $\mathbf{\mbox{\bf hit}_1}(p)$ will differ slightly from
$\mathbf{\mbox{\bf hit}_2}(p)$ and hence the weight factor is $\mathbf{w_{\rm hit}}(p) > 2$.

Noise maps from half-ring jackknifes are shown in the right-hand column of Fig.~\ref{hits}.  A detailed comparison of the jackknife noise estimates and other noise estimates (WNC, noise Monte Carlo; see next section) are presented in the LFI instrument paper \citep{planck2011-1.4}.

\subsection{Noise Monte Carlo simulations}
\label{sec:noiseMC}

To check the noise analysis, we produced Monte Carlo noise realizations on the ``Louhi'' supercomputer at ``CSC-IT Center for Science" in Finland. The simulation takes as input estimates of the white noise $\sigma_{\rm WN}$, knee frequency, and slope of the $1/f$ noise estimated from the TOD for each radiometer \citep{planck2011-1.4}, as well as satellite pointing information.  Flight pointing was reconstructed to machine accuracy using Planck Level-S simulation software \cite{reinecke2006}.  For each frequency channel,  we generated 101 Monte Carlo realizations of the noise, simulating white noise and correlated noise ($1/f$) streams separately.  Maps from these noise streams were produced with the map-making pipeline described in Sect.~\ref{sec:FrequencyMaps}.  For each simulated noise map, we computed the corresponding binned white noise maps defined in Eq.~(\ref{eq:white_noise_map}). The production of the binned white noise maps allows us to study the residual correlated noise, i.e., the difference between the total and binned white noise maps \cite{kurki-suonio2009}.  These Monte Carlo simulations were used to test and validate several approaches to noise estimation described in detail in the LFI instrument paper \citep{planck2011-1.4}.

\section{Colour correction}
\label{colcorr}

The power measured by LFI can be expressed as
\begin{equation}
P = \frac{G}{2}\int g(\nu)\Delta T_{RJ}(\nu)d\nu\, ,
\end{equation}
where $G$ is the overall gain, $g(\nu)$ is the bandpass, and $\Delta T_{RJ}$ is the Rayleigh-Jeans brightness temperature signal, in the case of LFI calibration procedure, due to the CMB dipole.   At a given frequency $\nu_0$, the overall gain $G$ is equal to $2P/\Delta T_{RJ}(\nu_0)$.  For small fluctuations around the mean CMB temperature $T_0$, the relation between intensity, $I$, Rayleigh-Jeans brightness temperature $T_{\rm RJ}$, and thermodynamic temperature $T$ is
\begin{equation}
\Delta I(\nu_0) = \frac{2 k_B\nu_0^2}{c^2}\Delta T_{\rm RJ}(\nu_0) = \left(\frac{\partial B(\nu_0,T)}{\partial T}\right)_{T_0}\Delta T\, .
\end{equation}
The differential black-body spectrum is
\begin{equation}
\left(\frac{\partial B(\nu,T)}{\partial T}\right)_{T_0} =
\frac{2k_B\nu^2}{c^2}e^{k\nu/k_BT_0}\left(\frac{h\nu/k_BT_0}
{e^{h\nu/k_BT_0}-1}\right)^2,
\end{equation}
\begin{equation}
\equiv \frac{2k_B\nu^2}{c^2}\eta_{\Delta T}(\nu)\, .
\end{equation}
The function $\eta_{\Delta T}(\nu)$ is the differential black-body spectrum in Rayleigh-Jeans units.  With our definition of the overall gain $G$, the bandpasses are normalised such that
\begin{equation}
\int g(\nu)\eta_{\Delta T}(\nu)d\nu = \eta_{\Delta T}(\nu_0)\, .
\end{equation}
Calibration data provide a nominal brightness temperature $\Delta \tilde{T}_{\rm RJ}=(2/G)P$; however, this is only exact for a monochromatic response. For a non-zero bandwidth, a colour correction $C(\alpha)$ is required to convert the brightness temperature for emission with a particular spectral index $\alpha$ to that of the map:
\begin{equation}
C(\alpha)\Delta T_{RJ}(\nu_0) = \Delta \tilde{T}_{RJ} = \eta_{\Delta T}(\nu_0)\Delta \tilde{T}\, .
\end{equation}
By definition, the colour correction is unity when the source observed has a CMB spectrum.  Within each LFI band, $g(\nu)$ is well-approximated by a power law with spectral index~$\alpha = 2- (h\nu_0/k_{\rm B}T)^2/6$.

The general expression for the colour correction is
\begin{equation}
C(\alpha) = \left[\frac{\eta_{\Delta T}(\nu_0)}{\int g(\nu)\eta_{\Delta T}(\nu)d\nu}\right]\int g(\nu)(\nu/\nu_0)^\beta d\nu\, ,
\label{cc}
\end{equation}
where we assumed a power-law spectrum with temperature spectral index $\beta = \alpha -2$.  The term in
square brackets is unity with our normalisation for $g(\nu)$, but has been included to show that $C(\alpha)$ depends only on the shape and not  the amplitude of the bandpass.  Thus $C(\alpha)$ is independent of $G$.

Each detector has a different bandpass, hence its own colour correction. We derive approximate colour corrections for band-averaged sky maps using bandpasses averaged over: (i)~the two detectors in each radiometer; (ii)~the two orthogonally-polarised radiometers behind each feed horn; and (iii)~the several feed horns in each frequency band.  In addition,  although the bandpass is mainly defined by the front-end \citep{bersanelli2010}, differences between back-end bandpasses on a single radiometer are measurable, e.g., in the form of $\beta$-dependent residuals in difference images.

Since the current sky maps have been  produced, both for pairs of horns and for several horns in each band, with calibrated data combined with equal weights, we have used an unweighted average of all the contributing bandpasses for our band-averaged corrections.  Using the bandpass models given in \cite{zonca2009} derived from the pre-launch calibration campaign, we evaluate  the integrals in Eq.(\ref{cc}) analytically for several spectral indices. The results are given in Table ~\ref{colorcorr}.

\begin{table*}
\begingroup
\newdimen\tblskip \tblskip=5pt
\caption{Colour corrections for different input power-law spectral indices}
\label{colorcorr}
\nointerlineskip
\vskip -2mm
\footnotesize
\setbox\tablebox=\vbox{
   \newdimen\digitwidth
   \setbox0=\hbox{\rm 0}
   \digitwidth=\wd0
   \catcode`*=\active
   \def*{\kern\digitwidth}
   \newdimen\signwidth
   \setbox0=\hbox{+}
   \signwidth=\wd0
   \catcode`!=\active
   \def!{\kern\signwidth}
\halign{\hbox to 1.5in{#\leaderfil}\tabskip=2em&
        \hfil#\hfil\tabskip=1.5em&
        \hfil#\hfil&
        \hfil#\hfil&
        \hfil#\hfil&
        \hfil#\hfil&
        \hfil#\hfil&
        \hfil#\hfil\tabskip=0pt\cr
\noalign{\doubleline}
\omit&\multispan7\hfil \sc Spectral Index $\alpha$\hfil\cr
\noalign{\vskip -4pt}
\omit&\multispan7\hrulefill\cr
\noalign{\vskip 0pt}
\omit\hfil\sc Detector\hfil&$-2.0$&$-1.0$&0.0&1.0&2.0&3.0&4.0\cr
\noalign{\vskip 3pt\hrule\vskip 5pt}
\multispan2{\bf 70\,GHz}\hfil\cr
\noalign{\vskip 2pt}
\hglue 2em LFI18 & 1.054 & 1.028 & 1.011 & 1.003 & 1.003 & 1.010 & 1.026\cr
\hglue 2em LFI19 & 1.170 & 1.113 & 1.066 & 1.026 & 0.994 & 0.969 & 0.949\cr
\hglue 2em LFI20 & 1.122 & 1.079 & 1.044 & 1.017 & 0.997 & 0.983 & 0.975\cr
\hglue 2em LFI21 & 1.087 & 1.053 & 1.028 & 1.010 & 1.000 & 0.996 & 0.998\cr
\hglue 2em LFI22 & 0.973 & 0.971 & 0.976 & 0.988 & 1.007 & 1.033 & 1.066\cr
\hglue 2em LFI23 & 1.015 & 1.004 & 0.999 & 0.998 & 1.003 & 1.012 & 1.026\cr
\noalign{\vskip 4pt}
\hglue 2em $\langle C\rangle_{70}$ & 1.070 & 1.041 & 1.021 & 1.007 & 1.001 & 1.001 & 1.007\cr
\noalign{\vskip 7pt}
\multispan2{\bf 44\,GHz}\hfil\cr
\noalign{\vskip 2pt}
\hglue 2em LFI24 & 1.028 & 1.015 & 1.007 & 1.002 & 1.000 & 1.003 & 1.009\cr
\hglue 2em LFI25 & 1.039 & 1.024 & 1.013 & 1.005 & 1.000 & 0.999 & 1.000\cr
\hglue 2em LFI26 & 1.050 & 1.032 & 1.017 & 1.007 & 1.000 & 0.997 & 0.997\cr
\noalign{\vskip 4pt}
\hglue 2em $\langle C\rangle_{44}$ & 1.039 & 1.024 & 1.012 & 1.004 & 1.000 & 0.999 & 1.002\cr
\noalign{\vskip 7pt}
\multispan2{\bf 30\,GHz}\hfil\cr
\noalign{\vskip 2pt}
\hglue 2em LFI27 & 1.078 & 1.049 & 1.026 & 1.010 &  1.000 & 0.996 & 0.998\cr
\hglue 2em LFI28 & 1.079 & 1.049 & 1.026 & 1.009 & 1.000 & 0.997 & 1.002\cr
\noalign{\vskip 4pt}
\hglue 2em $\langle C\rangle_{30}$ & 1.079 & 1.049 & 1.026 & 1.010 & 1.000 & 0.997 & 1.000\cr
\noalign{\vskip 5pt\hrule\vskip 3pt}}}
\endPlancktable
\endgroup
\end{table*}

At the current stage of the mission and data analysis, uncertainties in the colour corrections are much smaller than those of the gains $G$; however we aim to reduce the calibration error (using the orbital dipole) to below 0.2\%.  Two primary sources of error in $C(\alpha)$ will then need to be considered. The first is related to uncertainties in the bandpass model \citep{leahy2010, zonca2009}.  The second arises from the uneven sampling of individual sky pixels by the full set of detectors,  which causes pixel-to-pixel
variations in the colour correction.

\section{CMB removal}
\label{cmbremoval}

This section was developed in common with HFI \citep{planck2011-1.7} and is reported
identically in both papers.

In order to facilitate foreground studies with the frequency
maps, a set of maps was constructed with an estimate of the CMB
contribution subtracted from them.  The steps undertaken
in determining that estimate of the CMB map,
subtracting it from the frequency maps, and characterising the errors
in the subtraction are described below.

\subsection{Masks} \label{sec:masks}

Point source masks were constructed from the source catalogues
produced by the LFI pipeline for each of the LFI frequency channel
maps.  The algorithm used in the pipeline to detect the sources
was a Mexican-hat wavelet filter.  All sources detected with a
signal-to-noise ratio greater than 5 were masked with a cut of
radius $3\,\sigma \approx 1.27\, \mathrm{FWHM}$ of the effective
beam.  A similar process was applied to the HFI frequency maps
\citep{planck2011-1.7}.

Galactic masks were constructed from the 30\,GHz and 353\,GHz frequency channel
maps.  An estimate of the CMB was subtracted from the maps in order not to bias
the construction.  The maps were smoothed to a common resolution of 5\deg.
The pixels within each mask were chosen to be those with values above a threshold
value. The threshold values were chosen to produce masks with the desired fraction
of the sky remaining.
The point source and Galactic masks were provided as additional
inputs to the component separation algorithms.

\subsection{Selection of the CMB template} \label{sec:CMBselection}

Six component separation or foreground removal algorithms were applied
to the HFI and LFI frequency channel maps to produce CMB maps. They
are, in alphabetical order:
\begin{itemize}
\item{AltICA}:      Internal linear combination (ILC) in the map domain;
\item{CCA}:         Bayesian component separation in the map domain;
\item{FastMEM}:     Bayesian component separation in the harmonic domain;
\item{Needlet ILC}: ILC in the needlet (wavelet) domain;
\item{SEVEM}:       Template fitting in map or wavelet domain;
\item{Wi-fit}:      Template fitting in wavelet domain.
\end{itemize}
Details of these methods may be found in \cite{leach2008}.  These
six algorithms make different assumptions about the data, and may
use different combinations of frequency channels used as input.
Comparing results from these methods (see
Fig.~\ref{fig:cmbdispersion}) demonstrated the consistency of the
CMB template and provided an estimate of the uncertainties in the
reconstruction.  A detailed comparison of the output of these
methods, largely based on the CMB angular power spectrum, was used
to select the CMB template that was removed from the frequency
channel maps.  The comparison was quantified using a jackknife
procedure: each algorithm was applied to two additional sets of
frequency maps made from the first half and second half of each
pointing period.  A residual map consisting of half the difference
between the two reconstructed CMB maps was taken to be indicative
of the noise level in the reconstruction from the full data set.
The Needlet ILC (hereafter NILC) map was chosen as the CMB
template because it had the lowest noise level at small scales.

\begin{figure}
    \centering
    \includegraphics[width=8cm]{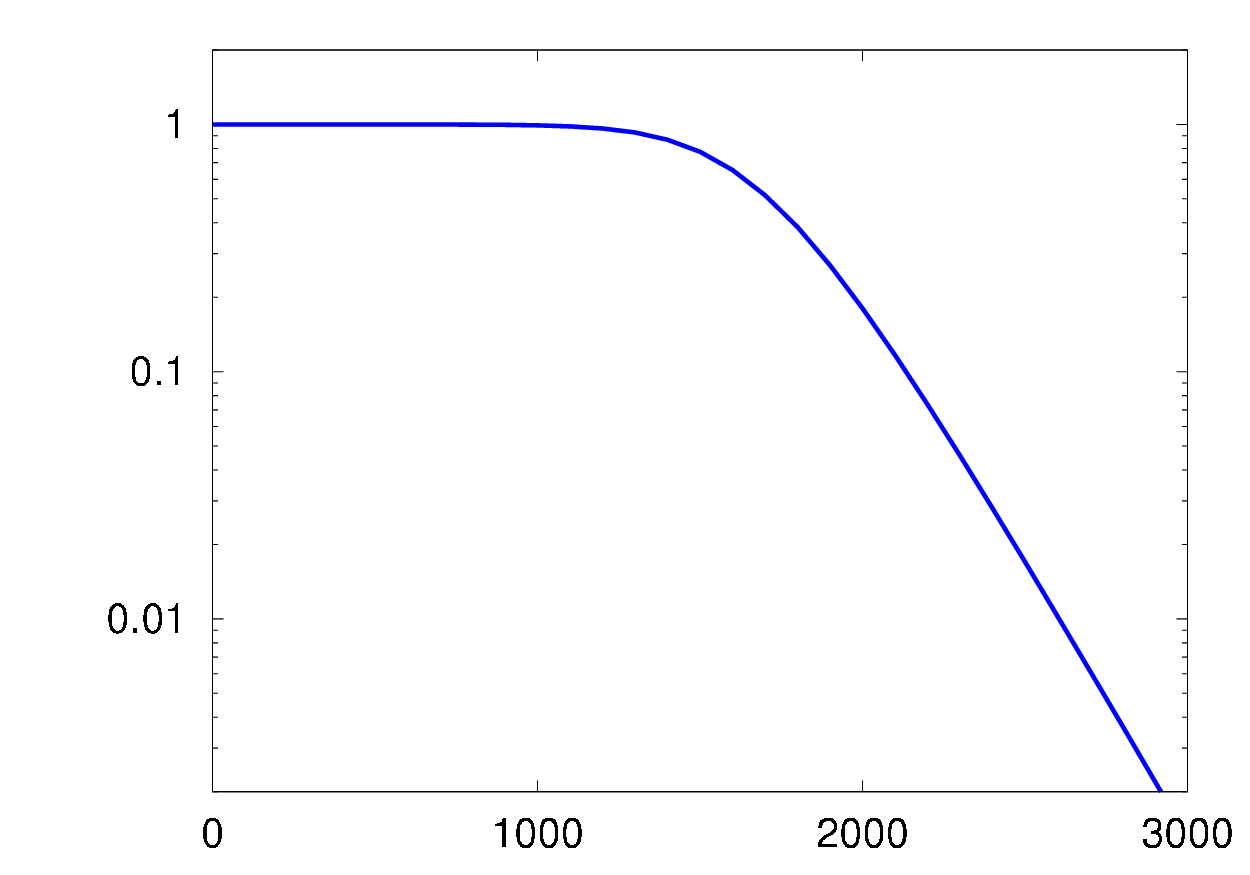}
    \caption{Wiener-like filter function, plotted versus multipole, which was applied to produce the template for CMB removal.}
    \label{fig:cmbremovalwienerfilter}
\end{figure}

The CMB template was removed from the frequency channel maps after
application of a filter in the spherical harmonic domain.  The filter
has a transfer function made of two factors.  The first
corresponds to the Gaussian beam of the channel to be cleaned; the
second is a transfer function attenuating the multipoles of the
CMB template that have low signal-to-noise ratio.  It is designed in
Wiener-like fashion, being close to unity up to
multipoles around $\ell = 1000$, then dropping smoothly to zero
with a cut-off frequency around $\ell=1700$ (see Fig.~\ref{fig:cmbremovalwienerfilter}).  All angular frequencies above $\ell=3900$ are completely suppressed.
This procedure was adopted to avoid doing more harm than good to the
small scales of the frequency channel maps where the signal-to-noise
ratio of the CMB is low.

\subsection{Description of Needlet ILC} \label{sec:nilc}

\begin{figure}
    \centering
    \includegraphics[width=8cm]{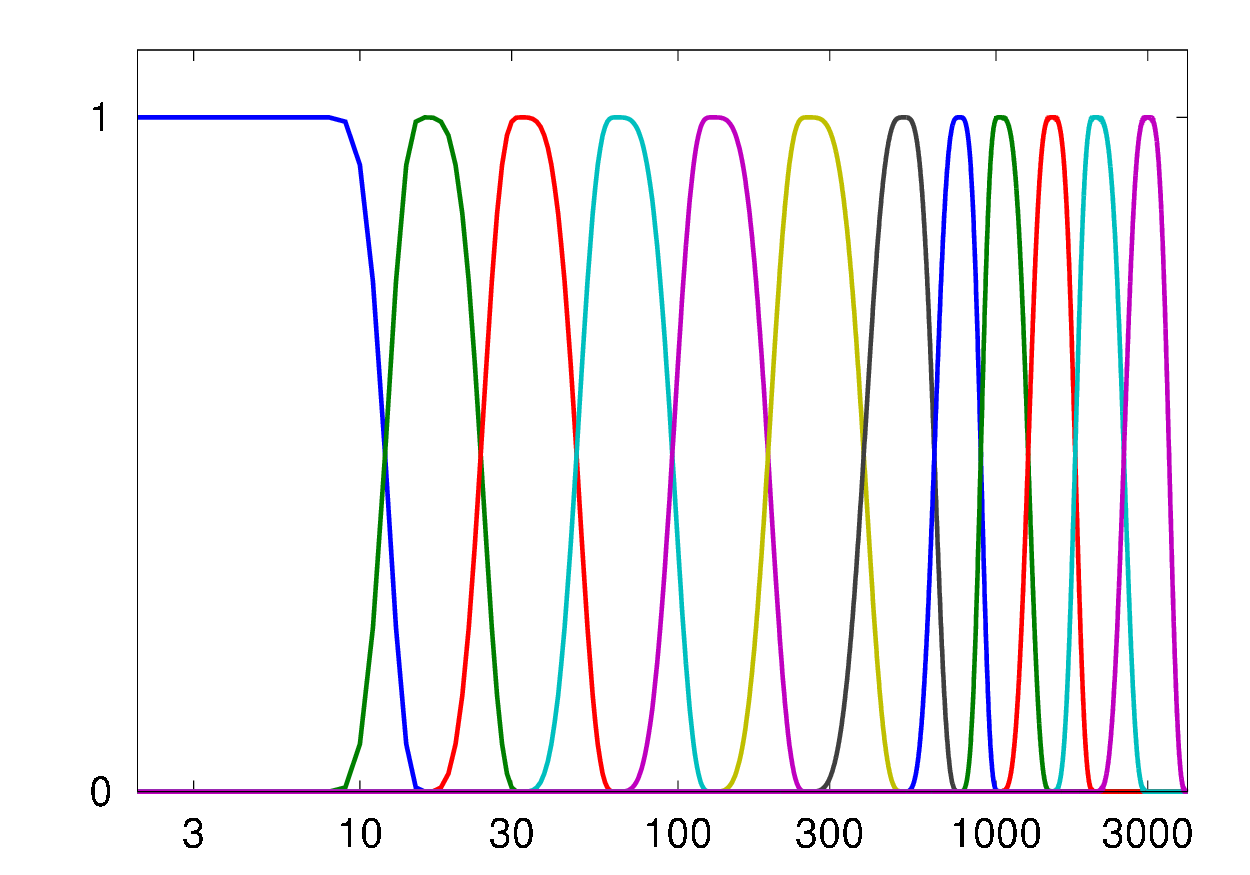}
    \caption{The bandpass filters, plotted versus multipole, that define the spectral domains
          used in the NILC.}
    \label{fig:needletbands}
\end{figure}

The NILC map was produced using the ILC method in the
``needlet'' domain.  Needlets are spherical wavelets that allow
localisation both in multipole and sky direction.  The input maps are
decomposed into twelve overlapping multipole domains (called
``scales''), using the bandpass filters shown in Fig.~\ref{fig:needletbands}
and further decomposed into regions of the sky.  Independent ILCs are
applied in each sky region at each needlet scale.  Large regions are used at
large scales, while smaller regions are used at fine scales.

\begin{figure}
    \centering
    \includegraphics[width=8cm]{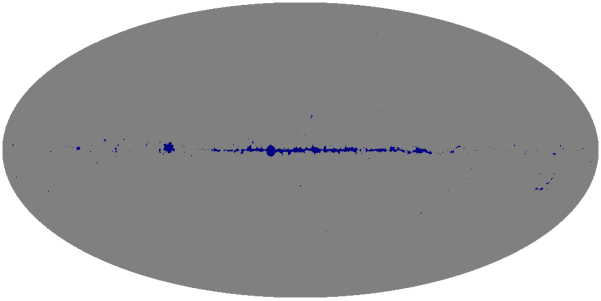}
    \caption{Galactic mask used with NILC.}
    \label{fig:nilcgalmask}
\end{figure}

The NILC template was produced from all six HFI channels, using
the tight Galactic mask shown in figure~\ref{fig:nilcgalmask},
which covers 99.36\% of the sky.  Additional areas are excluded on
a per-channel basis to mask point sources. Future inclusion of the
LFI channels will improve cleaning of low-frequency foregrounds
such as synchrotron emission from the CMB template.
Before applying NILC, pixels missing due to point source and Galactic
masking are filled in by a ``diffusive inpainting'' technique, which consists of
replacing each missing pixel by the average of its neighbours and iterating to
convergence.  This is similar to solving the heat diffusion equation in the
masked areas with boundary conditions given by the available pixel values at
the borders of the mask.  All maps are re-beamed to a common resolution of
5\arcmin.  Re-beaming blows up the noise in the less
resolved channels, but that effect is automatically taken
into account by the ILC filter.

The CMB template obtained after NILC processing is
filtered to have the `Wiener beam' shown in
Fig.~\ref{fig:cmbremovalwienerfilter}.  The ILC coefficients are saved
to be applied to the jackknife maps for performance evaluation as
described in Sect.~\ref{sec:NILCperf_internal}

\subsection{Uncertainties in the CMB removal} \label{sec:cmb_uncertainty}

The uncertainties in the CMB removal have been gauged in two
ways, firstly by comparing the CMB maps produced by the different
algorithms and secondly by applying the NILC coefficients to jackknife maps.

\begin{figure}
    \centering
    \includegraphics[width=8cm]{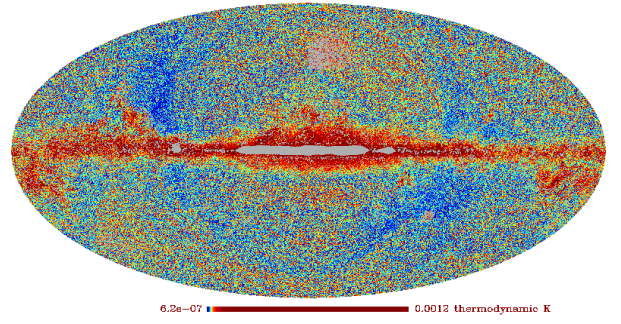}
    \caption{Estimate of the rms error in the CMB subtraction. The map is
        histogram-equalised to bring out the details.}
    \label{fig:cmbdispersion}
\end{figure}

\subsubsection{Dispersion of the CMB maps produced by
  the various algorithms.}

The methods that were used to produce the estimates of the CMB are
diverse.  They work by applying different algorithms (ILC,
template fitting, or Bayesian parameter estimation) in a variety
of domains (pixel space, Needlet/wavelet space, or spherical
harmonic coefficients).   Each method carries out its optimisation
in a different way and thus will respond to the foregrounds
differently. Dispersion in the CMB rendition by different methods
provides an estimate of the uncertainties in the determination of
the CMB, and thus in the subtraction process. The rms difference
between the NILC map and the other CMB estimates is shown in
Fig.~\ref{fig:cmbdispersion}.  As expected, the uncertainties are
largest in the Galactic plane where the foregrounds to remove are
strongest, and smallest around the Ecliptic poles where the noise
levels are lowest.

\subsubsection{CMB map uncertainties estimated by applying NILC filtering of jackknifes}
\label{sec:NILCperf_internal}

The cleanliness of the CMB template produced by the NILC filter
can be estimated using jackknives.  We apply the NILC filter to
the maps built from the first and last halves of the ring set. The
power distribution of the half-difference of the results provides
us with a reliable estimate of the power of the noise in the NILC
CMB template, (while previous results correspond to applying the
NILC filter to the half-sum maps from which they can be derived).

The jackknives allow estimates of the relative contributions of
sky signal and noise to the total data power. Assume that the data
are in the form $X = S + N$\ where $S$ is the sky signal and $N$
is the noise, independent of $S$. The total data power $\Var{X}$
decomposes as $\Var{X}  = \Var{S} + \Var{N}$. One can obtain
$\Var{N}$ by applying the NILC filter to half difference maps, and
$\Var{S}$ follows from $\Var{X} - \Var{N}$. This procedure can be
applied in pixel space, in harmonic space, or in pixel space
\emph{after} the maps have been bandpass-filtered, as described
next.

\begin{figure}
    \centering
    \includegraphics[width=8cm]{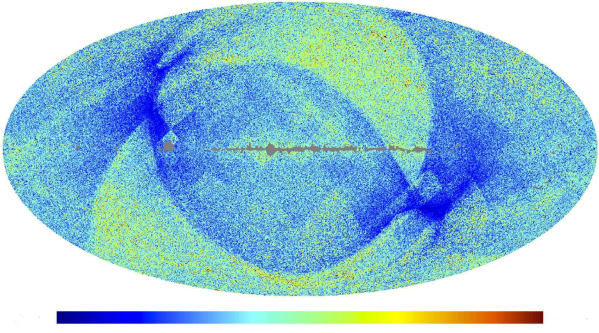}
    \caption{Local rms of the noise (estimated by jackknife) in the NILC
            CMB map.  The colour scale is from 0 to 30 $\mu$K per pixel at
            resolution $N_\mathrm{side}=2048$.}
    \label{fig:nilcnoisespatial}
\end{figure}

\begin{figure}
    \centering
    \includegraphics[width=8cm]{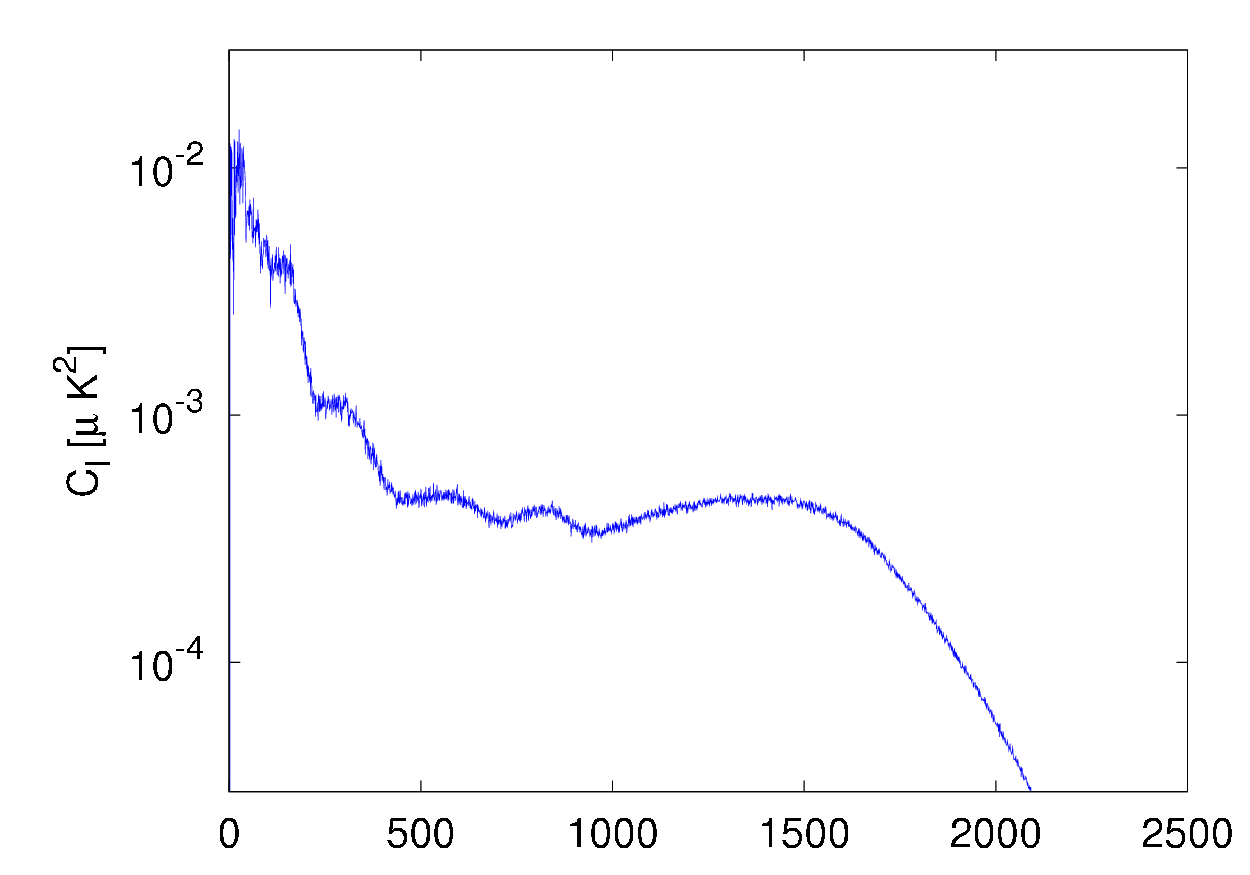}
    \caption{Angular spectrum in $\mu\mathrm{K}^2$ of the noise (estimated by
            jackknife) in the NILC CMB map.  It corresponds to 11
                $\mu\mathrm{K}$ per pixel.}
    \label{fig:nilcnoisespectral}
\end{figure}

We first used pixel space jackknifing to estimate the spatial
distribution of noise.  Figure~\ref{fig:nilcnoisespatial} shows a
map of the local rms of the noise.  We applied the NILC filter to
a half-difference map and we display the square root of its
smoothed squared values, effectively resulting in an estimate of
the local noise rms.
Using the same approach, we obtain an estimate of the angular
spectrum of the noise in the NILC map, shown in
Fig~\ref{fig:nilcnoisespectral}. That spectrum corresponds to an
rms $\left[(1/4\pi)\sum_\ell (2\ell+1)C_\ell\right]^{1/2}$ of
11\,$\mu$K per pixel.
The ``features'' in the shape of the noise angular spectrum at large
scale are a consequence of the needlet-based filtering (such features
would not appear in a pixel-based ILC map).  Recall that the
coefficients of an ILC map are adjusted to minimize the total
contamination by both foregrounds \emph{and} noise.  The strength of
foregrounds relative to noise being larger at coarse scales, the
needlet-based ILC tends to let more noise in, with the benefit of
better foreground rejection.

\begin{figure}
    \centering
    \includegraphics[width=8cm]{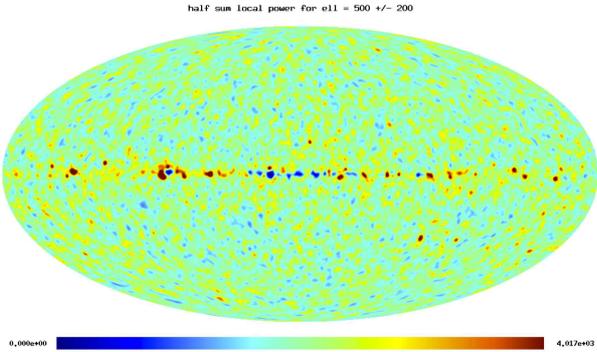}
    \caption{Local power of the NILC CMB template in the range $\ell=500\pm200$.}
    \label{fig:nilc_contamination}
\end{figure}

The half-difference maps offer simple access to the power
distribution of the residual noise in the estimated CMB template.
However, it is more difficult to evaluate other residual
contamination, since all fixed sky emissions cancel in half
difference maps. Any such large-scale contamination is barely
visible in the CMB template, since it is dominated by the CMB
itself.  However, contamination is more conspicuous if one looks
at intermediate scales. Figure~\ref{fig:nilc_contamination}
shows the local power of the CMB template after it is bandpassed to
retain only multipoles in the range $\ell=500\pm200$. This smooth
version of the square of a bandpassed map clearly shows where the
errors in the component separation become large
and so complicate some specific science analyses.


\begin{figure}
\begin{center}
    \includegraphics[angle=90,width=\columnwidth]{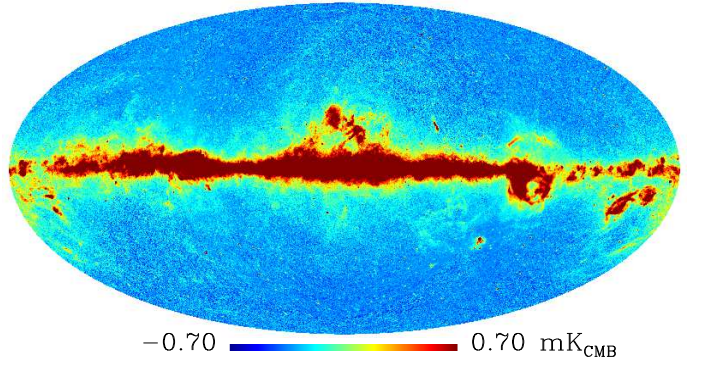}
    \includegraphics[angle=90,width=\columnwidth]{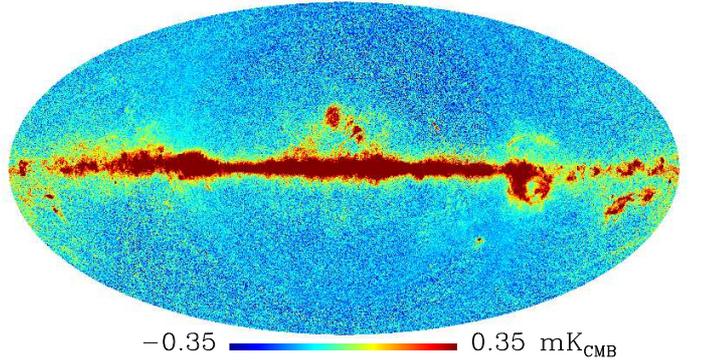}
    \includegraphics[angle=90,width=\columnwidth]{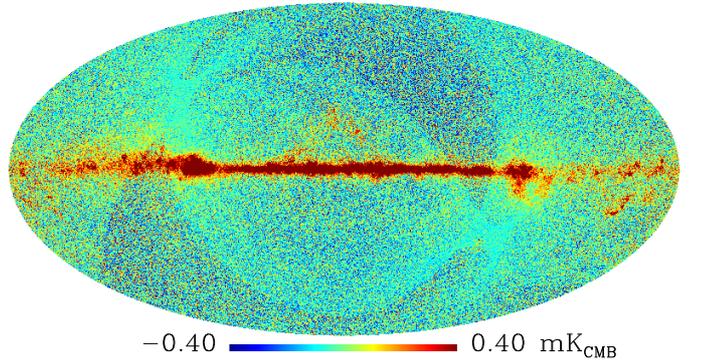}
\end{center}
\caption{CMB-removed channel maps. From top to bottom, 30, 44, and 70\,GHz. The main
galactic structures are clearly visible, as well as scanning strategy signatures at 44 and
70\,GHz. } \label{figure:cmbremoved}
\end{figure}

\section{Infrastructure overview}
\label{Infrast}

To organize the large number of data processing codes and data products, the DPC employs the \Planck\ Integrated Data and Information System (IDIS).  This allows flexible development of the processing pipeline, while ensuring complete traceability and reproducibility of data products.  For this,  the most relevant components of IDIS are the Data Management Component (DMC) and the Process Coordinator (ProC), developed at the MPA Planck Analysis Centre (MPAC) at the Max-Planck-Institute for Astrophysics in Garching. Access to both components as well as to the \Planck\ document and software management system is controlled via another IDIS component: the Federation Layer developed and maintained at the ESA ESTEC Research and Scientific Support Department (RSSD).

Here we describe the essential features of the IDIS data processing components and their use at the \hbox{DPC}. A more detailed description of these components and their capabilities will be given in a future paper.

\subsection{Data Management Component}

The DMC organizes the storage and access to DPC data products.  To combine optimal performance in data I/O with the data management capabilities of modern databases, scientific data are stored in files, while metadata identifying them are stored in a database.  The data files can only be modified in synchronization with the database, preventing concurrent access to data objects via locking mechanisms. The DMC software supports several database management systems of various complexity; the LFI DPC operates an Oracle 10g
database, which ensures good performance and stability.

The DMC provides a uniform Application Programming Interface (API) for Fortran, C, C++, and Java, hiding all specific database operations from the user, who is therefore not required to have database experience.  DMC data types are defined in the Data Definition Layer (DDL), which describes data and metadata structures.  The DDL supports inheritance of data types (e.g., a data type \texttt{polarized\_map} can be inherited from a data type \texttt{map}) as well as association of data types (i.e., one data type containing a reference to another).

In addition to the API, the DMC provides a Graphical User Interface (GUI), which supports user queries of the database and retrieval of information on the data. The GUI offers the user the ability to list the (meta-)data of specific objects and also to visualize the data in a simple way (although data can also be
exported to other powerful visualizing tools). The GUI also allows the display of history information on data objects, permitting the user to browse intermediate data products used in generating those objects, and the controlled deletion of data, observing dependencies of data types and maintaining the
history information for all remaining data.  For this, the DMC relies on additional metadata on the processing history of the data objects, which are generated by the Process Coordinator workflow engine (ProC).

\subsection{Pipeline management --- the ProC workflow engine}

The ProC is a generic engine to construct, verify, and execute computational workflows.  It comprises  computing modules and data flows between them.  The modules can be written in any programming language, provided they conform to simple I/O format requirements described in an XML module description file. These interface files specify the input and output objects, as well as the parameters of the individual programs, in terms of DMC data types as described above.

The ProC provides a \textit{Pipeline Editor} to support graphical construction of data processing workflows. It allows users to arrange and connect computing modules of a workflow in a clearly structured manner, and at the same time to configure the parameters of the algorithms used. It provides control structures
for data flow, for data object I/O and consistent parameter definition.

The execution of workflows is controlled by a forward chaining algorithm, which ensures that modules are executed as soon as all necessary data products and parameters are known. If the same version of a module has been executed with identical inputs and parameters, the ProC will skip the execution and use the data
product from the earlier execution for further processing.  The ProC maintains control of pipeline execution also on massive parallel computing environments. In the LFI implementation, the ProC communicates with the PBS (Portable Batch System) scheduling system to send jobs to the DPC cluster and to log their execution
status.

The ProC logs workflow executions on log files, which can also contain logging messages of the executed modules. Additionally, it creates so-called \textit{Pipeline-Run} and \textit{Module-Run} objects in the DMC, which are used to recover the generation history of data products (including versions of processing modules via MD5-sums). Besides the GUI, the ProC can also be executed from the command line.

At the LFI DPC, the ProC is used to execute the official pipeline producing \Planck\ data products.

\section{Discussion and conclusions}

We have described the status of the pipeline as it stands at the time of the ERCSC release and submission of the \Planck\ early papers.  All the algorithms run during this process have been verified, validated, and tested before launch and the start of operations using realistic simulations.  This allowed us to begin analyzing the data as soon as they were acquired from the first day of operations. The entire Level~1 pipeline suffered no significant problems, and all of the data were transformed efficiently from telemetry packets to timelines. At present, the Level 2 pipeline is capable of providing relative calibration to an overall statistical accuracy in the range 0.05--0.1\% and absolute calibration at around the 1\% level.  The beams are accurately characterised down to $-10$\,d\hbox{B}. We expect to improve many aspects in the near future.  Concerning the calibration, our intention is to reach the levels determined by the stability of the instrument.  For the beam reconstruction, our aim is to improve the characterisation of the far side lobes and to refine the entire beam reconstruction pipeline, with particular attention to polarization measurements.

\begin{acknowledgements}

  \Planck\ is too large a project to allow full acknowledgement of all
  contributions by individuals, institutions, industries, and funding
  agencies. The main entities involved in the mission operations are
  as follows. The European Space Agency operates the satellite via its
  Mission Operations Centre located at ESOC (Darmstadt, Germany) and
  coordinates scientific operations via the Planck Science Office
  located at ESAC (Madrid, Spain). Two Consortia, comprising around 50
  scientific institutes within Europe, the USA, and Canada, and funded
  by agencies from the participating countries, developed the
  scientific instruments LFI and HFI, and continue to operate them via
  Instrument Operations Teams located in Trieste (Italy) and Orsay
  (France). The Consortia are also responsible for scientific
  processing of the acquired data. The Consortia are led by the
  Principal Investigators: J.L. Puget in France for HFI (funded
  principally by CNES and CNRS/INSU-IN2P3) and N. Mandolesi in Italy
  for LFI(funded principally via ASI). NASA US Planck Project,
  based at JPL and involving scientists at many US institutions,
  contributes significantly to the efforts of these two Consortia. The
  author list for this paper has been selected by the Planck Science
  Team, and is composed of individuals from all of the above entities
  who have made multi-year contributions to the development of the
  mission. It does not pretend to be inclusive of all contributions.
  The \Planck -LFI project is developed by an International Consortium
  lead by Italy and involving Canada, Finland, Germany, Norway, Spain,
  Switzerland, UK, USA. The Italian contribution to \Planck\ is
  supported by the Italian Space Agency (ASI) and INAF.  This work was
  supported by the Academy of Finland grants 121703 and 121962. We
  thank the DEISA Consortium (www.deisa.eu), co-funded through the EU
  FP6 project RI-031513 and the FP7 project RI-222919, for support
  within the DEISA Virtual Community Support Initiative.  We thank CSC
  -- IT Center for Science Ltd (Finland) for computational resources.
  We acknowledge financial support provided by the Spanish Ministerio
  de Ciencia e Innovaci{\~o}n through the Plan Nacional del Espacio y
  Plan Nacional de Astronomia y Astrofisica.  We acknowledge The
  Max Planck Institute for Astrophysics Planck Analysis Centre (MPAC)
  is funded by the Space Agency of the German Aerospace Center (DLR)
  under grant 50OP0901 with resources of the German Federal Ministry
  of Economics and Technology, and by the Max Planck Society. This
  work has made use of the Planck satellite simulation package
  (Level-S), which is assembled by the Max Planck Institute for
  Astrophysics Planck Analysis Centre (MPAC) \cite{reinecke2006}. We
  acknowledge financial support provided by the National Energy
  Research Scientific Computing Center, which is supported by the
  Office of Science of the U.S. Department of Energy under Contract
  No. DE-AC02-05CH11231. Some of the results in this paper have been
  derived using the HEALPix package \cite{gorski2005}.
A description of the \Planck\ Collaboration and a list of its
members,  indicating which technical or scientific activities they
  have been involved in, can be found at
  \burl{http://www.rssd.esa.int/index.php?project=PLANCK&page=Planck_Collaboration}.

\end{acknowledgements}

\allearlypapers

\bibliographystyle{aa}
\bibliography{Planck_bib}

\raggedright

\end{document}